\documentclass[a4paper,fleqn,usenatbib]{mnras}

\usepackage{savesym}
\usepackage{amsmath}    
\savesymbol{iint}
\usepackage{txfonts}

\usepackage[T1]{fontenc}
\usepackage{ae,aecompl}

\usepackage{graphicx}	
\usepackage{float}
\usepackage{amssymb}	
\usepackage{caption}

\title[Resolving pulsar spin-down states]{Resolving discrete pulsar spin-down states with current and future instrumentation}
\author[Shaw, B. et al.]{
B. Shaw,$^{1}$\thanks{E-mail: benjamin.shaw@manchester.ac.uk}
B. W. Stappers,$^{1}$
and P. Weltevrede$^{1}$
\\
$^{1}$Jodrell Bank Centre for Astrophysics, The University of Manchester, Manchester, M13 9PL, United Kingdom\\
}

\pubyear{2018}

\begin{document}
\label{firstpage}
\pagerange{\pageref{firstpage}--\pageref{lastpage}}
\maketitle

\begin{abstract}
An understanding of pulsar timing noise offers the potential to improve the timing precision of a large number of pulsars as well as facilitating our understanding of pulsar magnetospheres.  For some sources, timing noise is attributable to a pulsar switching between two different spin-down rates $(\dot{\nu})$. Such transitions may be common but difficult to resolve using current techniques.  In this work, we use simulations of $\dot{\nu}$-variable pulsars to investigate the likelihood of resolving individual $\dot{\nu}$ transitions. We inject step-changes in the value of $\dot{\nu}$ with a wide range of amplitudes and switching timescales. We then attempt to redetect these transitions using standard pulsar timing techniques. The pulse arrival-time precision and the observing cadence are varied. Limits on $\dot{\nu}$ detectability based on the effects such transitions have on the timing residuals are derived. With the typical cadences and timing precision of current timing programs, we find we are insensitive to a large region of $\Delta \dot{\nu}$ parameter space which encompasses small, short timescale switches. We find, where the rotation and emission states are correlated, that using changes to the pulse shape to estimate $\dot{\nu}$ transition epochs, can improve detectability in certain scenarios.  The effects of cadence on $\Delta \dot{\nu}$ detectability are discussed and we make comparisons with a known population of intermittent and mode-switching pulsars. We conclude that for short timescale, small switches, cadence should not be compromised when new generations of ultra-sensitive radio telescopes are online.   
\end{abstract}




\section{Introduction}

The clock-like stability with which pulsars rotate has led to their adoption as precision timing tools.  Their large and stable moments of inertia yield precise intervals between the detection of successive pulses of radiation observed as the beam crosses the line of sight to the Earth.  In some cases, a simple model containing the star's spin frequency and first time derivative, the spin-down, can adequately predict times of arrival (TOAs) of pulses resulting in timing residuals that are normally distributed about zero with uncertainty on the TOAs dominated by receiver noise.  However, not all pulsars show such stability and subtracting a pulsar's timing model from the data, often reveals significant structure in the residuals, indicating that the pulsar's rotational properties, emission characteristics, or both are insufficiently modelled. 

There are two main classes of pulsar timing irregularity that interrupt otherwise stable rotation. \emph{Glitches}, which are more commonly seen in young pulsars, are characterised as a step increase in the rotation frequency (${\nu}$), often coincident with a change in the spin-down ($\dot{\nu}$).  In some cases, the pulsar returns to its initial, pre-glitch rotation frequency after some relaxation period. The process responsible for glitches is not well understood but is thought to be due to vortices in the superfluid interior, coupling to the inner crust of the neutron star, resulting in the rapid transfer of angular momentum to the surface  \citep{ai75}.   

The second type of irregularity is characterised by an apparent quasi-random wandering of the timing residuals relative to a simple spin-down model, often revealing some periodicity in the structure over long datasets \citep{hlk10}.  This phenomenon has become known as \emph{timing noise} and attempts to explain the underlying mechanism are numerous.  For example, it has been attributed to random noise processes intrinsic to the pulsar  \citep{ch80}, the unmodelled presence of planetary systems \citep{cor93}, interactions with asteroids, \citep{cs06}, heliospheric effects \citep{sfal97}, free precession \citep{pt96} and precession caused by the presence of a fossil accretion disk \citep{qxxw03}.  \cite{hlk10} studied the timing irregularities in 366 pulsars observed over more than 10 years with the Lovell telescope at Jodrell Bank Observatory.  They found strong correlations between the amplitude of timing noise and the spin-down rate and note that timing noise in younger pulsars appears to be dominated by glitch recoveries.  In many older pulsars however, a time-correlated structure is seen in the residuals.  This structure is often revealed to be quasi-periodic over long timescales, with the minima and maxima having different radii of curvature.

Many pulsars exhibit variability in their radio emission properties.  \emph{Mode-switching}, is an abrupt change in the average pulse profile in which it appears to switch between two or more well-defined shapes \citep{bmsh82}.  Emission modes may last from a few pulse periods up to many hours or days.  Pulse \emph{nulling}, is characterised by the sudden cessation of detectable radio emission from a pulsar, occurring on similar timescales to mode-switching \citep{bac70}.  In some cases, the pulsar's \emph{nulling fraction}, defined as the fraction of time spent in the null state, is as high as 95 per cent \citep{wmj07}.  A number of scenarios have been proposed to explain nulling behaviour.  Pulsars may simply cease to emit or may be mode-switching to an emission state that is below the detection threshold of the observing system \citep{elg+05} implying that, in some cases, nulling may merely be an extreme case of mode-switching.  Alternatively, the geometry of the pulsar emission region could change such that the beam contracts and no longer crosses Earth's line-of-sight \citep{tim10}.

The intermittent pulsar PSR B1931+24 ($\nu = 1.229$ Hz) switches abruptly (over < 10 seconds) between a short (5 - 10 days) \emph{on} mode and a longer \emph{off} mode (25 - 40 days). Its rotational properties were examined by \cite{klo+06} where they achieved a factor of 20 reduction in the amplitude of the timing residuals by using a timing solution in which the pulsar spins down 50 per cent slower when the pulsar is \emph{off}.  As the radio emission accounts for only a small fraction of the total spin-down power of the pulsar, the fact that the radio output is correlated with a spin-down variation suggests that the change to the emission is symptomatic of a process that is global rather than local to the emitting region.  One explanation for this is that a reduction or cessation of currents in the magnetosphere abates the radio emission and reduces the braking torques on the pulsar, resulting in a slower spin-down when the pulsar is \emph{off} (\cite{tim10} and \cite{lst12}). However, it remains unclear why the magnetospheric current distribution would undergo such changes.

The relation between emission properties and spin-down was further investigated by \cite{lhk+10}.  They examined the spin-down behaviour of 17 pulsars whose residuals are dominated by timing noise and found that the spin-down rate was quasi-periodically switching between two well-defined values.  They showed that for six of these pulsars, changes to the profile shape clearly traced changes to the spin-down rate, suggesting that sudden changes to the magnetospheric current distribution are responsible for changes to the intensity and/or shape of the pulse profiles. 

A timing solution which accounts for every rotation of a pulsar in the presence of an unstable spin-down must include terms for the epochs of such changes as well as their sizes.  Where spin-down changes are not instantaneous but occur over many rotations of the pulsar, appropriate higher order spin-down rate derivatives must also be included.  The precision with which changes to otherwise stable rotation can be resolved is dependent on a number of factors both intrinsic (such as the amplitude of the switches and the time interval between them) and extrinsic (such as TOA precision and length of time between successive observations) to the pulsar.  

An unmodelled change to the rotational properties of a pulsar will cause post-event TOAs to deviate from a best-fit model which accurately describes the pre-event TOAs.  The precision with which a TOA ($\sigma_{\phi}$) can be attained is related to the sensitivity of the observing system. Were a source monitored continuously, an unmodelled change to a pulsar's rotation would be discernible closer in time to the true event epoch where $\sigma_{\phi}$ is smaller (i.e. for a more sensitive telescope). In such a case, the TOA precision would be the primary source of uncertainty on the transition epoch.   Realistically, finite observing cadence imposes additional limitations on constraining the epochs of such timing events, independent of TOA precision. The next generation of radio telescopes, such as the Five hundred metre Aperture Spherical Telescope (FAST) \citep{srk+09} and the Square Kilometre Array (SKA) \citep{sks+08} will boast unprecedented sensitivity on completion.  However, the cadence at which any given pulsar is observed will be limited by the large number of pulsars that are routinely timed as well as the fraction of telescope time allocated to pulsar timing.  The SKA will be able to increase the cadence by splitting the array into sub-arrays, but in these modes sensitivity is forfeited in favour of cadence and so a trade-off exists between the number of TOAs attainable and the precision with which they are attained. Other new facilities, such as UTMOST \citep{bjf+17} and CHIME \citep{n17} will be capable of daily monitoring of a large number of pulsars, when online. 


In this work we undertake a study to understand the detection limitations of spin-down changes in pulsar timing residuals. We inject instantaneous step changes of the frequency derivative into simulated TOAs and attempt to detect these changes using standard pulsar timing techniques (see \cite{ld13} for a review).  We vary properties both intrinsic and extrinsic to the pulsar and show how these variations affect the detectability of spin-down changes.  These properties include observing cadence, TOA precision, the time a pulsar spends in any one spin-down mode (the switching timescale) and the magnitude of the change in frequency derivative.    In addition, we take a less empirical approach and derive detectability limits based on the effects that sets of $\dot{\nu}$ transitions have on the timing residuals of a pulsar. 

\section{Limits on detectable changes to emission and $\dot{\nu}$} \label{limits}
\begin{figure*}
   \includegraphics[width=\columnwidth]{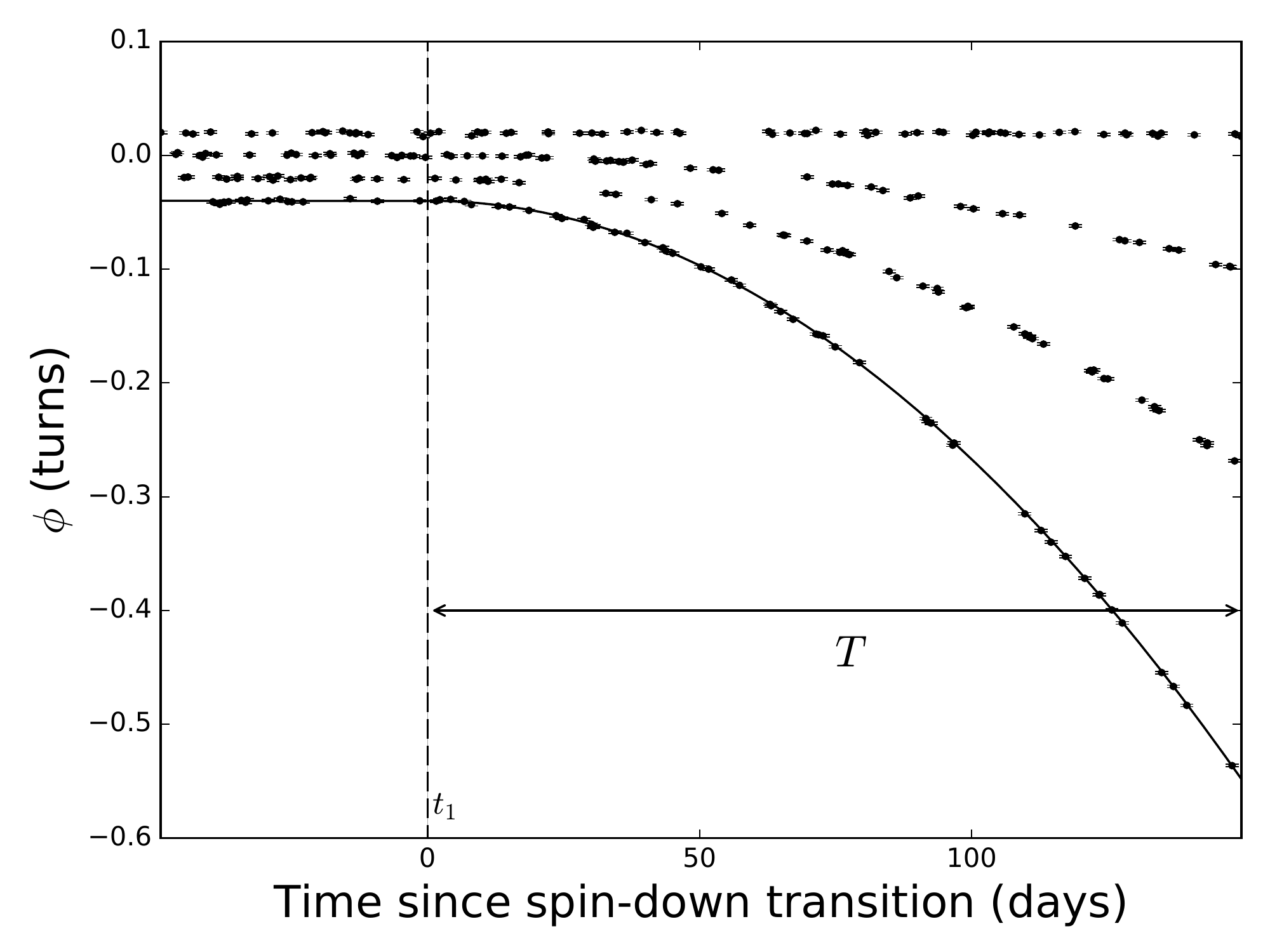}
   \includegraphics[width=\columnwidth]{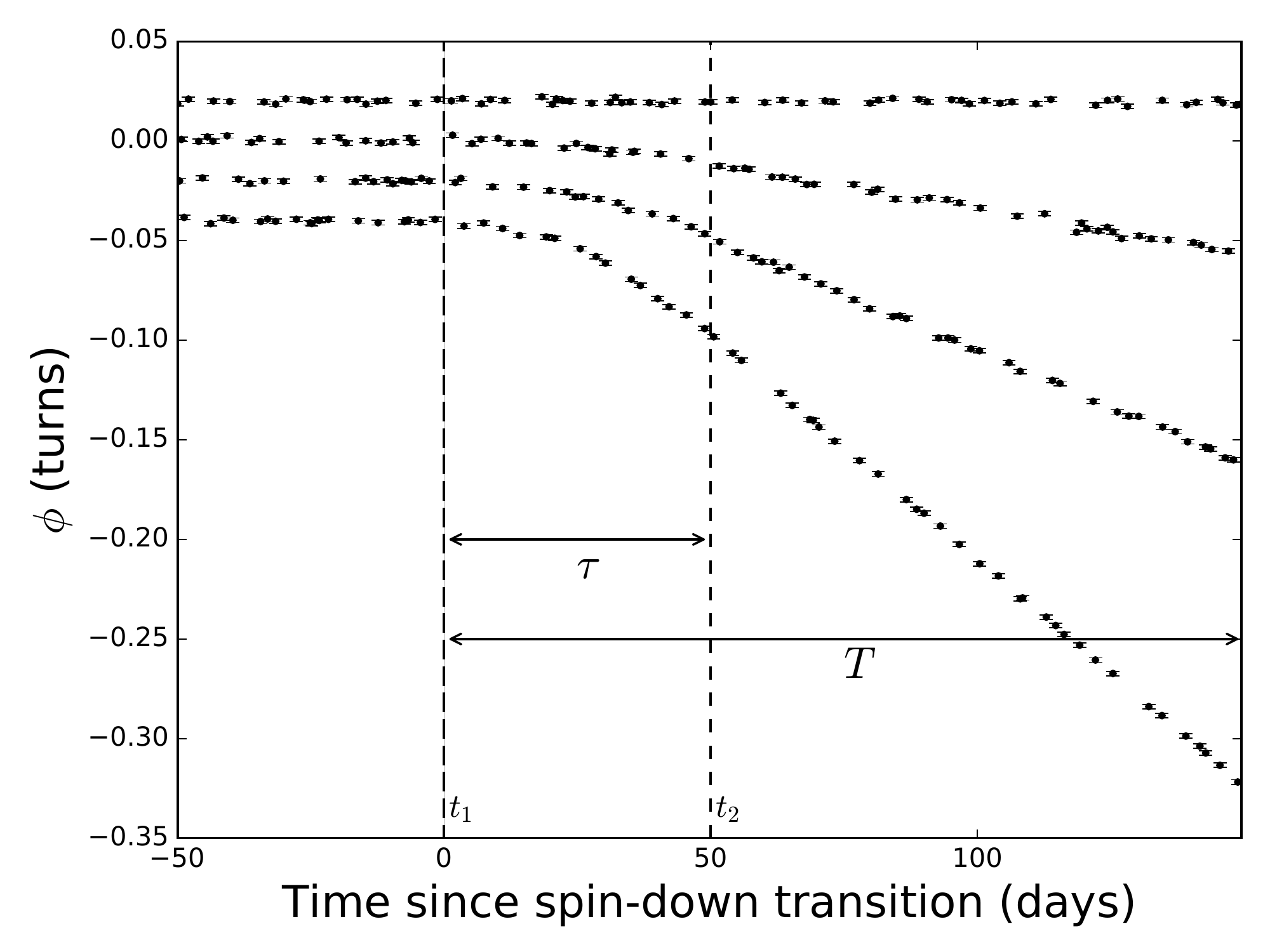}
   \caption{Left: 200 days of simulated residuals from a pulsar undergoing unmodelled spin-down increases of 0.1 per cent, 10 per cent, 25 per cent and 50 per cent of the initial $\dot{\nu}$ from top to bottom respectively. The quantity $\Delta \phi$ is the fraction of one pulsar rotation by which TOAs are arriving early or late.  The initial rotational parameters are based on the intermittent pulsar PSR B1931+21 (see Table \ref{properties}). The epoch of the $\dot{\nu}$-transition is denoted by the dashed vertical line.  White noise with an amplitude of 1 ms is added to each of the randomly spaced TOAs. A small vertical offset has been added to each set of residuals so they are visually distinguishable. The solid line is the analytic function given by equation \ref{phaseoffset} for a $\Delta \dot{\nu} = 50$ per cent occurring at the same epoch. Right: The $\dot{\nu}$-transitions occur at the same epoch $t_1$ and have the same amplitudes as those in the left panel. In this case, a second reverse transition of the same amplitude occurs after $\tau$ days at $t_2$ (the rightmost dotted line). Following this transition, the pulsar has reverted to the same value of $\dot{\nu}$ which it started prior to $t_1$.}
   \label{residualsforunmodelledchange}
\end{figure*}

In this section we derive an analytical limit on the sensitivity of changes to $\dot{\nu}$ in a pulsar that can be timed to an average TOA precision $\bar{\sigma}_{\phi}$.  In cases where the epochs of $\dot{\nu}$-changes can be inferred from mode-switching behaviour, it is equally important to understand how well profile variability can be constrained in time. 

In principle, a telescope that observes a source continuously, would reveal profile changes on a pulse-to-pulse timescale, provided that a) sufficient signal to noise (S/N) is available and b) the pulses are consecutively stable.  Although the first of these conditions is met for a handful of pulsars, individual pulses are successively variable and so in the limit of infinite cadence and sensitivity,  the resolution of the epoch of a change to the profile shape is limited to the timescale over which a stable integrated pulse profile is formed - typically many hundreds of pulses.  In reality however, pulsars are not continuously observed.  For example, the majority of pulsars timed at Jodrell Bank Observatory are observed for up to $\sim$700 s approximately twice per month. Unless a clear profile switch occurs during an observation, these limitations allow us to constrain switch epochs only as far as the timescale between individual observations.

The precision to which the epochs of abrupt changes to the rotational behaviour of a pulsar, such as glitches and $\dot{\nu}$ transitions, can be determined is also subject to limitations.  A pulsar's spin frequency at some epoch $t$ can be expressed as a Taylor expansion of its frequency derivatives, in the form

\begin{equation}
    \label{frequencytaylor} \nu = \nu_0 + \dot{\nu}_{0}(t-t_0) + \frac{1}{2} \ddot{\nu}_{0}(t-t_0)^2 + ......\mathrm{,}
\end{equation}

\noindent where $\nu_0$ is the spin-frequency at some reference epoch $t_0$, and $\dot{\nu}_{0}$ and $\ddot{\nu}_0$ are the first and second order derivatives of the spin-frequency at $t_0$.  In pulsars where a correlation exists between the frequency derivative and the profile shape, we can in principle constrain the epoch at which a change to the spin-down occurs by monitoring the pulse shape. This then allows us to fit timing residuals with a model that contains a term for a spin-down transition at or near that epoch, thereby subtracting from the residuals the timing noise caused by such events.  If a pulsar's spin evolution is well modelled, the timing residuals will be normally distributed about zero with an RMS that is consistent with the individual TOA errors. If an abrupt change to the spin occurs at $t_1$ and is not included in the timing model, then the expected amplitude of a timing residual $\phi$ at epoch $t$, is given by,

\begin{equation}
\label{phaseoffset} \phi = - \Delta \nu (t - t_1) - \Delta \dot{\nu} \frac{(t - t_1)^2}{2} + ... \mathrm{,} \; \; (t > t_1) \mathrm{,}
\end{equation}

\noindent where $\Delta \nu$ and $\Delta \dot{\nu}$ are changes to the magnitudes of $\nu$ and $\dot{\nu}$ respectively.  For all calculations throughout this work we define a $\dot{\nu}$-transition as a step-change in the spin frequency derivative $\dot{\nu}$ of a pulsar with no associated instantaneous change in the spin frequency $\nu$ allowing us to discriminate between spin-down changes and glitches.  We assume that all changes to $\dot{\nu}$ are instantaneous with no smooth gradient between the spin-down rates in each mode. 

\subsection{The effect of a single transition}

In Figure \ref{residualsforunmodelledchange} (left panel) we show the effect of a sudden increase in the frequency derivative at an epoch denoted by the dashed vertical line. After this the initial timing model no longer accurately describes its rotational parameters as TOAs are arriving increasingly earlier than predicted. In order for the effects of a switch in the frequency derivative to become discernible, the deviation of the residuals must be larger than the weighted RMS of the residuals before the transition epoch by the time the dataset has ended after time $T$.  The upper set of residuals in Figure \ref{residualsforunmodelledchange} indicates that a 0.1 per cent change in $\dot{\nu}_0$ is undetectable in the time span shown.

In principle we can use Equation \ref{phaseoffset} to determine whether or not a transition of amplitude $\Delta \dot{\nu}$ will have a significant effect on the timing residuals within some timespan, by computing the expected value for the $\chi^2$ of a dataset of timing residuals which contains such a transition. A change in $\dot{\nu}$ with no associated instantaneous change in $\nu$ (and no changes to other higher order terms) allows us to express Equation \ref{phaseoffset} as 

\begin{equation}
    \label{phase} \phi(t) = \frac{|\Delta \dot{\nu}|}{2} (t - t_1)^2.
\end{equation}
    
\noindent For a transition of amplitude $|\Delta \dot{\nu}|$ in a pulsar observed with cadence $C$ for time $T$ after the transition, one can predict the expected $\chi^2$ as follows (see Appendix A),

\begin{equation}
    \label{chi2_1} \langle \chi^2 \rangle = \frac{|\Delta \dot{\nu}|^2 T^5}{20 \bar{\sigma}^2_{\phi} C}.
\end{equation}

\noindent $T$, $C$ and $\bar{\sigma}$ are expressed in seconds and $\Delta \dot{\nu}$ in Hz/s. The number of data points after the event (related to the degrees of freedom) is $T / C$.  By establishing a significance level of 99.7 per cent, we can calculate a critical value of the $\chi^2$ for $T/C$ post-event data points.  We refer to this critical value as $\chi^2_\mathrm{crit}$.  If $\langle \chi^2 \rangle > \chi^2_\mathrm{crit}$, the amplitudes of the residuals, due to the transition, have sufficiently departed from the timing model's predicted values that the model can be, in principle, significantly improved by inclusion of the transition parameters. In other words the transition is \emph{potentially} detectable. 

\begin{figure*}
    \includegraphics[width=\columnwidth]{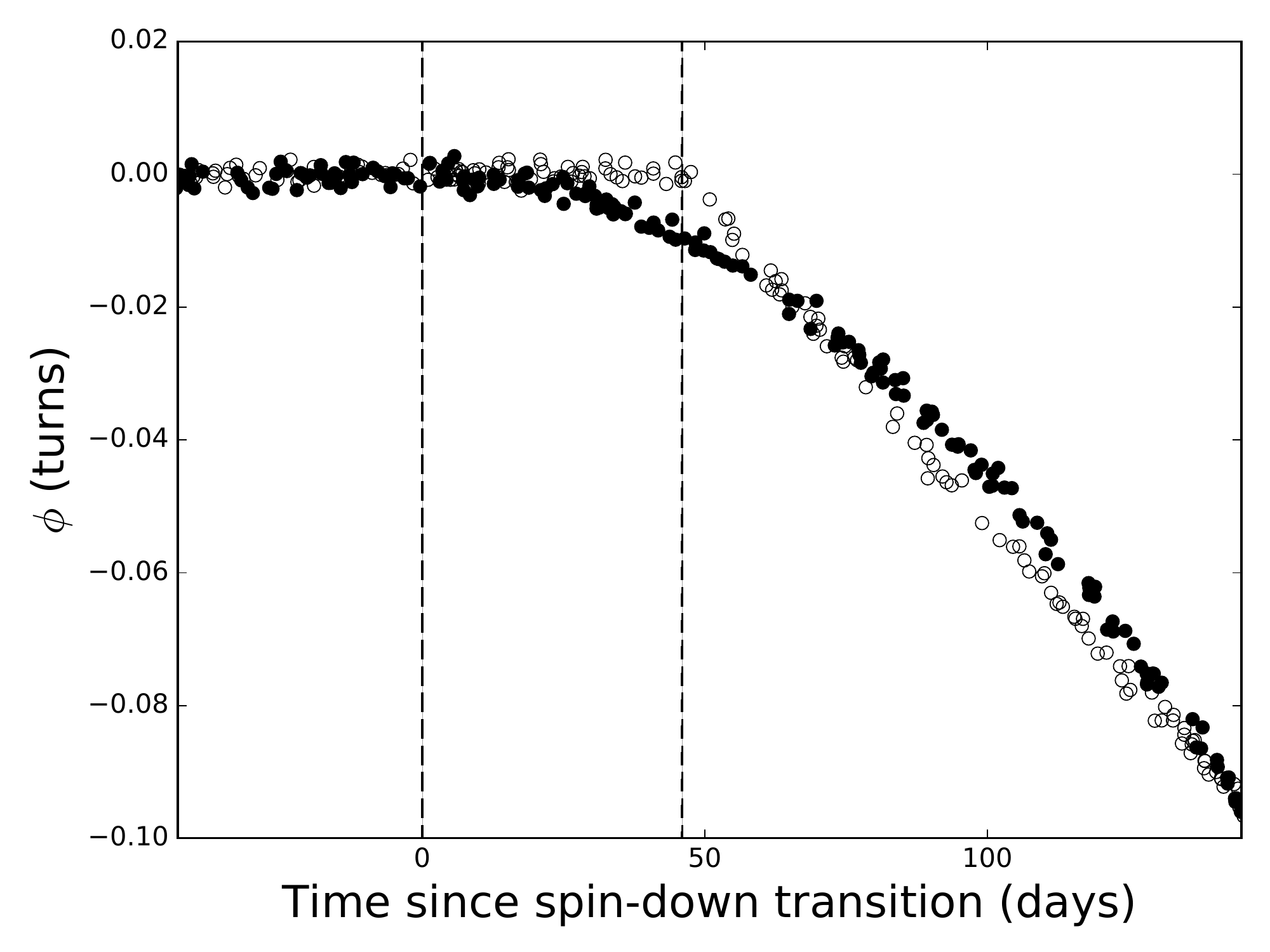}
    \includegraphics[width=\columnwidth]{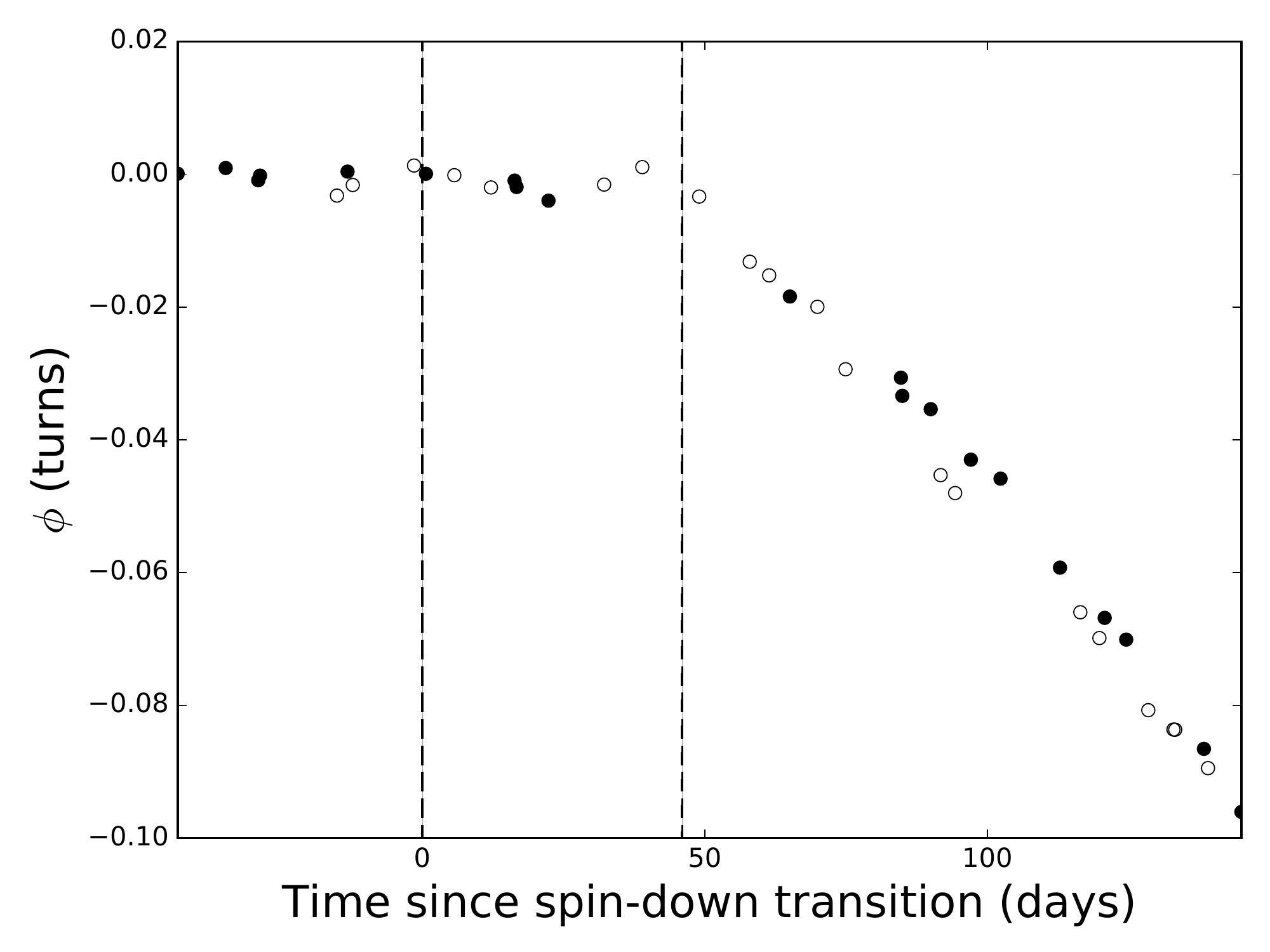}
    \caption{An example of a glitch signature and a spin-down change in the same pulsar's pre-fit timing residuals. Timing residuals are simulated over a timespan of 200 days. The pulsar begins with $\nu$ and $\dot{\nu}$ of 1.23 Hz and $1.22 \times 10^{-14}$ Hz s\textsuperscript{-1} respectively. In one case (filled circles), a 10 per cent change in $\dot{\nu}$ occurs at the epoch denoted by the left-most dashed line.  In the other case (open circles) a glitch occurs (right-most dashed line) with a $\Delta \nu/\nu \sim 1 \times 10^{-8}$.  The left panel shows a comparison of the signatures of these events when the pulsar is observed daily (200 TOAs). In the right panel the pulsar is observed once per 10 days (20 TOAs), in which case the two signatures become indistinguishable in the time-span shown.}
    \label{glitchswitch}
\end{figure*}

Observing cadence can play a role in how residuals in the presence of timing irregularities are interpreted (see Figure \ref{glitchswitch}).  Low and irregular cadence observations near a $\dot{\nu}$ transition epoch may result in the residuals mimicking a glitch thereby leading one to misinterpret the behaviour of the pulsar.  In the left panel of Figure \ref{glitchswitch} the pulsar is observed on average once per day for 200 days.  A $\dot{\nu}$-transition occurs at 0 days (filled circles) and this is clearly distinguishable from the signature of a glitch (modelled as a step change in $\nu$) which occurs at $\sim$45 days (open circles). However, when observed with an average cadence of once per 10 days over the same timespan (right panel), the nature of the event is less clear as the glitch signature is indistinguishable from that of the $\dot{\nu}$-transition. Simulations could, in principle, determine whether glitches of sufficient size become distinguishable from $\dot{\nu}$ transitions, independent of cadence, however this is beyond the scope of this work.

\subsection{The effect of two transitions} \label{two_trans_limit}

The right panel of Figure \ref{residualsforunmodelledchange} shows the effect of two transitions in a dataset, the second of which (at $t_2$) is the opposite sign of the first (at $t_1$) and occurs after some time $\tau$ after $t_1$. In this case, the quadratic signature described by Equation \ref{phase} is only valid between the two transitions.  Between $t_2$ and the end of the dataset, the residuals follow a linear function with a non-zero gradient due to the spin-frequency $\nu$ at $t_2$ being discrepant from the value predicted by the model. At $t_2$ the discrepancy in $\nu$ has a value $\Delta \nu = |\Delta \dot{\nu}| \tau$.  The expected value for the $\chi^2$ due to two transitions (see Appendix A) is given by


\begin{equation}
    \label{chi2_2} \langle \chi^2 \rangle = \frac{|\Delta \dot{\nu}|^2 \tau^2}{60 \bar{\sigma}^2_{\phi} C} \left( -2\tau^3 + 15\tau^2 T - 30\tau T^2 + 20 T^3 \right), \quad (T > \tau),
\end{equation}

\noindent where $T$ is the time between the first event and the end of the dataset.  As before, the expected number of data points after $t_1$ is $T / C$.  Note that when $T = \tau$ (i.e., no observations take place after $t_2$) we recover Equation \ref{chi2_1}.

\begin{figure}
    \includegraphics[width=\columnwidth]{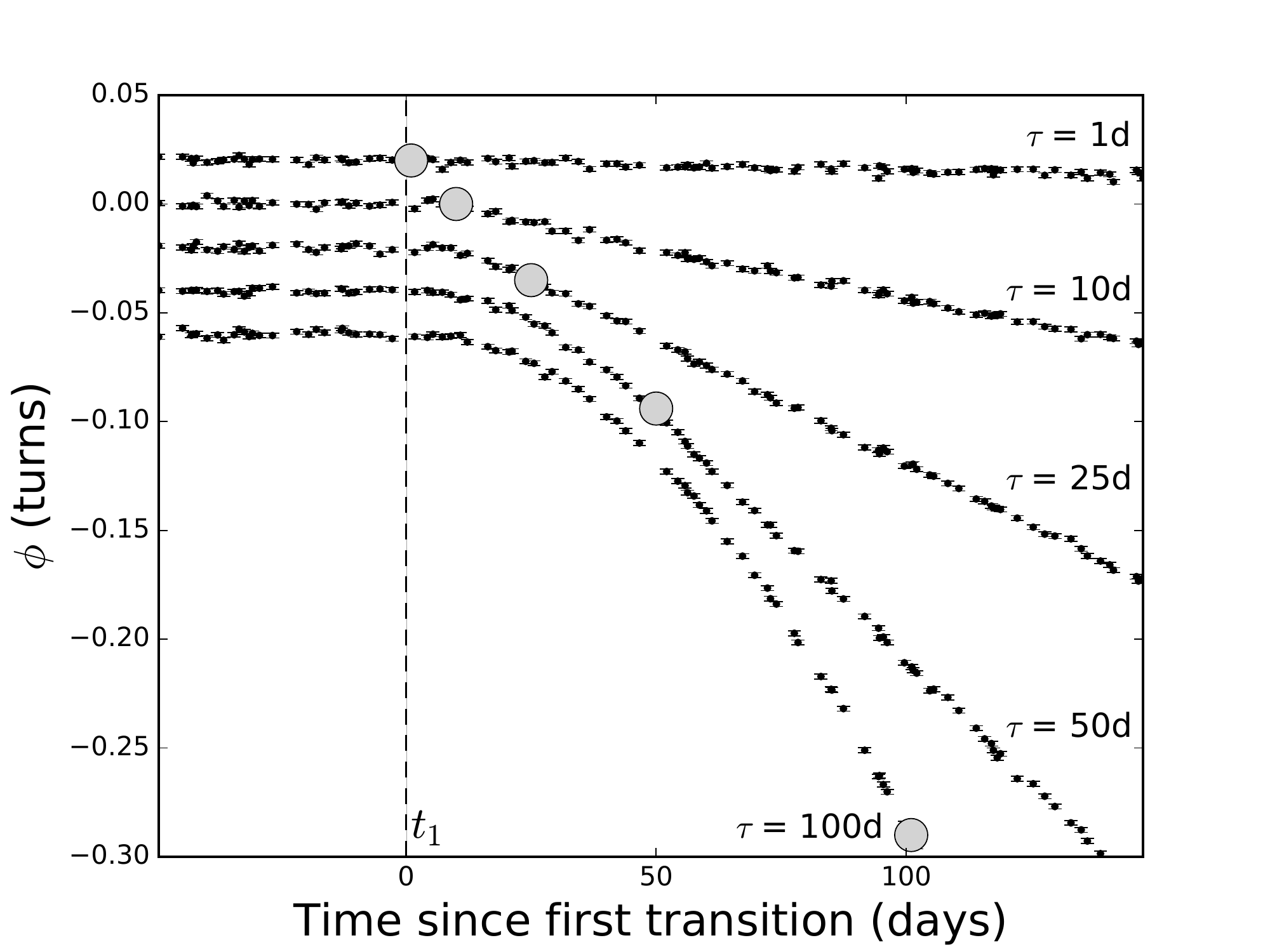}
    \caption{The effect of two transitions on 200 days of timing residuals for a $\dot{\nu}$-variable pulsar. The pulsar's initial rotation parameters are the same as those in Figure \ref{glitchswitch}. After $t_1$ days (vertical dashed line), the pulsar undergoes a 50 per cent change in the value of $\dot{\nu}$ and switches back to the initial $\dot{\nu}$ at the $t_2$ epochs denoted by the grey circles. The durations of the new spin-down states are shown next to each set of residuals. For each dataset there are 100 TOAs corresponding to an average cadence $C = 2$ days.  As Figure \ref{residualsforunmodelledchange} otherwise.}
    \label{variable_tau}
\end{figure}

There are scenarios in which $\tau$ is sufficiently small as to not cause a large departure in the residuals by the time $t_2$. However, $\Delta \dot{\nu}$ and/or $T - \tau$ may be large enough to cause a considerable post-$t_2$ departure. In such cases, the residuals may suggest a glitch has occurred rather than a closely spaced pair of $\dot{\nu}$ transitions.  This is illustrated in Figure \ref{variable_tau} which shows the effect of various values of $\tau$ on a 200 day dataset in which the first transition takes place after $t_1$ days.  For the highest values of $\tau$ ($\tau = 100$ and 50 days),  the post-$t_1$ residuals show the unambiguous quadratic signature of a $\dot{\nu}$-change followed by the expected linear signature post-$t_2$, due to the resultant period discrepancy. Conversely, for $\tau = 10$ and 25 days, the residuals indicate the possibility of a permanent change to the period and could be adequately modelled as such. For $\tau = 1$ day, continuous spin parameters are sufficient to model the pulsar's behaviour over the 200 days.  Therefore one may define two limits - one based on all of the time covered after $t_1$ (Equation \ref{chi2_2}) and one based only on the time covered between the two transitions. The second of these indicates the minimum $\tau$ for which a transition becomes discernible by the time $t_2$ with no contribution from any post $t_2$ data points. In this case, the number of data points from which $\chi^2_{\mathrm{crit}}$ is computed is $\tau/C$ as opposed to $T/C$ which is only applicable when data points after $t=\tau$ are considered.  (See Appendix A for details).

\subsection{The effect of three transitions} \label{three_trans_limit}

We can extend Equation \ref{chi2_2} to describe three transitions occurring in a dataset such that $\Delta \dot{\nu} (t_1) = \Delta \dot{\nu}  (t_3) = -\Delta \dot{\nu} (t_2)$ and each event is separated from the next by $\tau$. In this case, at $t_3$, the residuals adopt a quadratic signature once again, as the $\dot{\nu}$ value is not correct with respect to the timing model. Therefore the extension of Equation \ref{chi2_2} to three transitions (see Appendix A) is 


\begin{equation}
\begin{aligned}
    \label{chi2_3} \langle \chi^2 \rangle = & \frac{1}{\bar{\sigma}^2_{\phi} C} \Bigg( \frac{17}{15} |\Delta \dot{\nu}|^2 \tau^5 + \frac{9}{4}  |\Delta \dot{\nu}|^2 \tau^2 (T - 2\tau)  \\
                                            &  + \frac{5}{6} |\Delta \dot{\nu}|^2 \tau^2 (T^3 - 8\tau^3) - \frac{3}{2} |\Delta \dot{\nu}|^2 \tau^3 (T^2 - 4\tau^2)  \\
                                            & - |\Delta \dot{\nu}|^2 \tau (0.25T - 4\tau^4) + \frac{|\Delta \dot{\nu}|^2}{20} (T^5 - 32\tau^5) \Bigg),
                                            &  (T > 2\tau)
\end{aligned}
\end{equation}

If $T - t_3 \gg \tau$, the triple-transitioning behaviour could be modelled by the inclusion of a single $\Delta \dot{\nu}$ event in the timing solution. In other words, three close transitions can be modelled as one.  It is possible for $\tau$ to be extremely short such that the residuals between $t_1$ and $t_3$ are not affected but the rapid rise in the amplitude of the residuals after $t_3$ yields a large value of $\langle \chi^2 \rangle$. For this reason, for low $\tau$ Equation \ref{chi2_3} becomes dominated by the last transition only and in the limit that $\tau \rightarrow 0$, Equation \ref{chi2_1} is recovered.  

It should be noted that the above procedures for predicting $\langle \chi^2 \rangle$ values simply offer a method for determining whether or not one or more $\dot{\nu}$ transitions will affect the timing residuals within some timescale.  Where residuals are affected by $\dot{\nu}$ transitions, there is clearly scope for an improved timing solution but, especially where $\Delta \dot{\nu}$ and/or $\tau$ are small, it is not necessarily the case that individual transitions are resolvable.

\section{Simulating $\dot{\nu}$ transitions}

To study the detection limitations of abrupt frequency derivative changes, we simulate timing residuals of pulsars which undergo a number of $\dot{\nu}$-changes.  We then attempt to fit for the transition parameters. The fact that a $\dot{\nu}$ change is detectable is predicated on the fact that, at the epoch of the change, the timing residuals will quickly deviate from a model that accurately described the pulsar rotation prior to the event, thereby increasing the $\langle \chi^2 \rangle$ of the timing residuals. A simulated set of timing residuals is considered to contain detectable spin-down changes if the inclusion of transition parameters in the timing solution results in a clear and significant improvement in the $\chi^2$ of the timing residuals.

\subsection{The model}
\label{model}

The first step in the simulation of a pulsar that undergoes $\dot{\nu}$ variations is the selection of a time-span over which the TOAs are to be generated.  In this work, datasets lasting 500 days are created in which we simulate observing cadences ranging from 1 to 28 days.  In these simulations, either two or three epochs, $t_i$, are selected at which to place spin-down transitions.  The transition epochs $t_i$ are randomly selected within the central 200 days, ensuring at least 150 days elapses between the limits of the dataset and the nearest transition.  This is to ensure that the pulsar spends sufficient time in its initial and final spin-down modes.   After the earliest selected transition epoch $t_1$, we introduce an instantaneous increase to the spin-frequency derivative. This causes the value of $\dot{\nu}$, after $t_1$ to assume the value $\dot{\nu}_\mathrm{0} + \Delta \dot{\nu}$.   In other words, at $t_1$, the pulsar switches from its initial spin-down state to one in which it is spinning down more slowly.  The second spin-down transition at $t_2$ is always of equal magnitude to the first, though opposite in direction. This means that after the second switch, the pulsar returns to its initial spin-down rate. Where simulated, a transition at $t_3$ has the same magnitude and direction as the transition at $t_1$.

The size of the switch, $\Delta \dot{\nu}$ is also a randomly selected value ranging from 0 to 100  per cent of the initial $\dot{\nu}$.  We note that it is the size, rather than the direction of the switch which causes TOAs to depart from a pulsar's timing solution. If the pulsar switches to a smaller frequency derivative, the residuals will simply become more positive with time, rather than increasingly negative as shown in Figure \ref{residualsforunmodelledchange}.  When three transitions, $t_i$, are simulated, they are spaced such that the time between $t_1$ and $t_2$ is equal to the time between $t_2$ and $t_3$.  The resulting model is of a pulsar whose spin-down rate is switching, instantaneously, between two well defined values, 2 or 3 times within a dataset.  The selected parameters are compiled into an ephemeris. For the pulsar's initial rotation parameters we arbitrarily use those of PSR B1931+24, though in principle any initial $\nu$ and $\dot{\nu}$ are equivalent. We assume there is no other source of timing noise besides the introduced spin-down transitions.

\begin{figure*}
   \includegraphics[width=2\columnwidth]{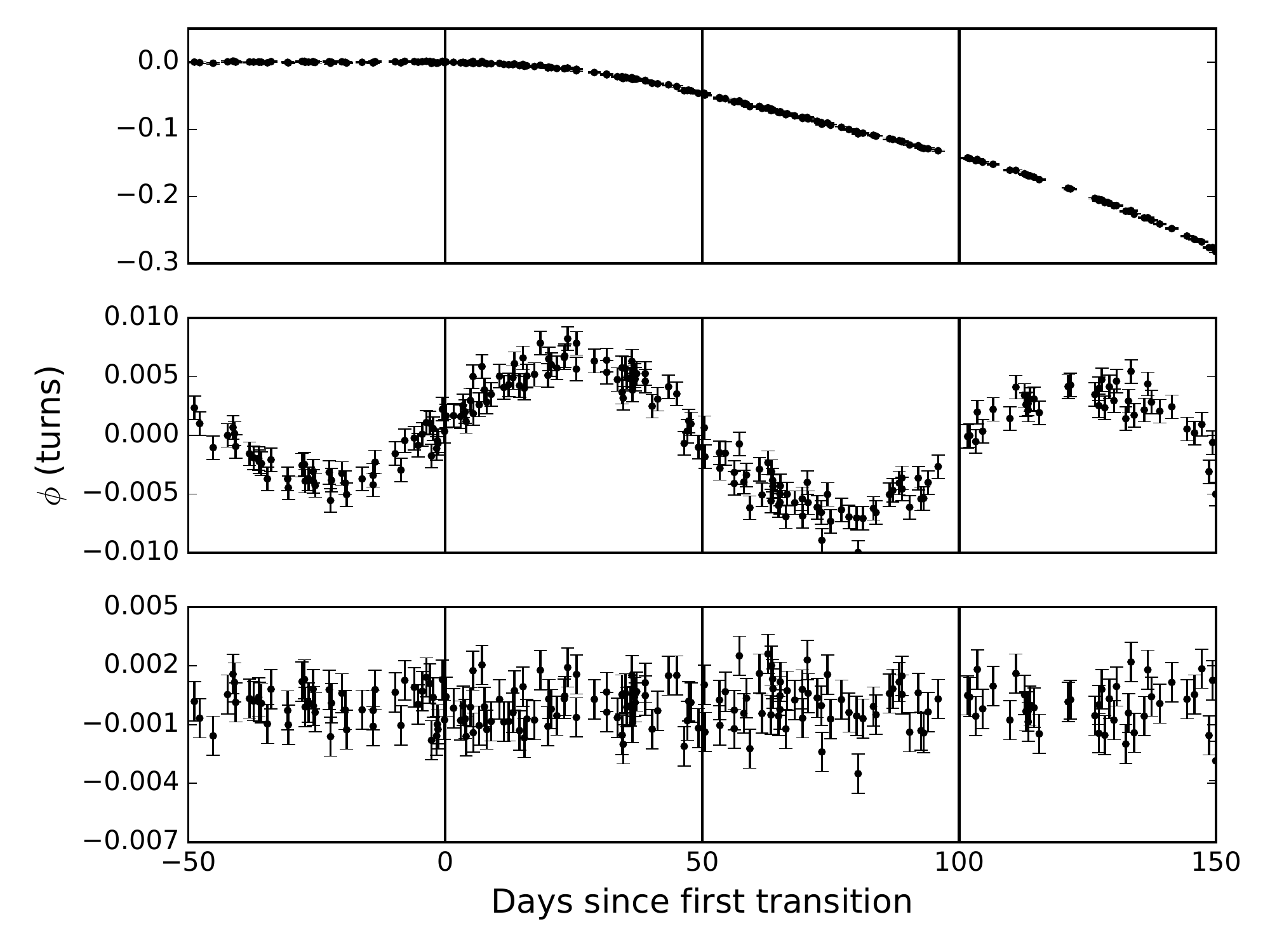}
   \caption{An example of the timing residuals obtained from a set of TOAs which includes three abrupt $\Delta \dot{\nu} = \pm 5.0 \times 10^{-15}$ Hz s\textsuperscript{-1} changes at the epochs, $t_i$, denoted by the black solid lines.  The dataset consists of 200 randomly spaced TOAs spanning 200 days.  White noise with an amplitude of 1 ms is added to the TOAs.  The initial $\dot{\nu} = -1.0 \times 10^{-14}$ Hz s\textsuperscript{-1}.  The upper and centre panels show the pre- and post-fit residuals respectively for a model which contains a single $\nu_0$ and $\dot{\nu}_0$ across the dataset. The RMS of the pre-fit residuals is 112 ms.  The RMS of the post-fit residuals is $\sim$4 ms - roughly 4 times greater than the mean errorbar size.  This fit yields a reduced $\chi^2$ ($\chi^2_{\mathrm{r,no trans}}$) of 15.6.  The lower panel shows the residuals when the transition parameters are included in the model.  In this case the RMS of the residuals is $\sim$1 ms, in line with the errorbar size and the reduced $\chi^2$ ($\chi^2_{\mathrm{r,trans}}$) is close to 1, indicating a good timing solution.}  
   \label{trans_notrans}
\end{figure*}

We generate a set of simulated arrival times from this ephemeris for the selected timespan, for a given cadence.  The generated TOAs are randomly spaced in time to simulate the irregularity with which pulsars are typically observed.  White noise with an amplitude of 1 ms is added to the TOAs as this is typical of many normal pulsars.  We first fit\footnote{Using a simplex method \citep{nm65} for $\chi^2$ minimisation.} a reference model containing $\nu_0$ and a single $\dot{\nu_0}$ to these data using a function of the form of equation \ref{frequencytaylor}. For details of the fitting procedure used throughout this work, see \cite{wje11}.  The  reduced $\chi^2$ ($\chi^2_{\mathrm{r}}$) of the timing residuals obtained by fitting only for these constant rotational parameters is recorded and is used as a reference value $\chi^2_{\mathrm{r,no trans}}$. As the dataset contains timing irregularities that are not modelled, $\chi^2_{\mathrm{r,no trans}}$ is expected to be further from unity than if the model had included them, as demonstrated in Figure \ref{trans_notrans}.  The pre-fit TOAs (top panel) are well described by the model and remain in a flat distribution about $\phi = 0$ until the first transition $t_1$  (left-most vertical line).  The residuals then become increasingly negative in a quadratic signature between $t_1$ and $t_2$ as $\dot{\nu}$ is now greater than the value in the model. At $t_2$ the pulsar switches back to its initial $\dot{\nu}$.  From then the $\dot{\nu}$ in the model is correct however there is now a discrepancy in $\nu$ and the residuals increase linearly with time.  At $t_3$, the pulsar switches again to the unmodelled value of $\dot{\nu}$ after which the quadratic increase in the residuals resumes until the end of the dataset.  Fitting for a single $\nu_0$ and $\dot{\nu_0}$ to these data results in timing noise (centre panel) reminiscent of those noted in \cite{hlk10}, and the RMS of the timing residuals is much greater than the mean TOA error.  This fit results in a $\chi^2_{\mathrm{r,no trans}}$ that is far from unity.  In the lower panel, the transitions are included in the timing model and the RMS is consistent with the mean errorbar size and $\chi^2_{\mathrm{r,trans}}$ is close to unity.

\begin{center}
\begin{table}

    \captionof{table}{Basic parameters used in all simulations.}
    \begin{tabular}{ | p{3.5cm} | p{3.5cm} | }
    \hline
    Parameter & Allowed simulation values \\ \hline
    $\nu_0$ & 1.229 Hz \\
    $\dot{\nu_0}$ & $-1.21 \times 10^{-14}$ Hz s\textsuperscript{-1} \\
    $\Delta \dot{\nu}$ & $0 \leq \Delta \dot{\nu} \leq 100$  per cent \\
    Time Baseline, $T$ & 500 days \\ 
    Switching timescale, $\tau$ & $0 \leq \tau \leq 200$ days \\
    $n_{\mathrm{simulation}}$ & 100,000 \\
    $n_{\mathrm{trials}}$ ($t_i$ unknown) & 10 \\
    $n_{\mathrm{trials}}$ ($t_i$ known) & 1 \\
    Dataset length & 500 days \\
    Cadences simulated & 1,2,7,14,28 days \\
    \hline
    \end{tabular}
    \label{properties}
\end{table}
\end{center}

To blindly search for the missing transition parameters a set of $n_{\mathrm{trials}}$ trial timing solutions is created.  These trials contain the pulsar's spin-frequency and initial first derivative as well as trial values for the unknown transition parameters.  Initially, we consider the case where neither the epochs of transitions, $t_i$, nor their amplitudes $|\Delta \dot{\nu}|$, are known.  In this case, 10 trials are used.  In initial tests, we found that this reduces the possibility that the fitting procedure finds a local minimum on the $\chi^2$ surface, and thereby fails to find the optimum solution.   

For each trial we fit for the unknown transition parameters, $t_i$ and $\Delta \dot{\nu}$.  As we have imposed that all changes to $\dot{\nu}$ in a given simulation are of equal magnitude we fit only for a single |$\Delta \dot{\nu}$| as well as the appropriate number of $t_i$, thereby minimising the number of free parameters.  Once a best-fit for these parameters is determined, a further fit is undertaken that includes $\nu_0$ and $\dot{\nu_0}$.  The RMS of the timing residuals for each trial is evaluated and recorded and the trial that returns the lowest value of the RMS is selected. For each cadence simulated, we apply this technique to 100,000 simulations whose transition parameters, $\tau$ and $\Delta \dot{\nu}$ populate the parameter space detailed in Table 1.

We later consider the case where transition epochs can be estimated in advance by using pulse shape changes to estimate $t_i$. In this case, the same transition parameters are included in the fit ($t_1$, $t_2$, $\Delta \dot{\nu}$, $\nu_0$ and $\dot{\nu_0}$), however the $t_i$ values in the trial timing solutions are not randomly selected but initially set to the estimated transition epochs, inferrable from profile variations.  We refer to these throughout this work as \emph{emission-inferred transition epochs}.  For each transition, the TOAs which immediately precede and follow the actual transition epoch are recorded.  The time exactly half way between these TOAs is used as the initial estimate for the transition epoch. This corresponds to the case in which spin-down transitions can be inferred by monitoring changes to the pulse shape from one observation to the next. 

The parameter space over which we simulate pulsar spin-down transitions is populated as follows.  $\Delta \dot{\nu}$ values are linearly distributed across the range of values noted in Table \ref{properties}.  The selection of the $\tau$ value for each simulations is achieved by the random selection of transition epochs within the dataset.   The interval between these values forms the switching timescale, $\tau$.  Consequently, there are a larger number of simulations for which $\tau$ is short as there are more combinations of epochs that yield a shorter $\tau$.  This results in sparser sampling towards the upper end of the switching timescale range.  This could be mitigated with supplementary sampling resulting in a larger number of high $\tau$ scenarios, however we expect transition detectablity to be limited by shorter timescale switching behaviour and so this is not a concern.

To ensure at least 99.7 per cent confidence in any detection made, we require $\Delta \dot{\nu}_{\mathrm{fit}} \geq 3\sigma_{\Delta \dot{\nu}_{\mathrm{fit}}}$ and $\tau_{\mathrm{fit}} \geq 3\sigma_{\tau_{\mathrm{fit}}}$, where $\sigma$ are the errors on the associated fit parameters.  We also require that $\chi^2_{\mathrm{r,trans}}$ is closer to unity than the $\chi^2_{\mathrm{r,no trans}}$.  To yield a detection, $\chi^2_{\mathrm{r,trans}}$ must be between 0.5 and 1.5. If fitting only for a constant $\dot{\nu}$ yields $0.5 < \chi^2_{\mathrm{r,no trans}} < 1.5$, we do not consider a solution including transitions to be a detection as the timing model cannot sufficiently be improved on. 

Although we simulate $0 < \tau \leq 200$ days within a 500 day dataset, the results are scalable to other scenarios by considering the form of Equation \ref{phase}. Consider a 100 day dataset consisting of 100 TOAs ($C = 1$ day), that contains two $\Delta \dot{\nu}$ transitions, $\tau = 20$ days apart as shown in the right panel of Figure \ref{residualsforunmodelledchange}. By multiplying the dataset length by a factor $s$ (in this case $s=5$), we arrive at the 500 day dataset simulated here.  Correspondingly $\tau$ must also scale up by the same factor to 100 days, as does the time covered since the first transition, $T$. Consequently, the cadence reduces by a factor of $s=5$ to one observation per 5 days.  Due to the $\tau^2$ dependence of Equation \ref{phase}, to preserve the effect on the residuals, $\Delta \dot{\nu}$ must be reduced by a factor of $1/s^2$.  If these steps are followed, the $\langle \chi^2 \rangle$ from these two datasets is identical.


\section{Simulation results} 
\subsection{Two transitions} \label{two_trans}

\begin{figure*}
   
    \includegraphics[width=\columnwidth]{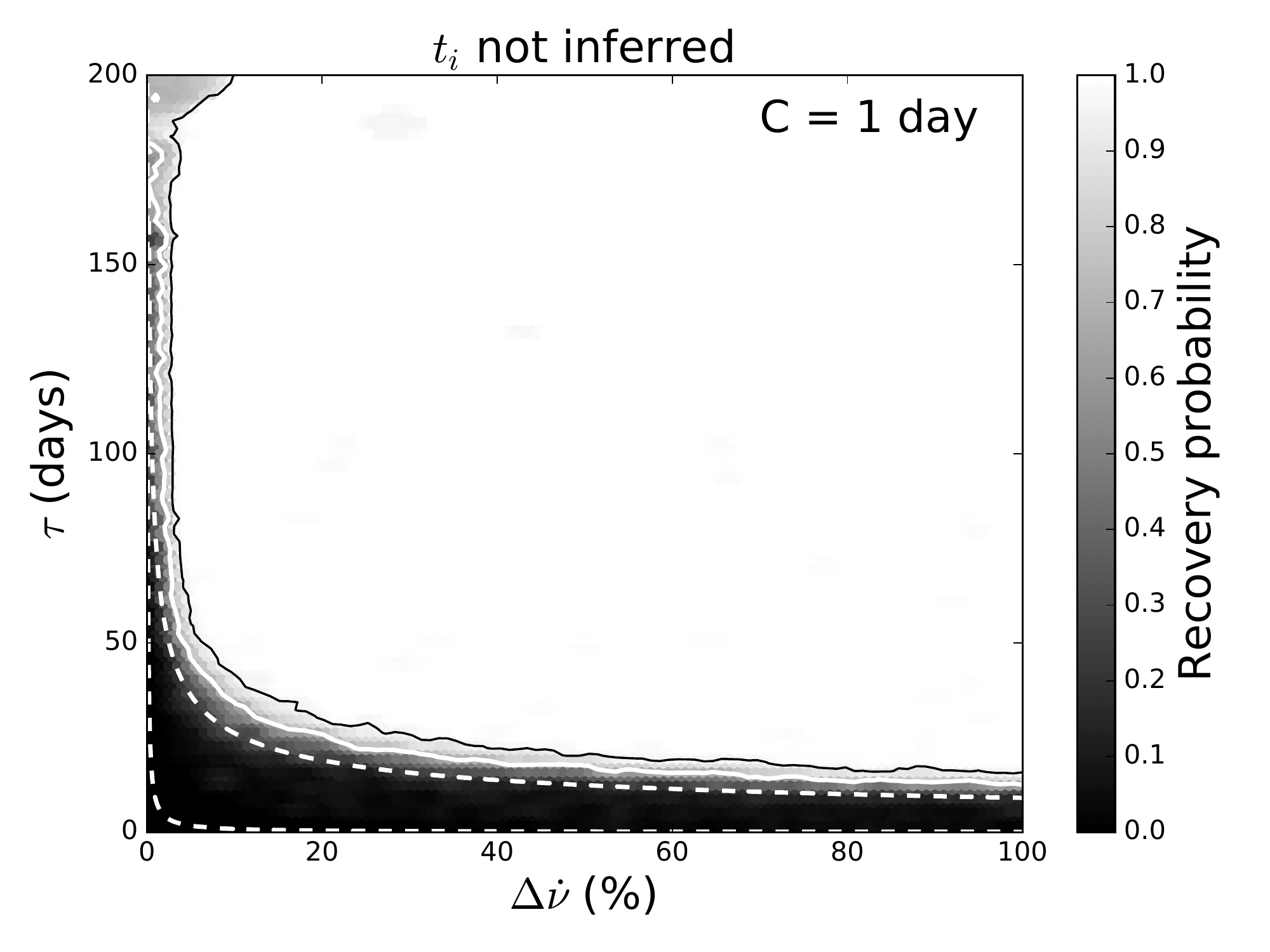}
    \includegraphics[width=\columnwidth]{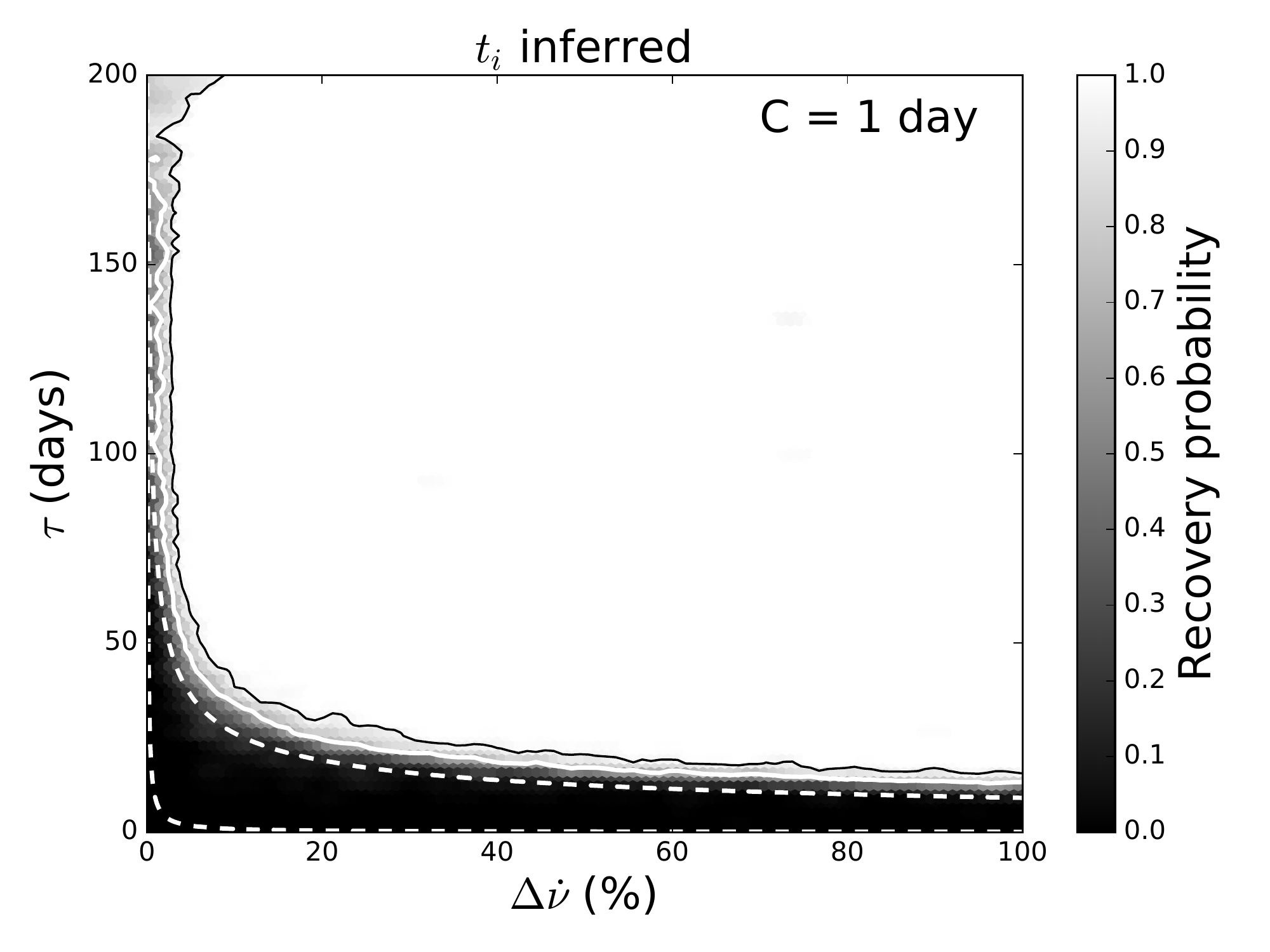}\\[-1ex]
    
    \includegraphics[width=\columnwidth]{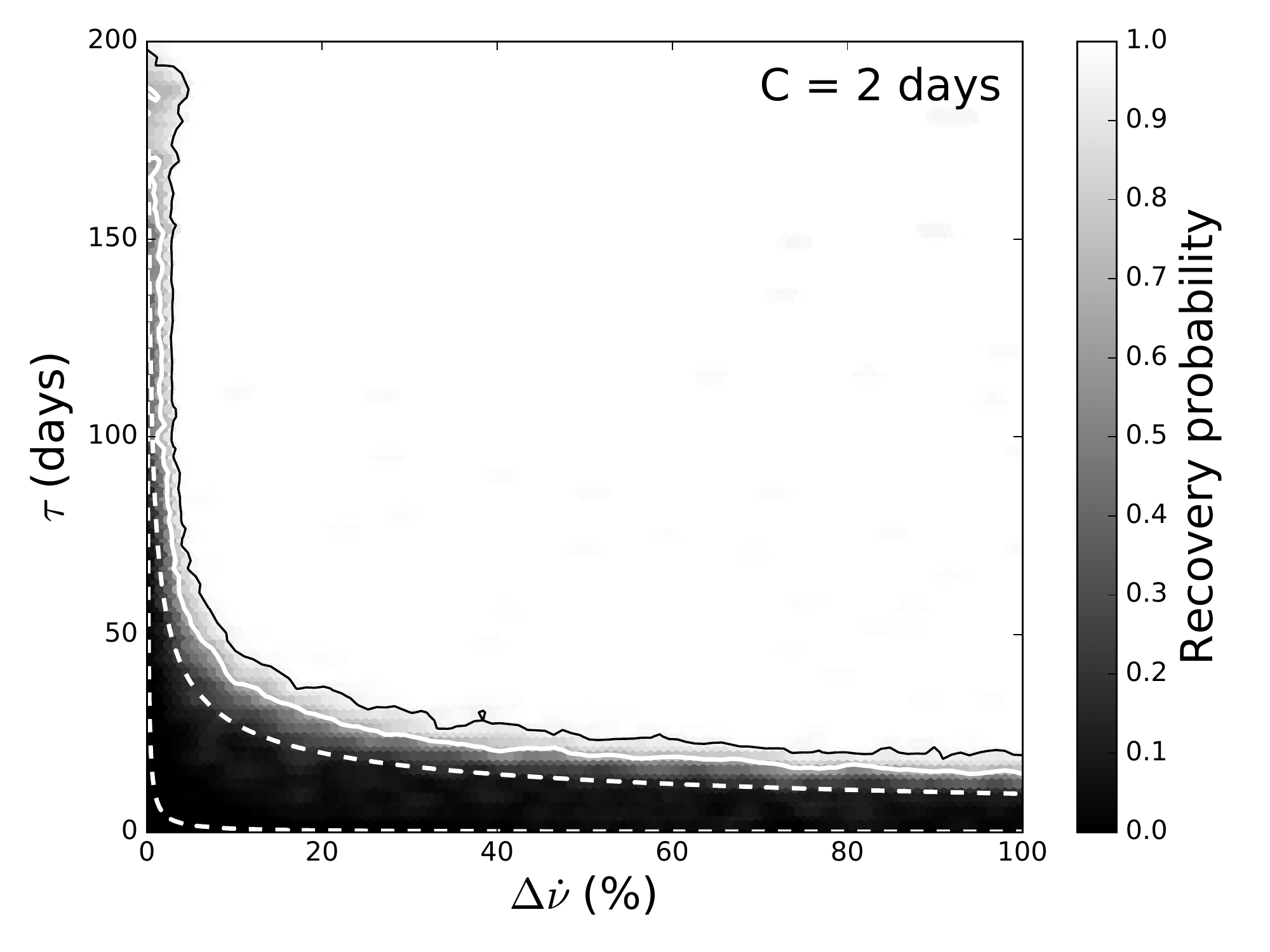}
    \includegraphics[width=\columnwidth]{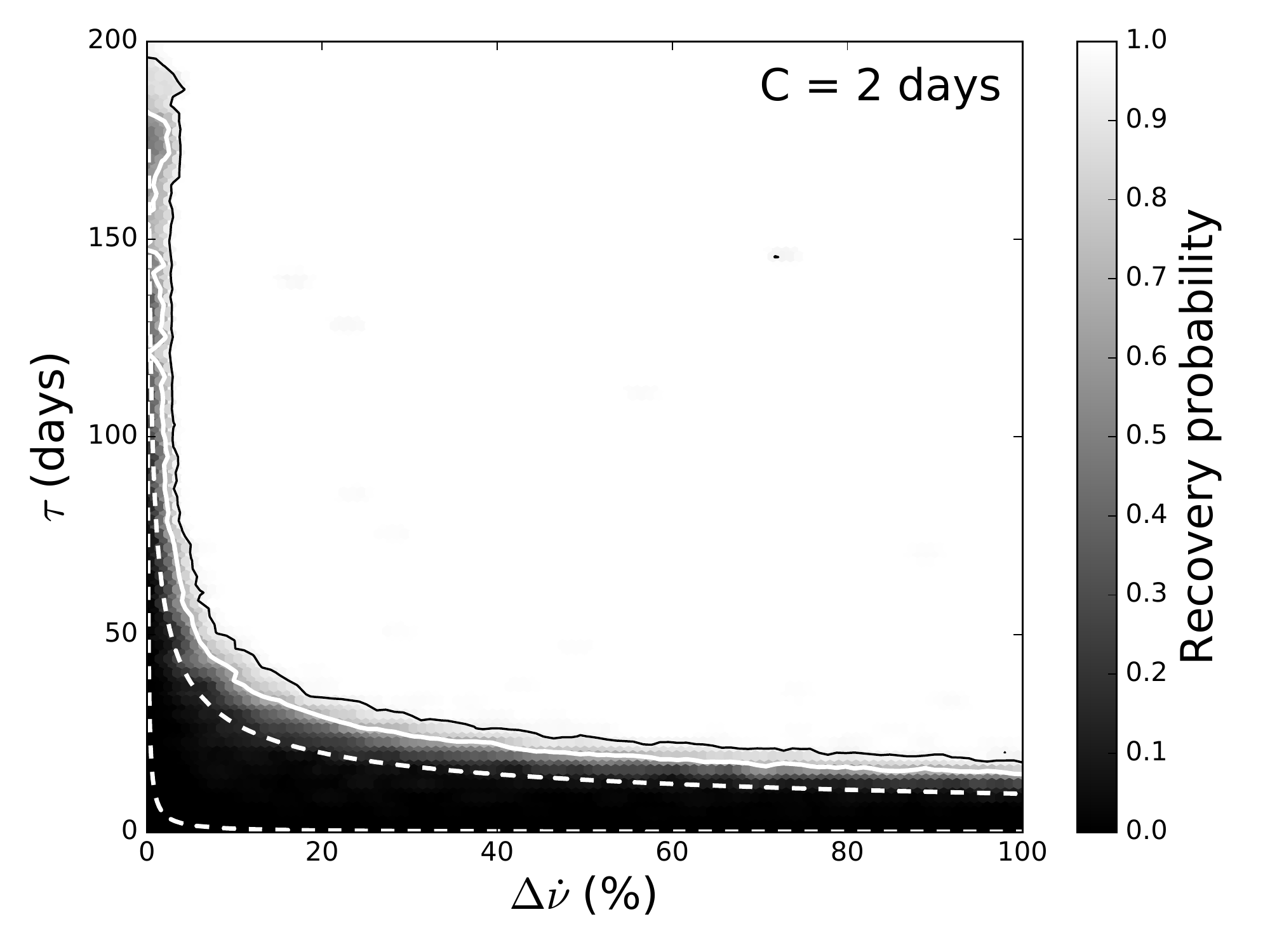}\\[-1ex]

    \includegraphics[width=\columnwidth]{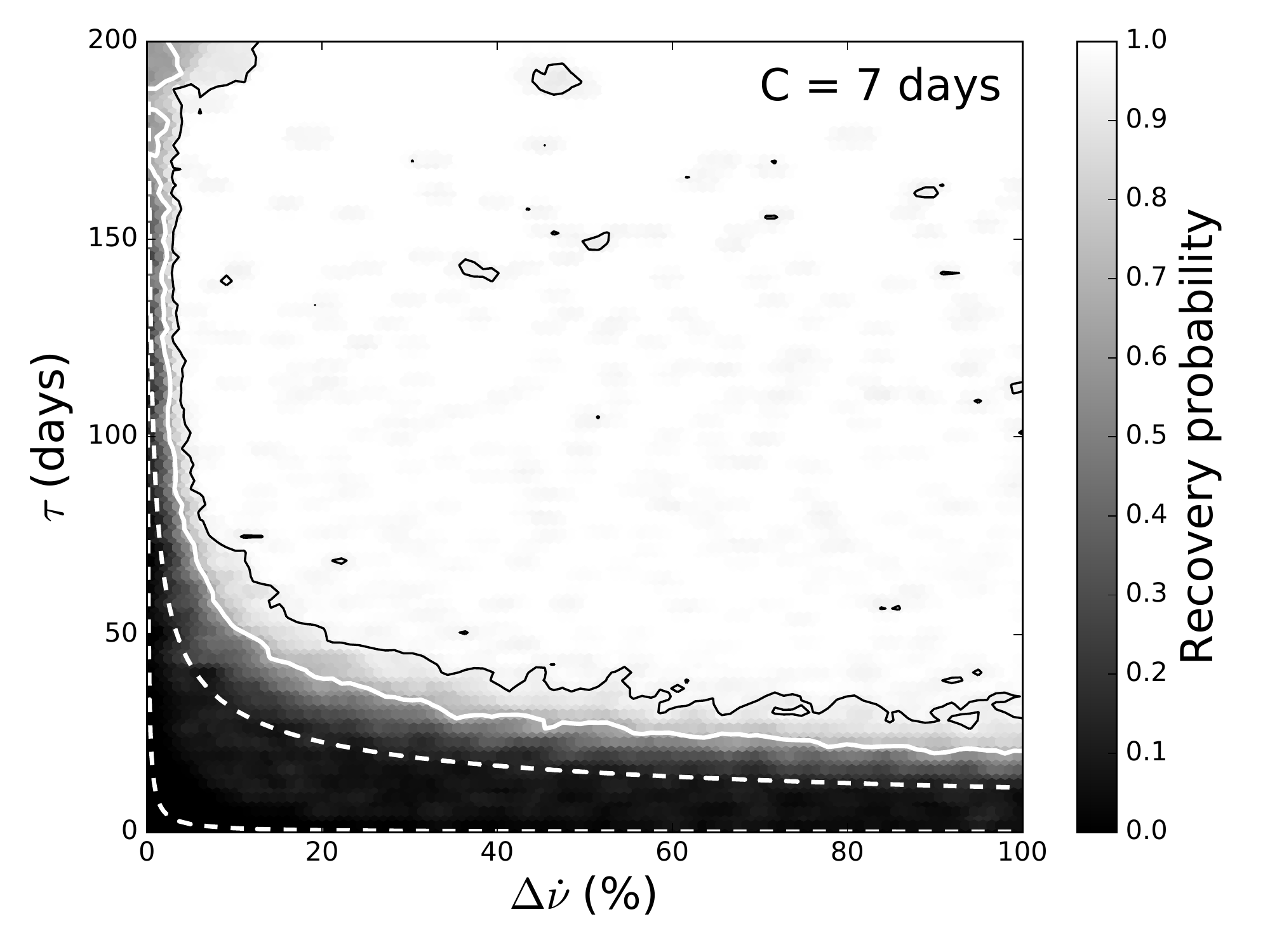}
    \includegraphics[width=1\columnwidth]{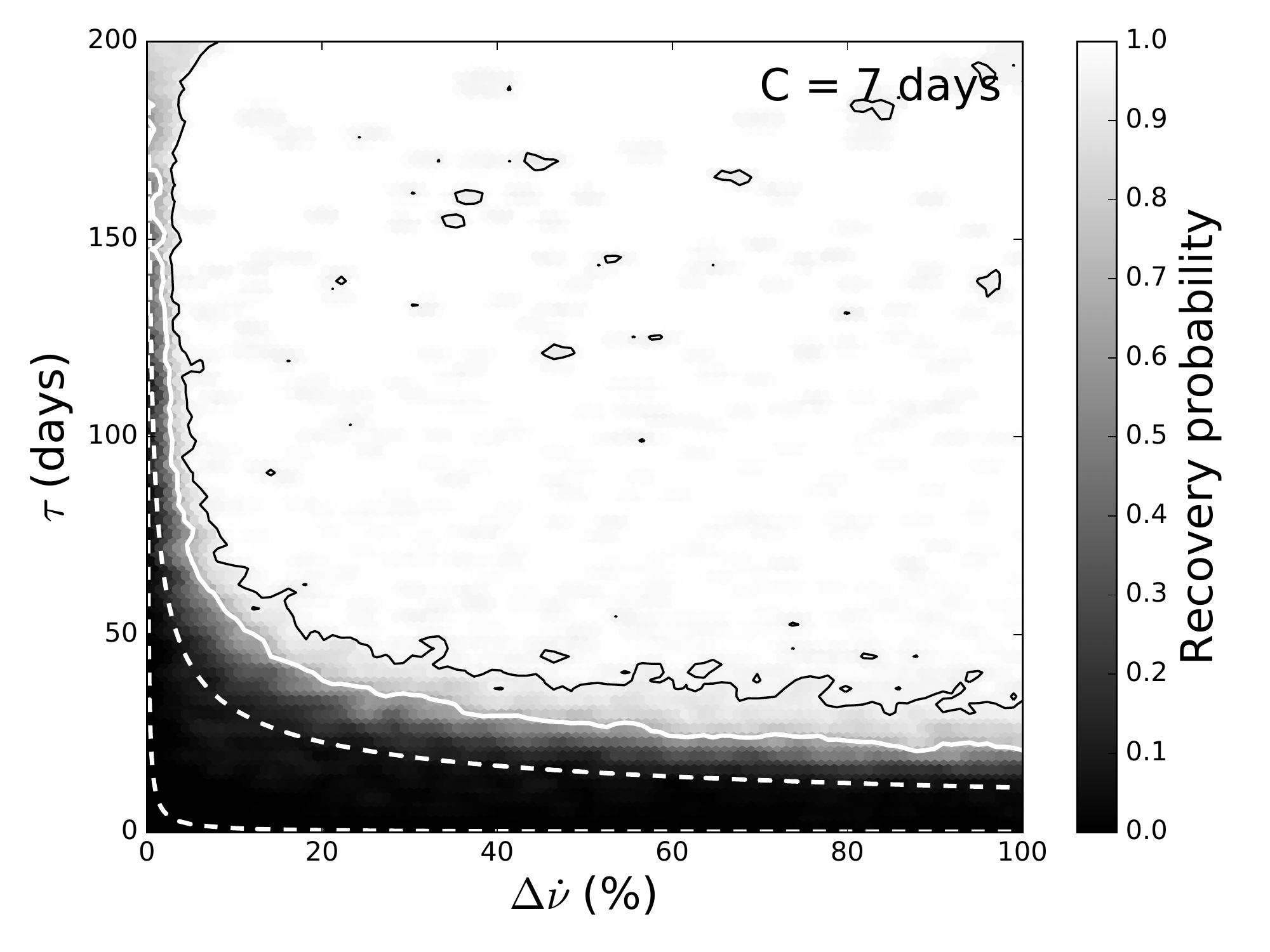}
    \caption{Heatmaps showing the fraction of simulations for which a recovery was made over the parameter space sampled for observing cadences of 1 (top), 2 (centre,) and 7 (bottom) days. Two transitions are simulated within each dataset.  The left panels correspond to simulations in which no emission-inferred transition epochs $t_i$ are used in the fit. In the right panels, initial estimates on $t_i$ are provided, as described in the text.  Brighter regions denote a higher detection probability. The solid white and black lines denoted the 68 per cent and 95 per cent detection probability contours respectively. The lower dashed lines denote the minimum $\tau$ for a given $\Delta \dot{\nu}$ for which the residuals are affected by the transitioning behaviour as estimated in Equation \ref{chi2_2}. The upper of the dashed lines, denotes the value of $\tau$ for which the behaviour between events can be expected to be resolved.} 
    \label{rec_heatmaps_double_mode_no_prior}
\end{figure*} 

\begin{figure*}
   
    \includegraphics[width=\columnwidth]{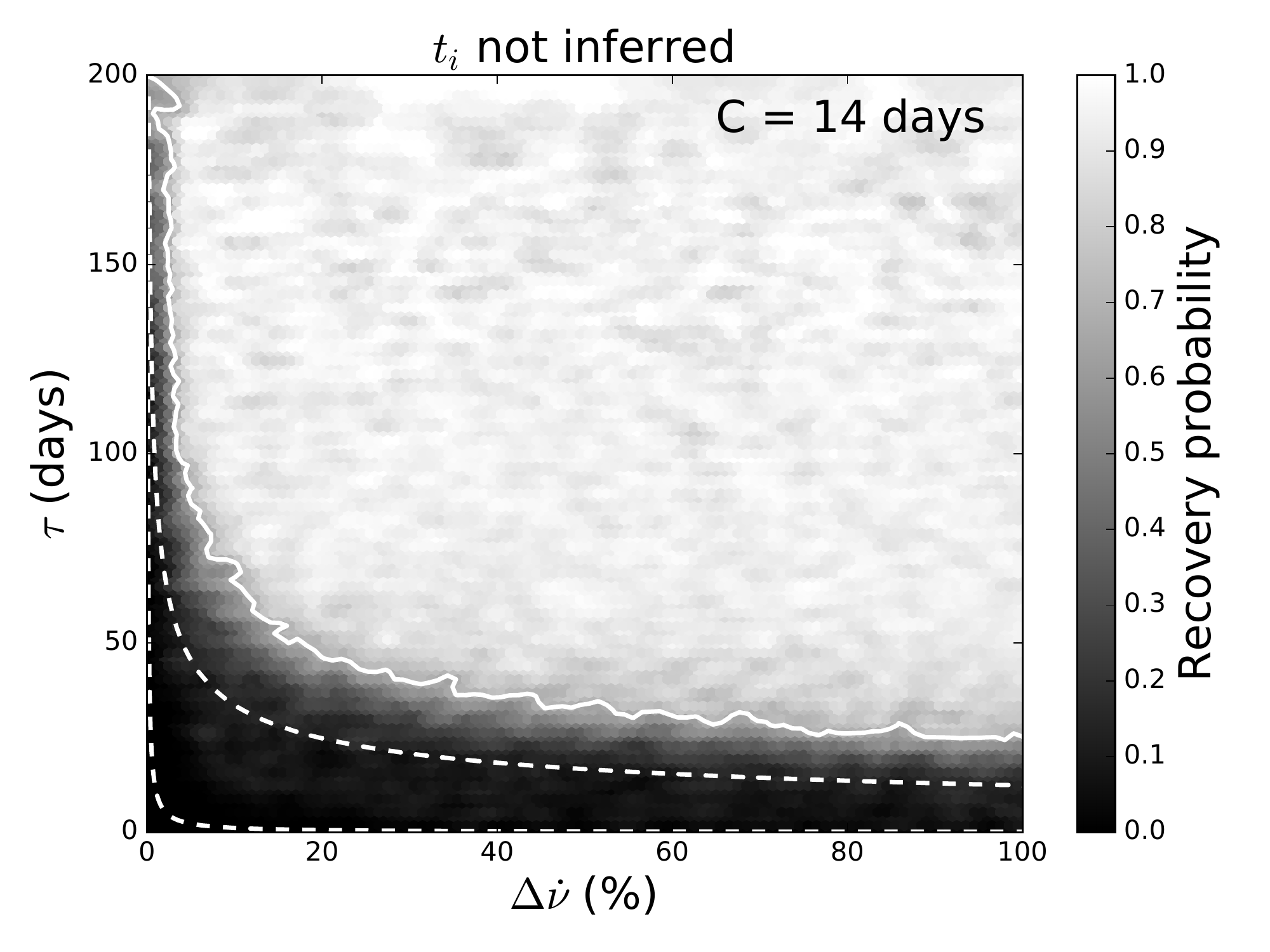}
    \includegraphics[width=\columnwidth]{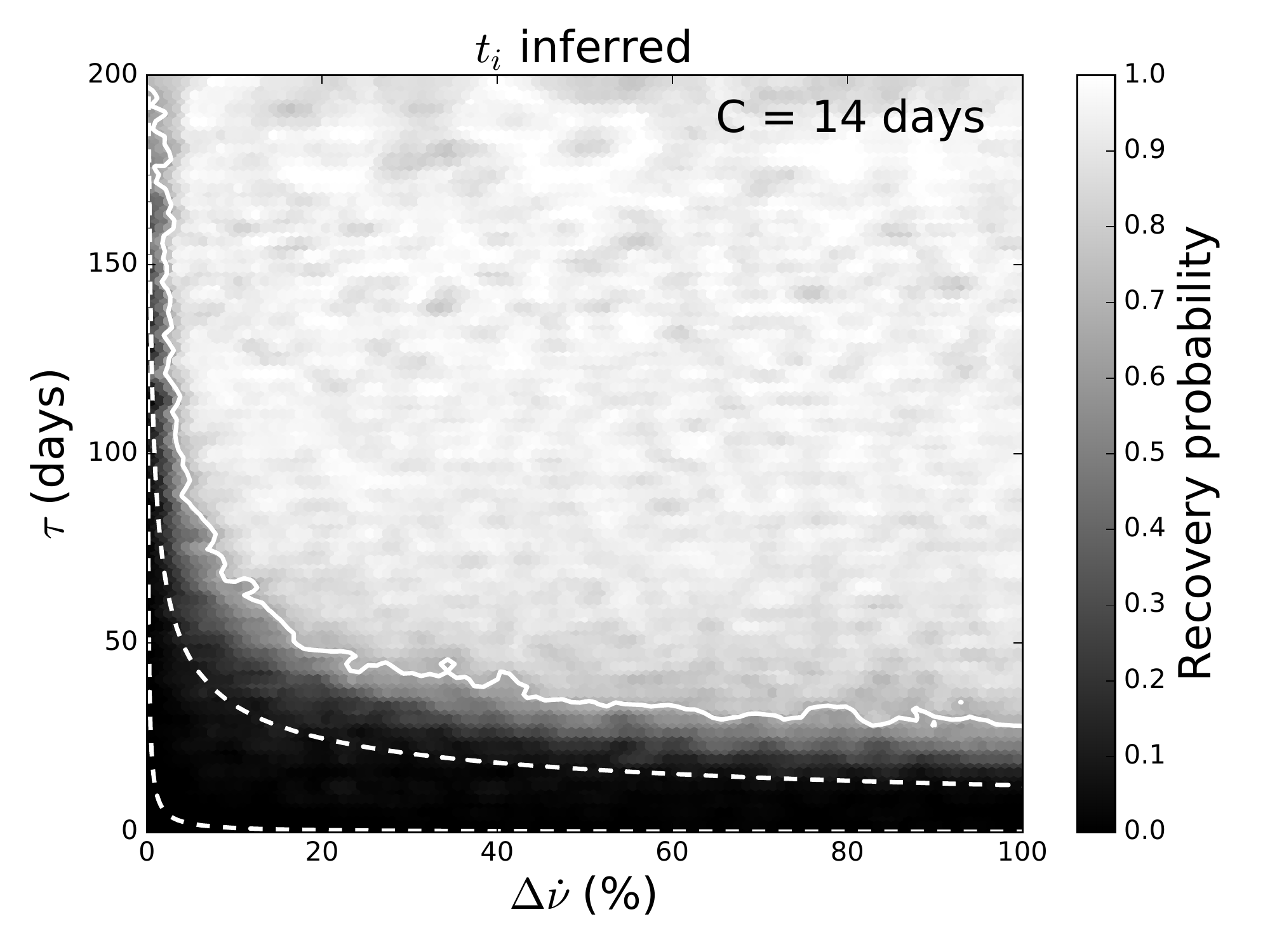}\\[-1ex]
    
    \includegraphics[width=\columnwidth]{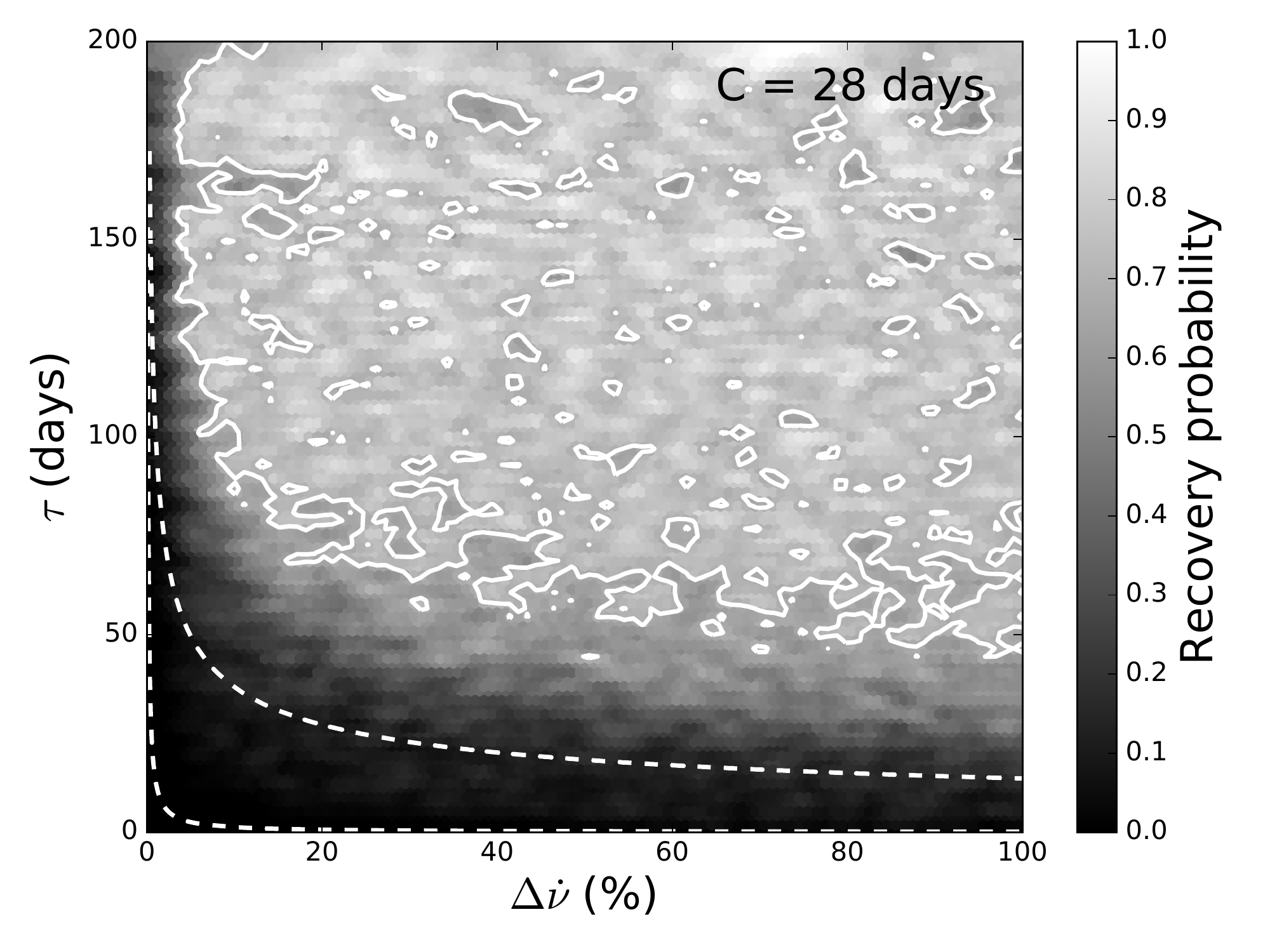}
    \includegraphics[width=\columnwidth]{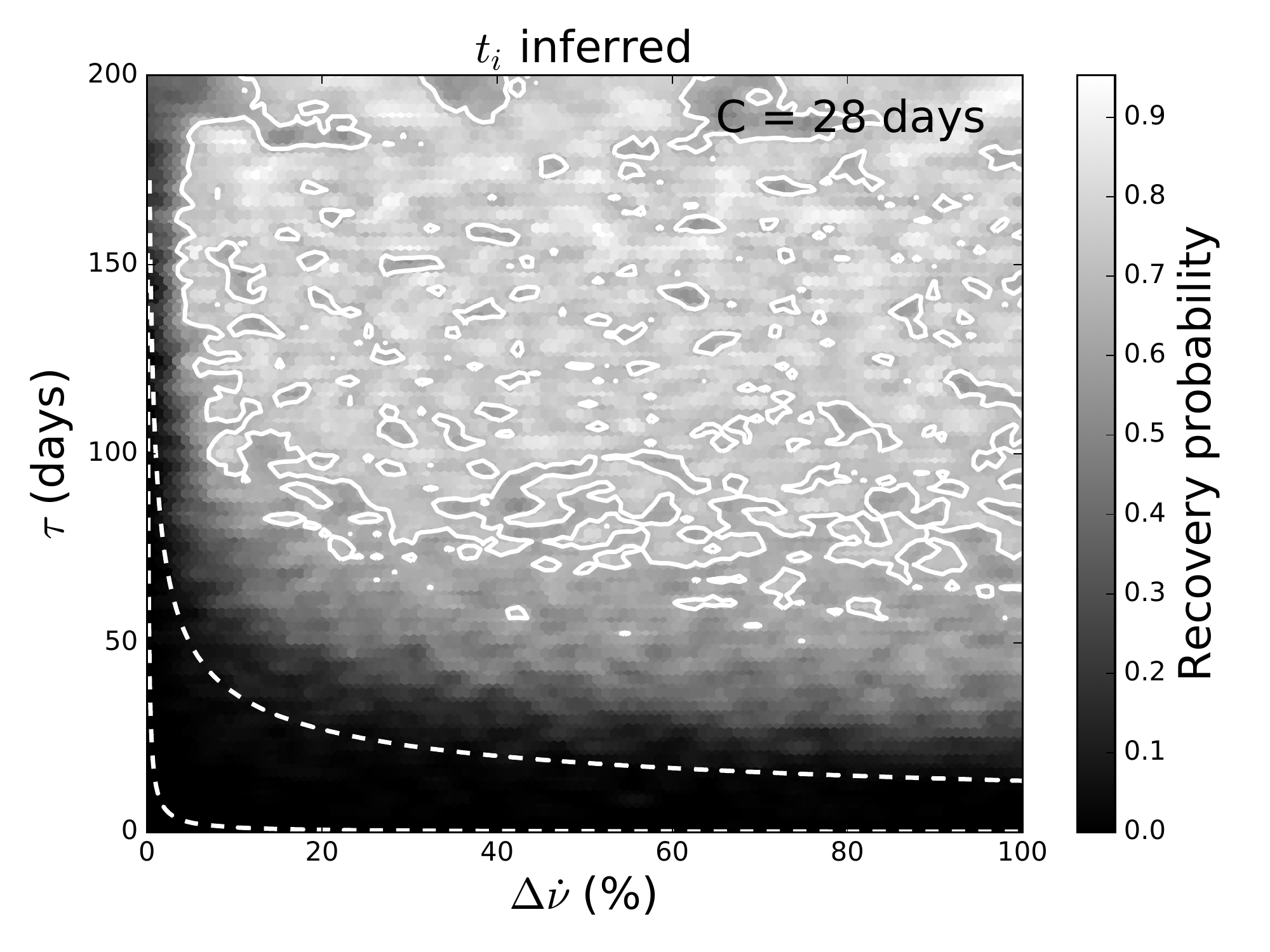}\\[-1ex]

    \caption{As Figure \ref{rec_heatmaps_double_mode_no_prior} but for observing cadences of once per 14 days (top) and once per 28 days (bottom).}
    \label{rec_heatmaps_double_mode_no_prior_2}
\end{figure*} 

We first simulate the scenario in which two $\dot{\nu}$ transitions occur in a dataset.  Table \ref{properties} details the parameter space over which transitions are simulated.  As we insert two transitions into the central 200 days of the dataset, the maximum switch timescale $\tau$ we can simulate here is 200 days however these transition parameters are scalable to larger durations. 

Figures \ref{rec_heatmaps_double_mode_no_prior} and \ref{rec_heatmaps_double_mode_no_prior_2} show heatmaps representing the probability of detecting $\dot{\nu}$ transitions in a dataset as a function of their amplitudes and the elapsed time between the two transitions, for a range of cadences. To construct these maps we split the $\Delta \dot{\nu} - \tau$ plane into individual cells. In each cell we evaluate the ratio of detections to the total number of simulations in that cell.  Each ratio is assigned the centre coordinates of its cell.  We then interpolate between the grid points to create a smooth detection probability surface. 

At the highest cadences ($C = 1,2$ days) the detection probability rises very rapidly with $\tau$, noted from the close proximity of the 68 and 95 per cent probability contours. Where a transitioning pulsar is observed daily, its transitions are resolvable with more than 68 per cent likelihood if $\tau$ exceeds $\sim$10 days for the strongest amplitudes simulated ($\Delta \dot{\nu} \sim100$ per cent). For smaller amplitude transitions, $\tau$ must be greater to maintain detectability.  A small region in the upper left corner (representing very low amplitude transitions separated by $\sim$200 days), shows a decrement in the detection probability.  This is because in this region of the $\Delta \dot{\nu} - \tau$ parameter space, the pulsar is in the new spin-down state for $\sim$50 per cent of the dataset. As a consequence modelling the pulsar's rotation with a single, continuous frequency derivative is more likely to yield a $\chi^2_{\mathrm{r,notrans}} \approx 1$. In initial tests where the time baseline was shorter, this effect was augmented, confirming that is it a feature of the dataset length and does not represent an intrinsic drop in detectabililty.

As cadence decreases, the datasets with transitions are more often able to be fit sufficiently well without transitions. For example at approximately monthly cadence there are no pairs of $\Delta \dot{\nu} - \tau$ values for which transition detection is better than 95 per cent likely.  At the typical cadence of once per 14 days (Figure \ref{rec_heatmaps_double_mode_no_prior_2}, upper plots), the detectability of transitions with $\tau < 30$ days is less than 68 per cent likely for all $\Delta \dot{\nu}$.  We note that the upper dashed lines, representing the minimum $\tau$ for which the first transition becomes instantaneously detectable, roughly follows the 10 per cent detection probability contour at all cadences.

The use of profile variations to estimate transition epochs does not, at face value, offer significant advantages over not doing so, especially at high cadence.  Figures \ref{rec_heatmaps_double_mode_no_prior} and \ref{rec_heatmaps_double_mode_no_prior_2} do however show differences between the two techniques. At low cadence, detectability extends to lower $\tau$ \emph{without} a-priori estimates on $t_i$. However this is because we are only able to infer $t_i$ when the average time between observations is shorter than the switching timescale $\tau$ to ensure that at least one observation takes place when the pulsar is in each of its spin-down states.  Therefore, for $C=28$ days, there are fewer opportunities to use mode-switching as a proxy for spin-down transitions at low $\tau$.  In cases where observations do occur in each spin-down state, the cadence is sufficiently low such that our estimates of $t_i$ are no more efficacious than random trial values.   

\begin{figure*}
    \includegraphics[width=0.85\columnwidth]{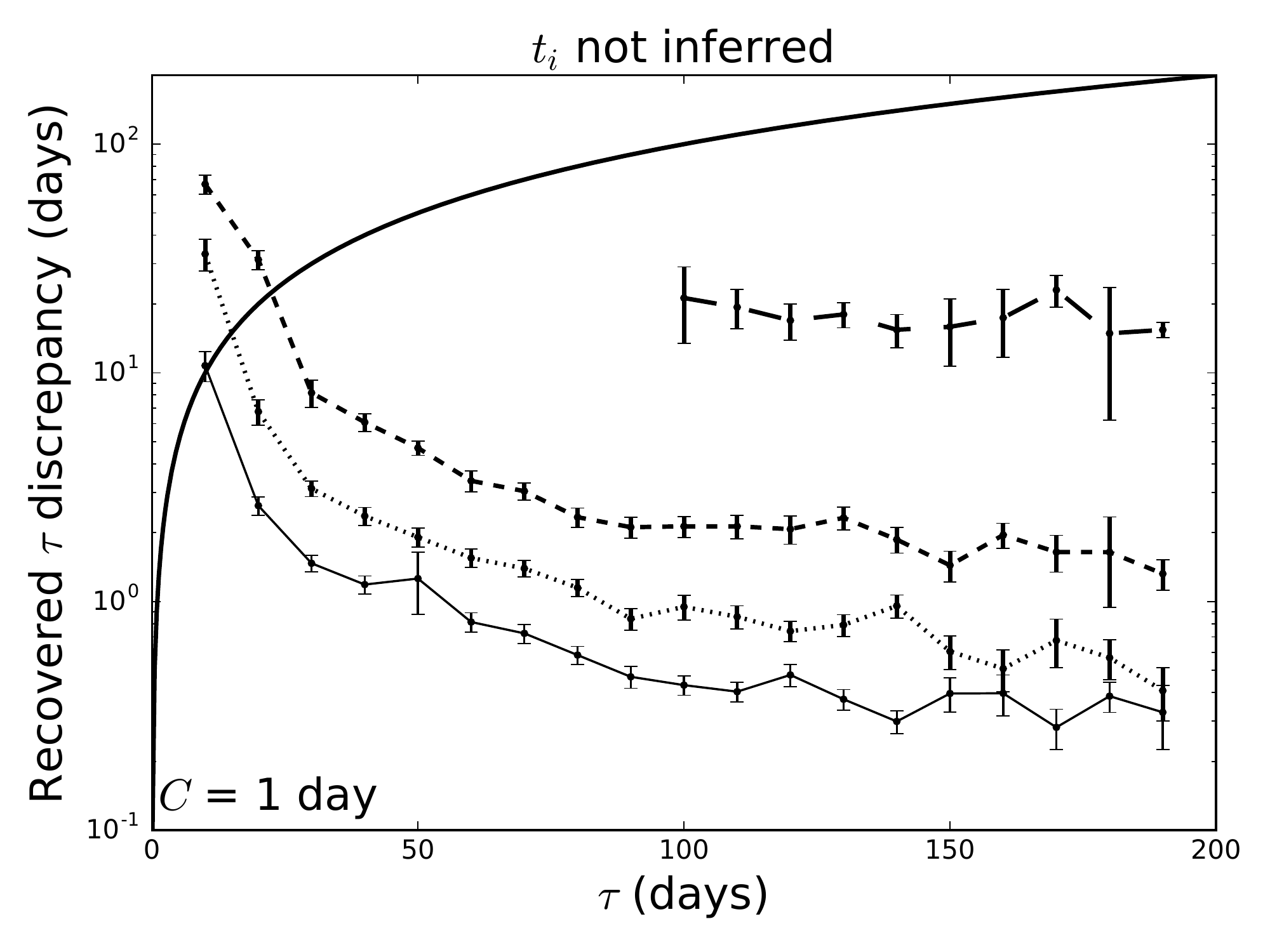}
    \includegraphics[width=0.85\columnwidth]{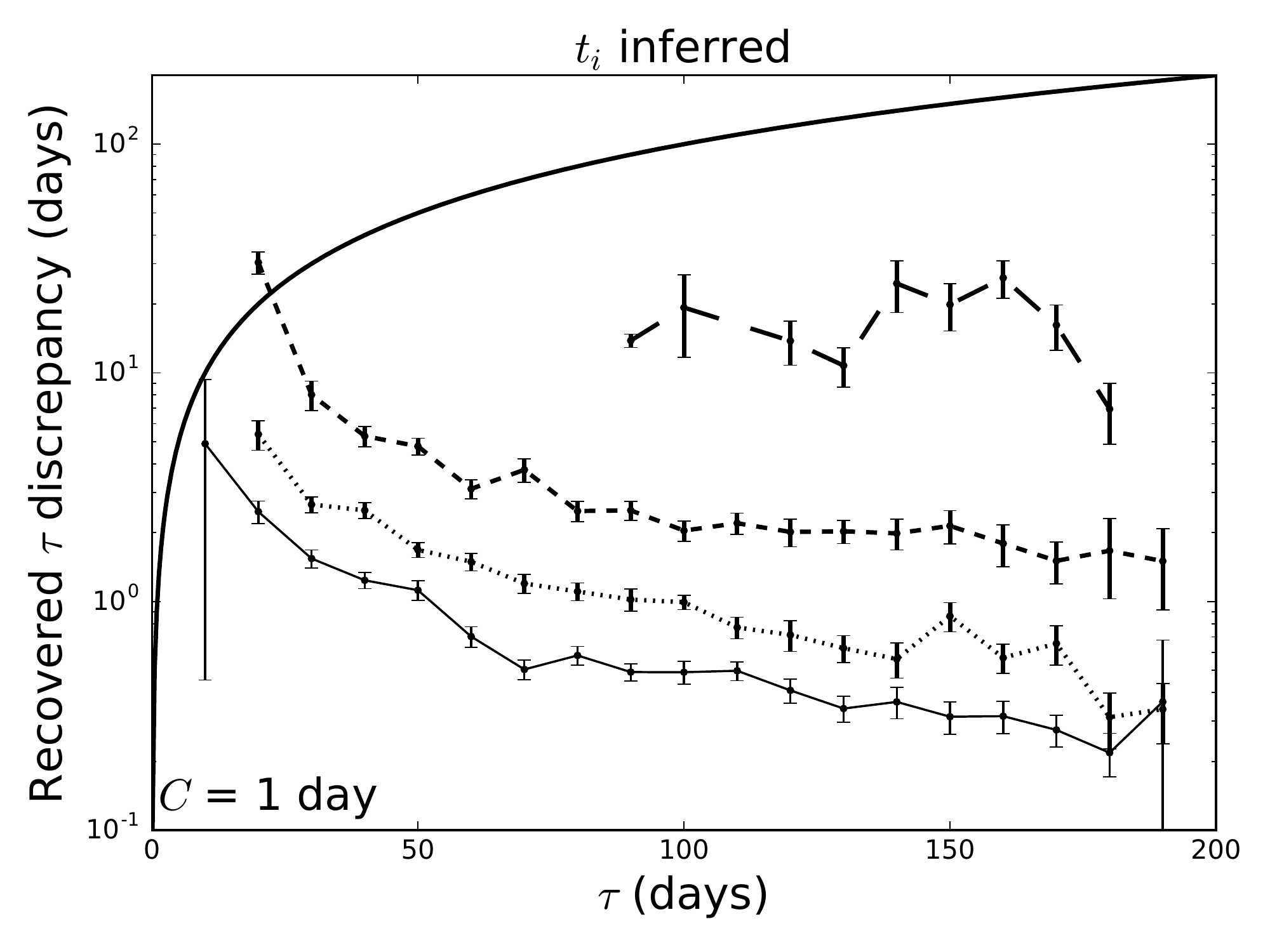} \\
    \includegraphics[width=0.85\columnwidth]{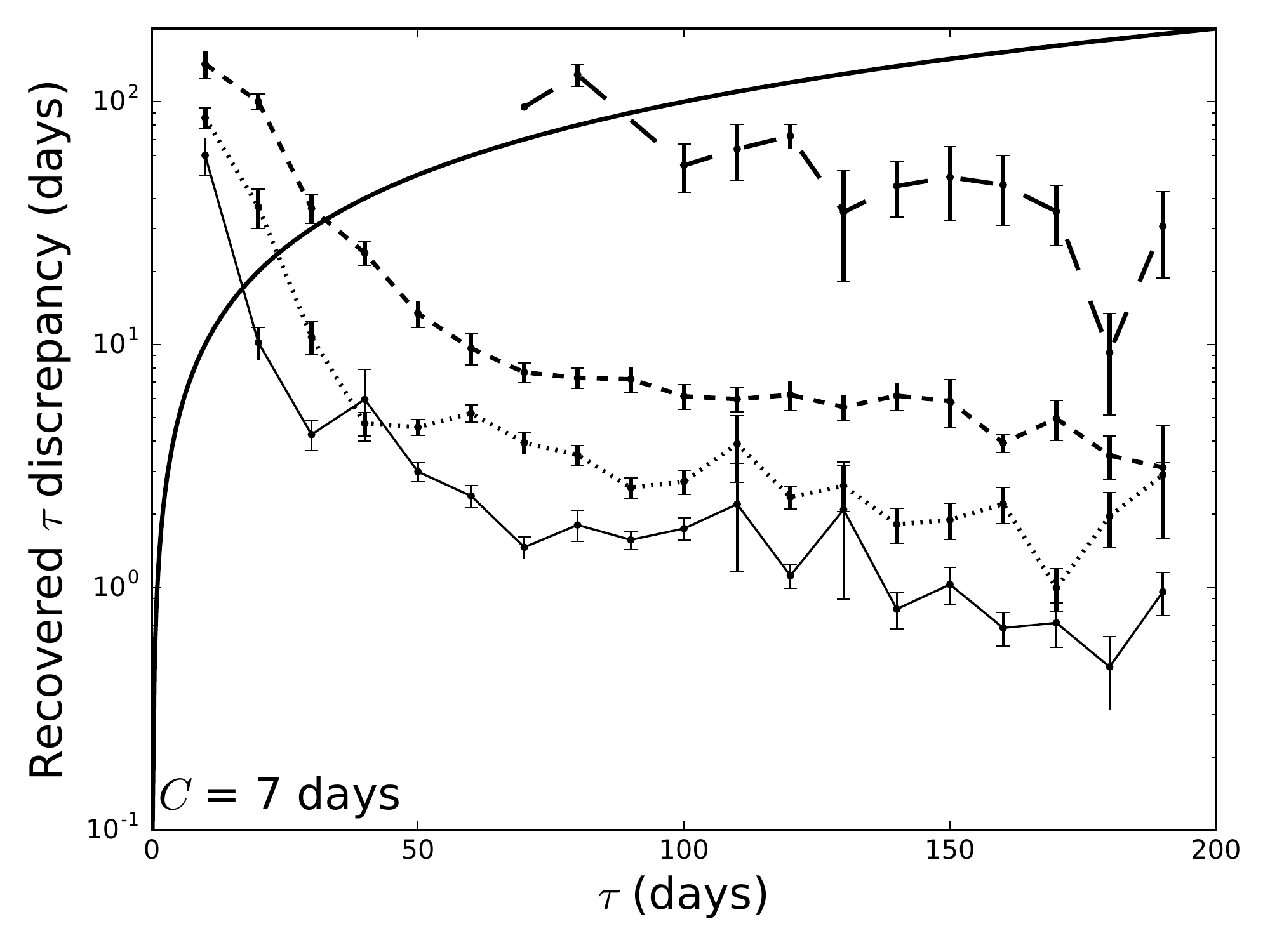}
    \includegraphics[width=0.85\columnwidth]{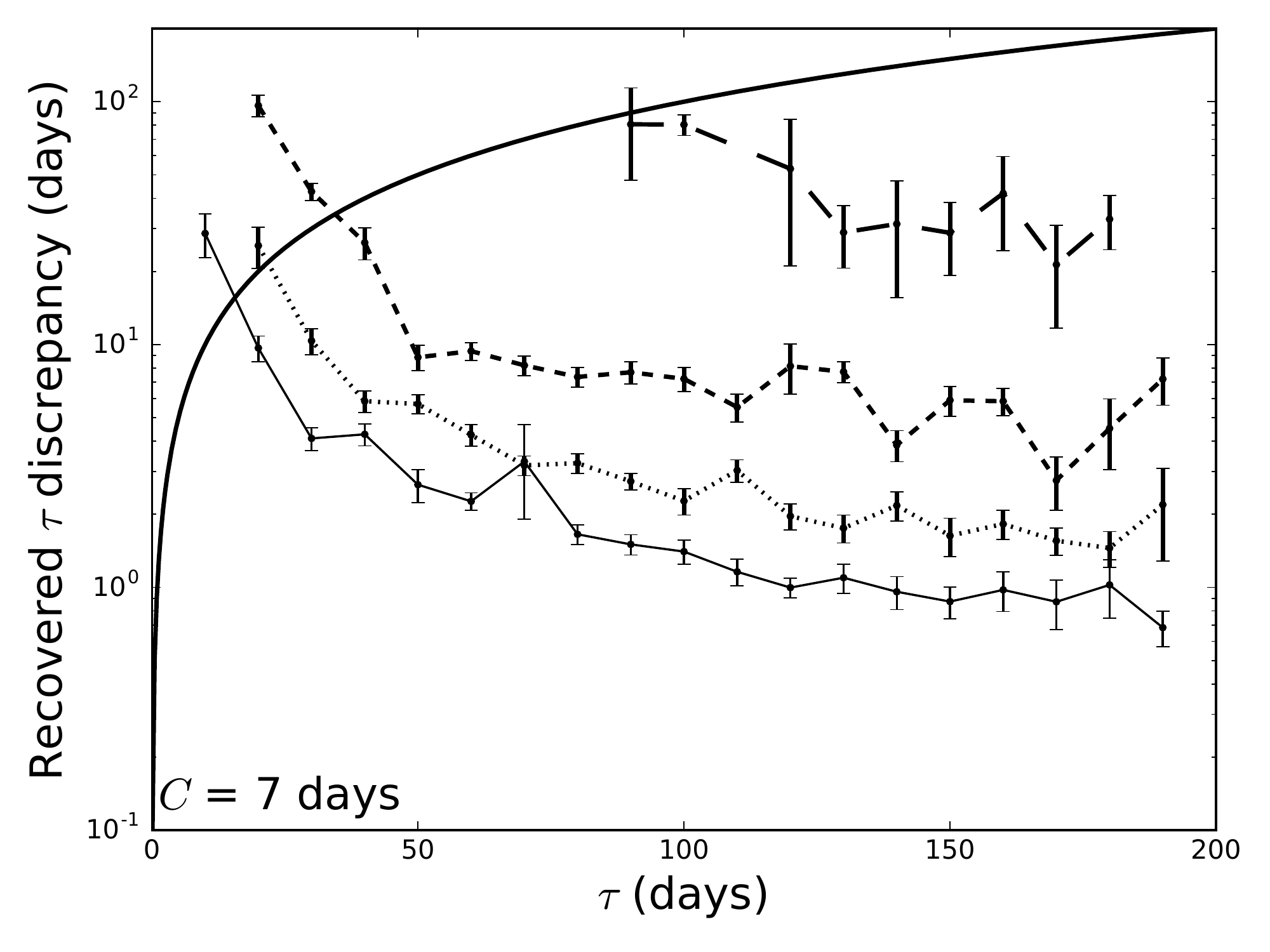} \\
    \includegraphics[width=0.85\columnwidth]{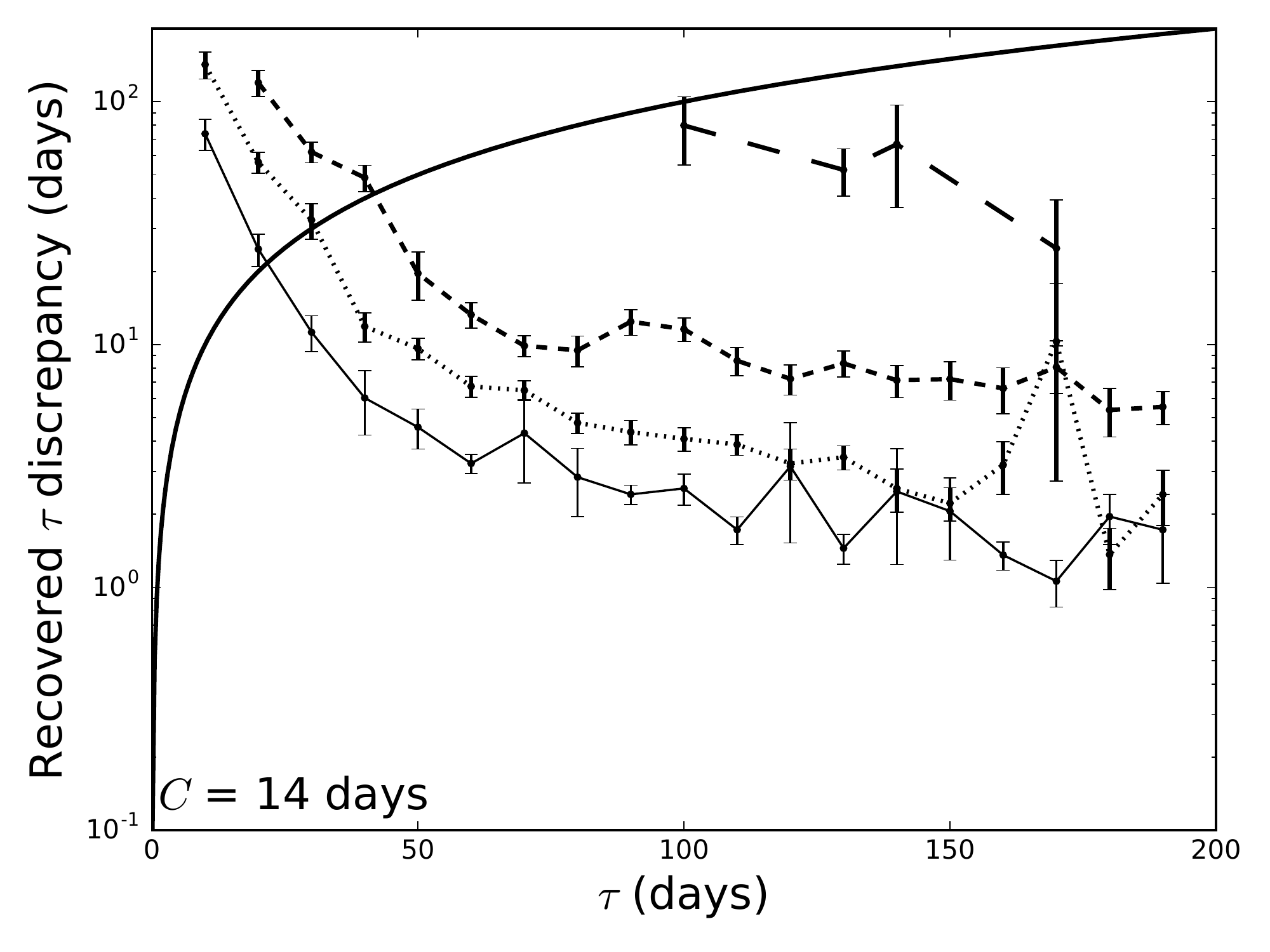}
    \includegraphics[width=0.85\columnwidth]{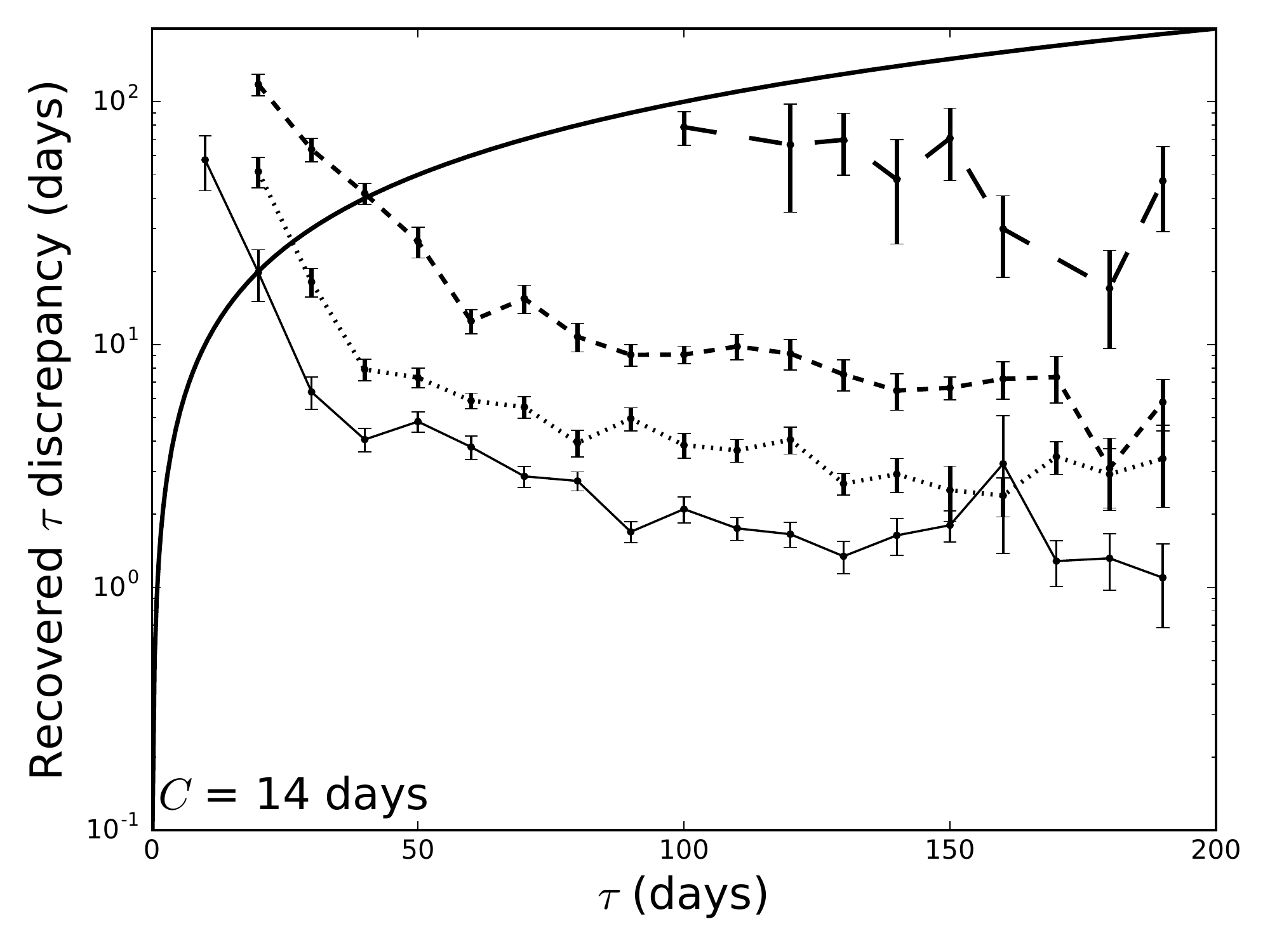} \\
    \includegraphics[width=0.85\columnwidth]{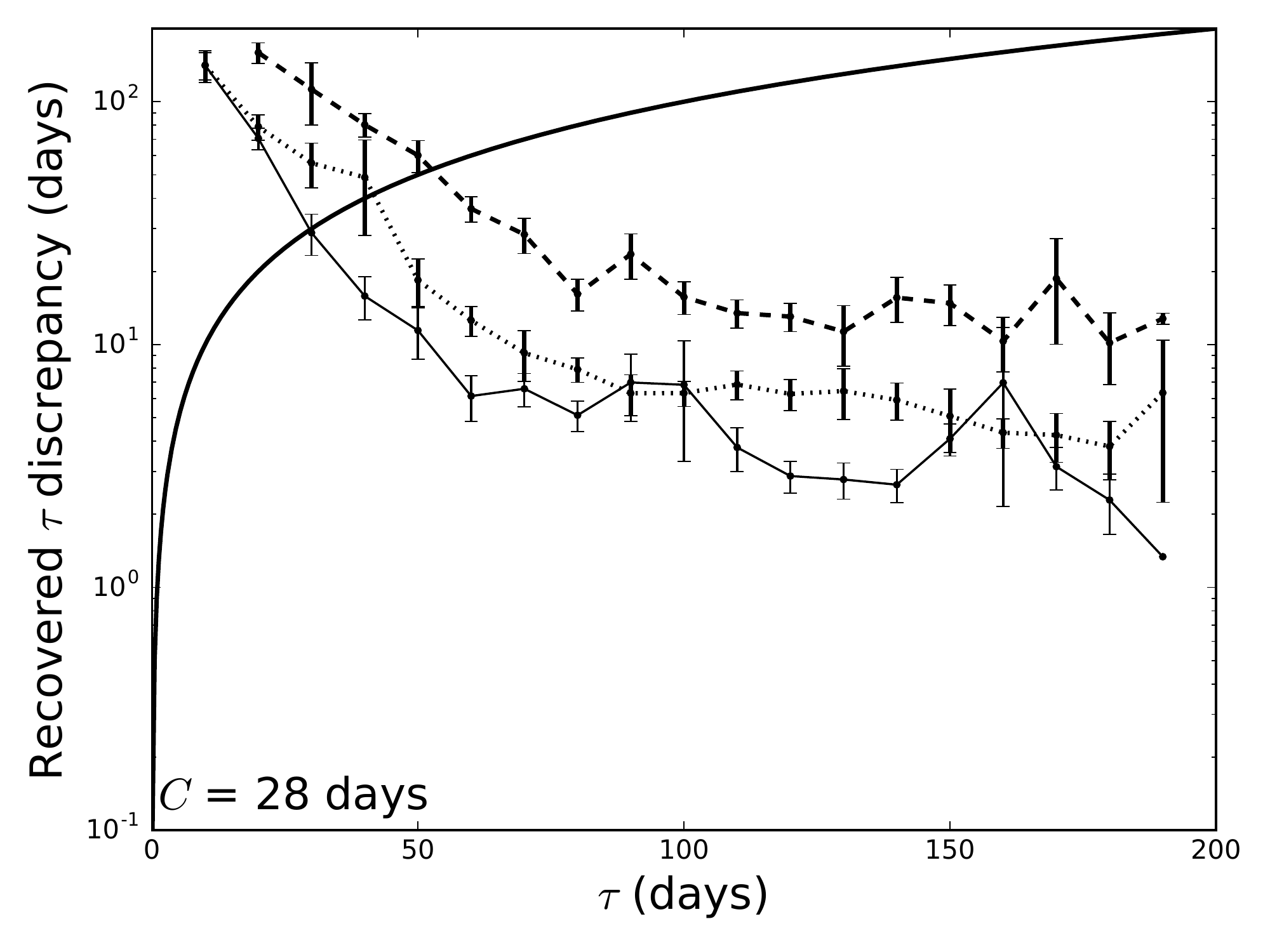}
    \includegraphics[width=0.85\columnwidth]{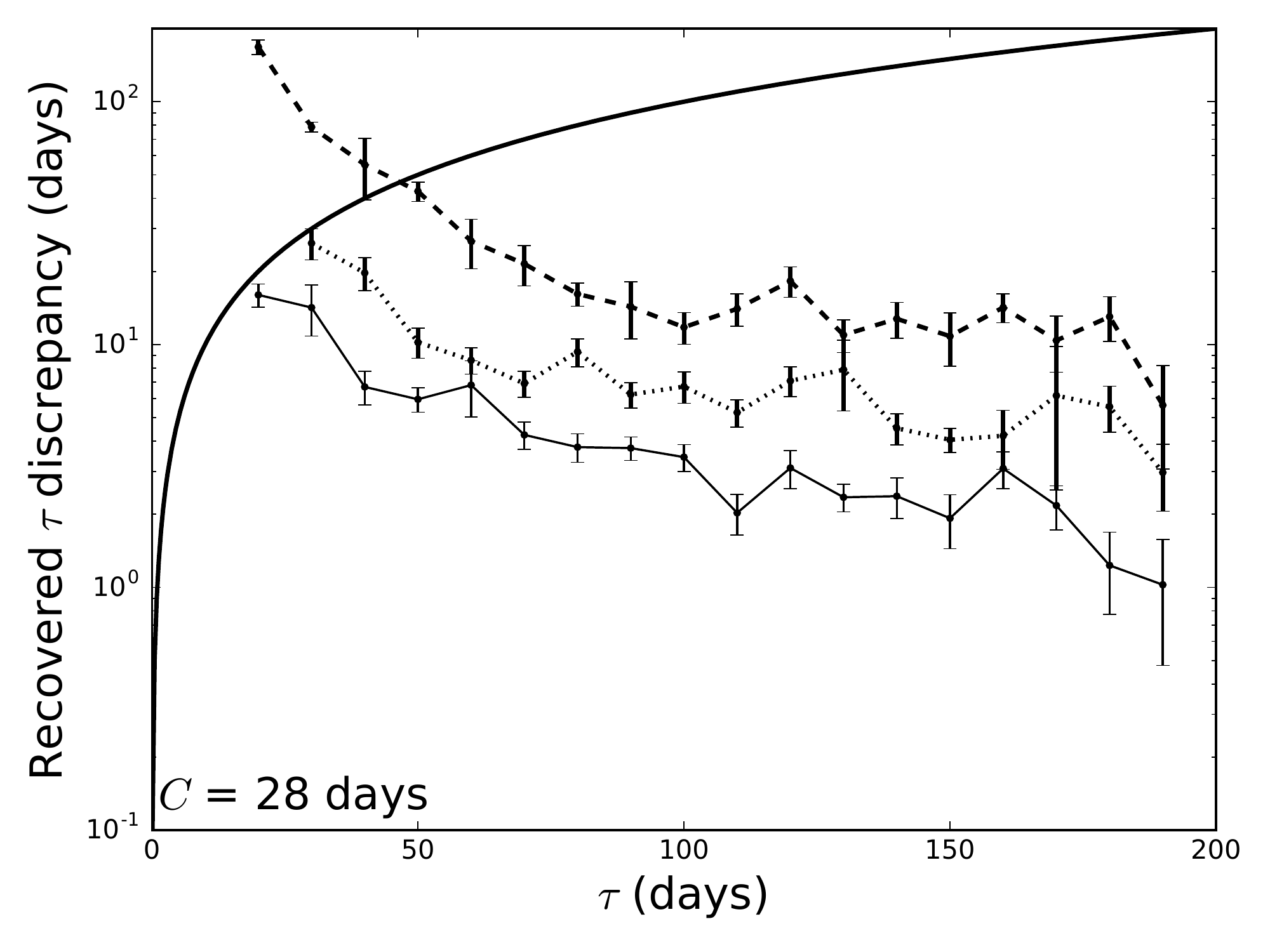} \\
    \caption{Average discrepancies in the recovered values of $\tau$ as a function of the simulated $\tau$ values for $\Delta \dot{\nu}$ values of 1 per cent (large dashes), 10 per cent (small dashes), 25 per cent (dotted) and 50 per cent (thin solid). In the right panels emission-inferred transition epochs are used, in the left they are not. The thick black line denotes the boundary at which the discrepancy in $\tau$ is equal to the value simulated. For C = 28 days, no detections were made for $\Delta \dot{\nu} = 1$ per cent.}
    \label{ddiscs}
\end{figure*}

Although without detectable profile changes we appear to detect transition pairs at lower $\tau$, the fit parameters we recover are highly discrepant from those simulated. This is shown in Figure \ref{ddiscs} in which we undertook a further 1000 simulations for specific fixed values of $\Delta \dot{\nu}$ (1, 10, 25 and 50 per cent). For each $\Delta \dot{\nu}$ we calculate the mean difference between the simulated and recovered values of $\tau$ ($\Delta \tau$) by averaging over all simulations that fall into $\tau$ windows that are 10 days wide. The resulting average differences are shown for cadences of 1, 7, 14 and 28 days. Clearly, $\tau$ can be recovered closer to the true value when $\tau$ is large and with decreasing cadence the average difference increases. For example, a $\Delta \dot{\nu} = 10$ per cent transition to a new spin-down mode that lasts 100 days can be recovered to within $\Delta \tau =3$, 10, 20 and 30 days for $C$ = 1, 7, 14, 28 days. We note that no detections were made for $\Delta \dot{\nu} = 1$ per cent with $C = 28$ days.

Figure \ref{ddiscs} also shows that not only do we not detect in areas of the parameter space corresponding to smaller transitions when using emission-inferred transition epochs, but also the recovered values of the transition parameters are generally equally discrepant from those simulated. There is, however, a reduction in the number of spurious detections at low $\tau$ for which $\Delta \tau$ exceeded the simulated value, the boundary for which is denoted by the solid black lines.  Therefore estimates of $t_i$ from mode-switching behaviour neither improves the probability of detecting smaller or shorter term timing irregularities, nor does it improve the precision with which such transitions can be resolved. However, "detections" of transition parameters that model the residuals well but do not represent the true transitioning behaviour of the pulsar's rotation are avoided when $t_i$ is estimatable in advance.

\subsubsection{Other TOA precisions}

\begin{figure*}
    \includegraphics[width=1.0\columnwidth]{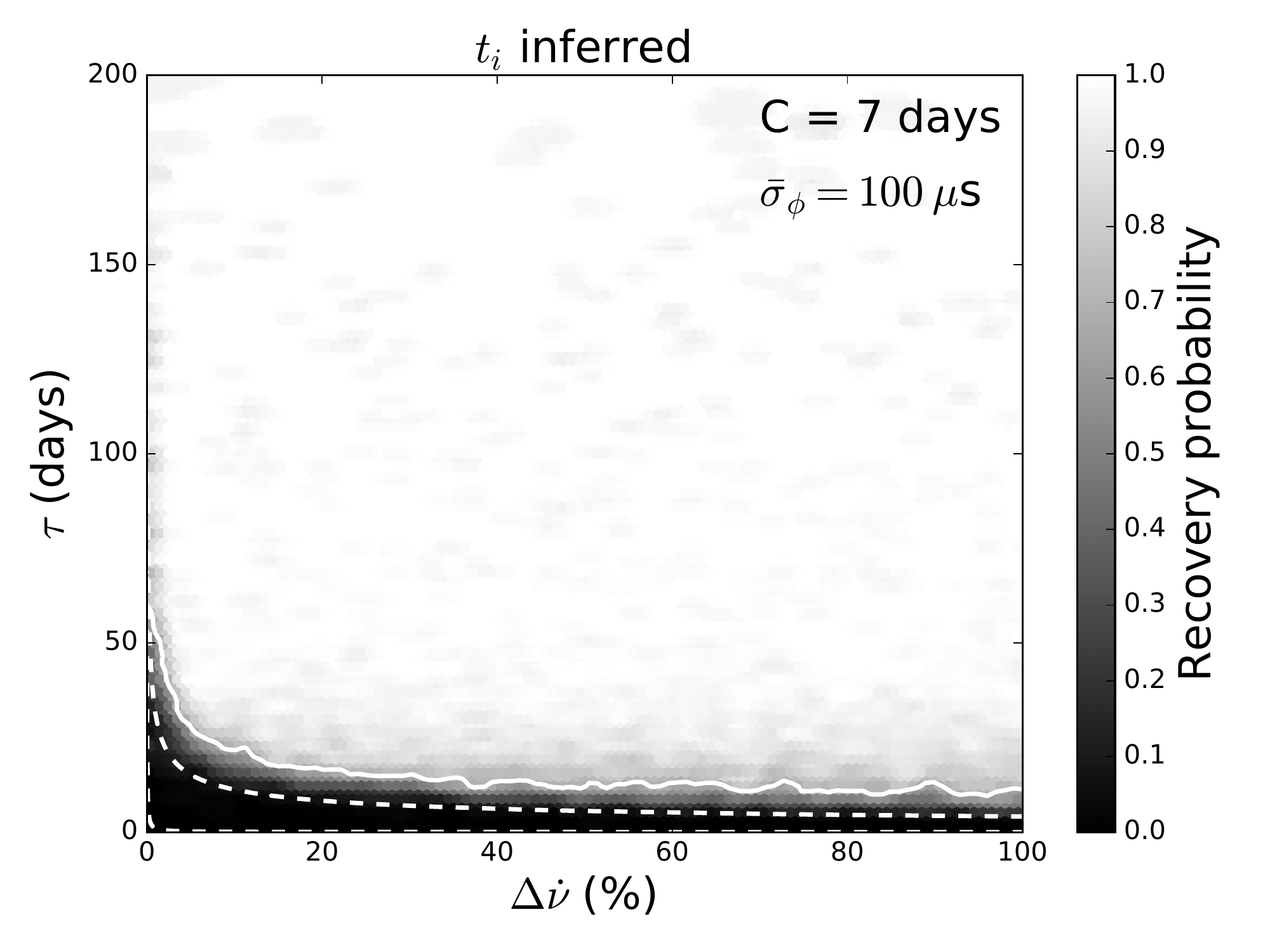}
    \includegraphics[width=1.0\columnwidth]{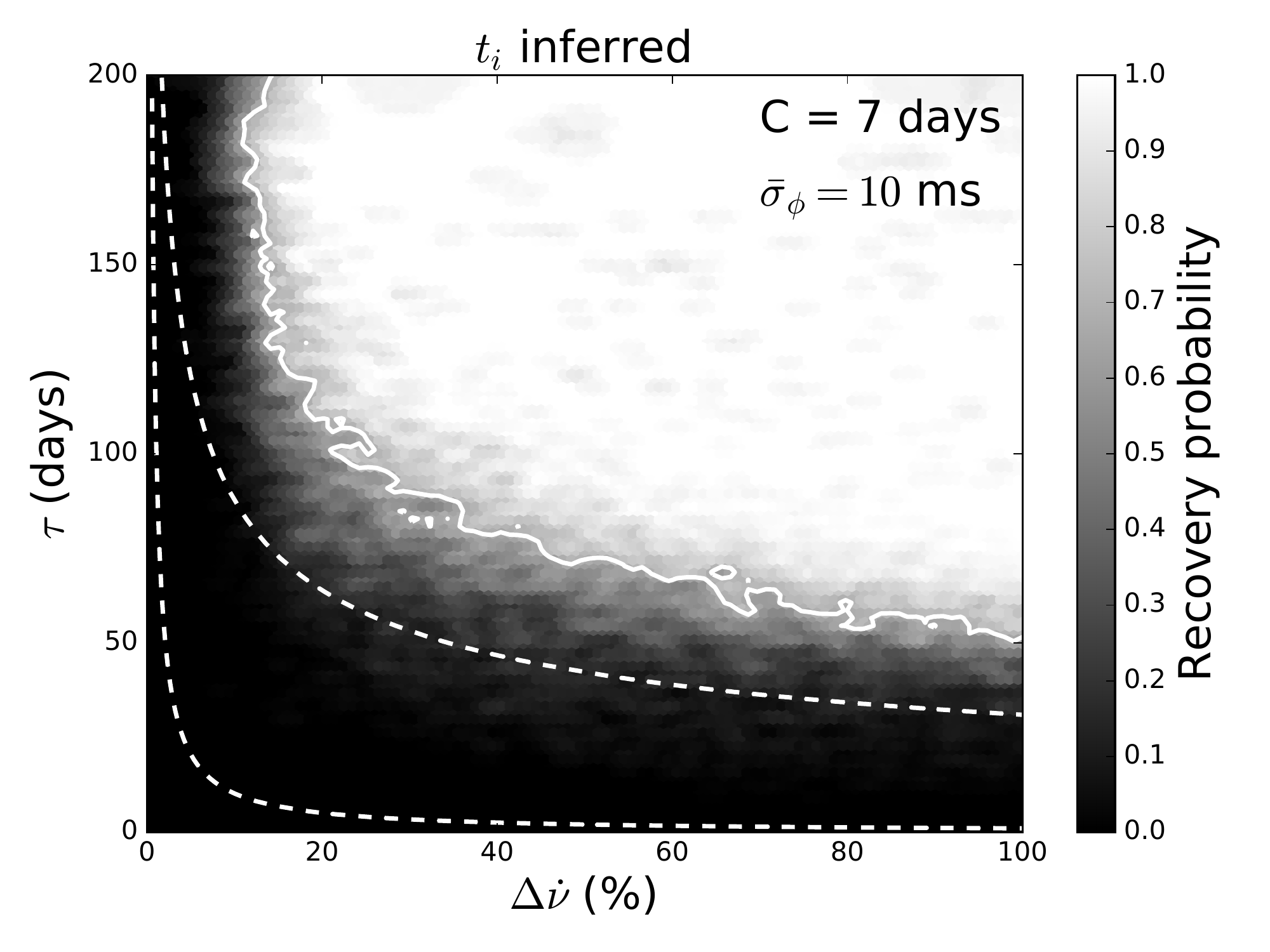}
    \caption{Heatmaps representing the detection probability for two-transitions for $\bar{\sigma}_{\phi} = 100 \: \mu$s (left) and  $\bar{\sigma}_{\phi} = 10$ ms (right). The observing cadence simulated is once per 7 days. In both cases emission-inferred transition epochs \emph{were} used in the fits.  As Figure \ref{rec_heatmaps_double_mode_no_prior} otherwise.}
    \label{other_sens}
\end{figure*}

We also examine the detectability of spin-down transition pairs with other values of the available TOA precision.  Figure \ref{other_sens} shows detection probability heatmaps where TOA precision $\bar{\sigma}_\phi = 100\:\mu$s (left panel) and $\bar{\sigma}_\phi = 10$ ms (right panel); i.e., a factor of 10 better and worse than the above simulations respectively. Clearly TOA precision has a much greater effect on detection probability than observing cadence. This is not surprising due to the $1/\bar{\sigma}_\phi^2$ dependence of the value of $\langle \chi^2 \rangle$ described in Section \ref{limits}. Where $\bar{\sigma}_\phi = 100\:\mu$s, transitions become detectable at notably lower $\tau$ compared to the $\bar{\sigma}_\phi = 1$ ms case.  For example, in the latter case (see Figure \ref{rec_heatmaps_double_mode_no_prior}, lower right panel), $\Delta \dot{\nu} = 10$ per cent transitions require $\tau$ to be 50 days or greater to have a greater than 68 per cent probability of detection.  Improving the TOA precision by a factor of 10 reduces this $\tau$ requirement to 25 days - a factor of two improvement. Conversely, where $\bar{\sigma}_\phi = 10$ ms there are no simulations for which $\Delta \dot{\nu} = 10$ per cent results in an unambiguously greater than 68 per cent detection probability. Even at the highest values of $\Delta \dot{\nu}$, no transitions are detected for $\tau < 50$ days, rendering us insensitive to a large fraction of the parameter space in which transitions exist.  We note that when sensitivity is improved, we are still limited by the cadence.  The $\bar{\sigma}_\phi = 100\:\mu$s simulations show that, even for the greatest $\Delta \dot{\nu}$,  $1\sigma$ detection probability is possible only when $\tau$ is in excess of $\sim$ 15 days - roughly twice the cadence, showing that detection is possible only when there is more than one TOA between the two transitions.

\subsection{Three transitions}

\begin{figure*}
   
    \includegraphics[width=\columnwidth]{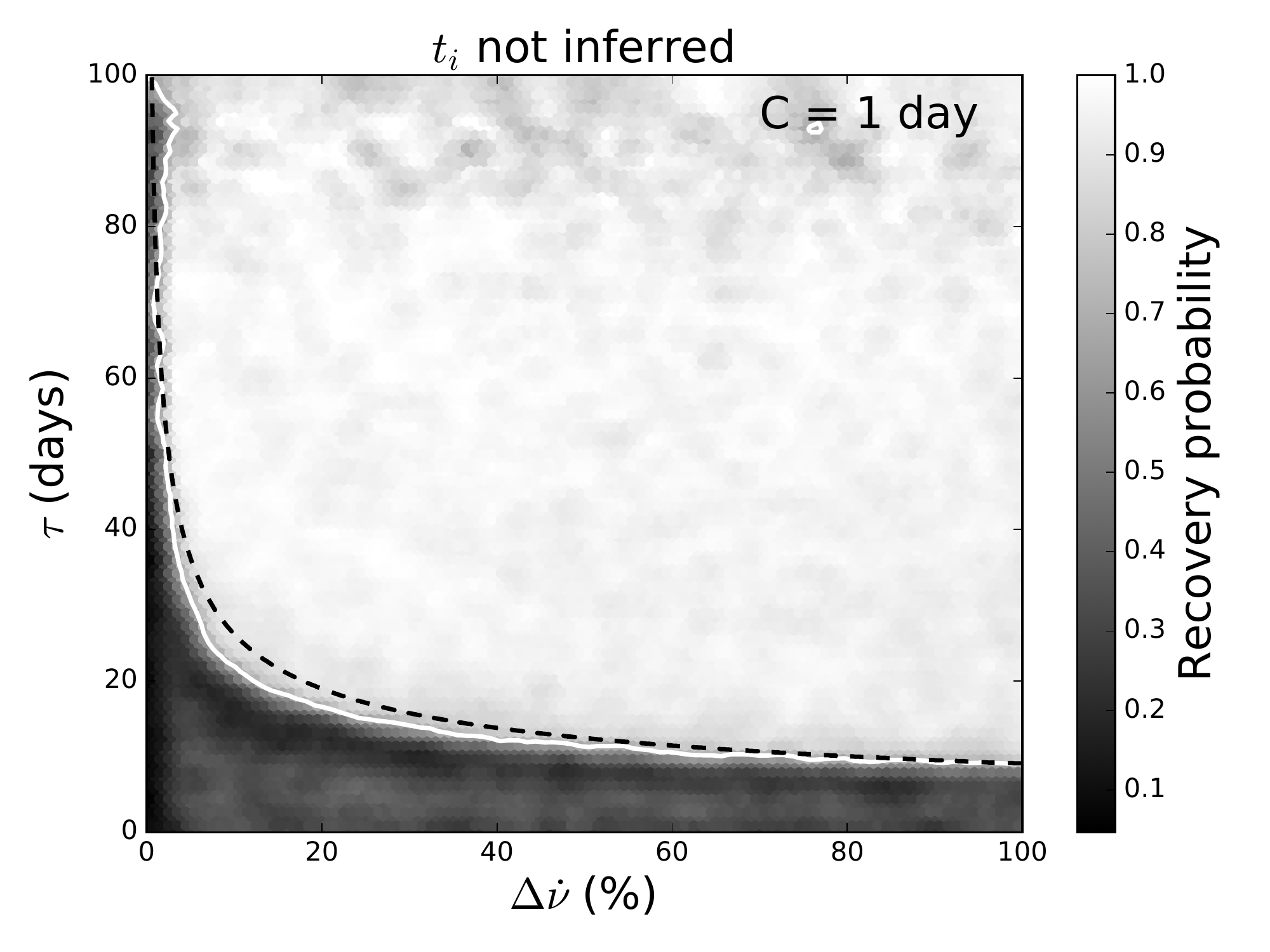}
    \includegraphics[width=\columnwidth]{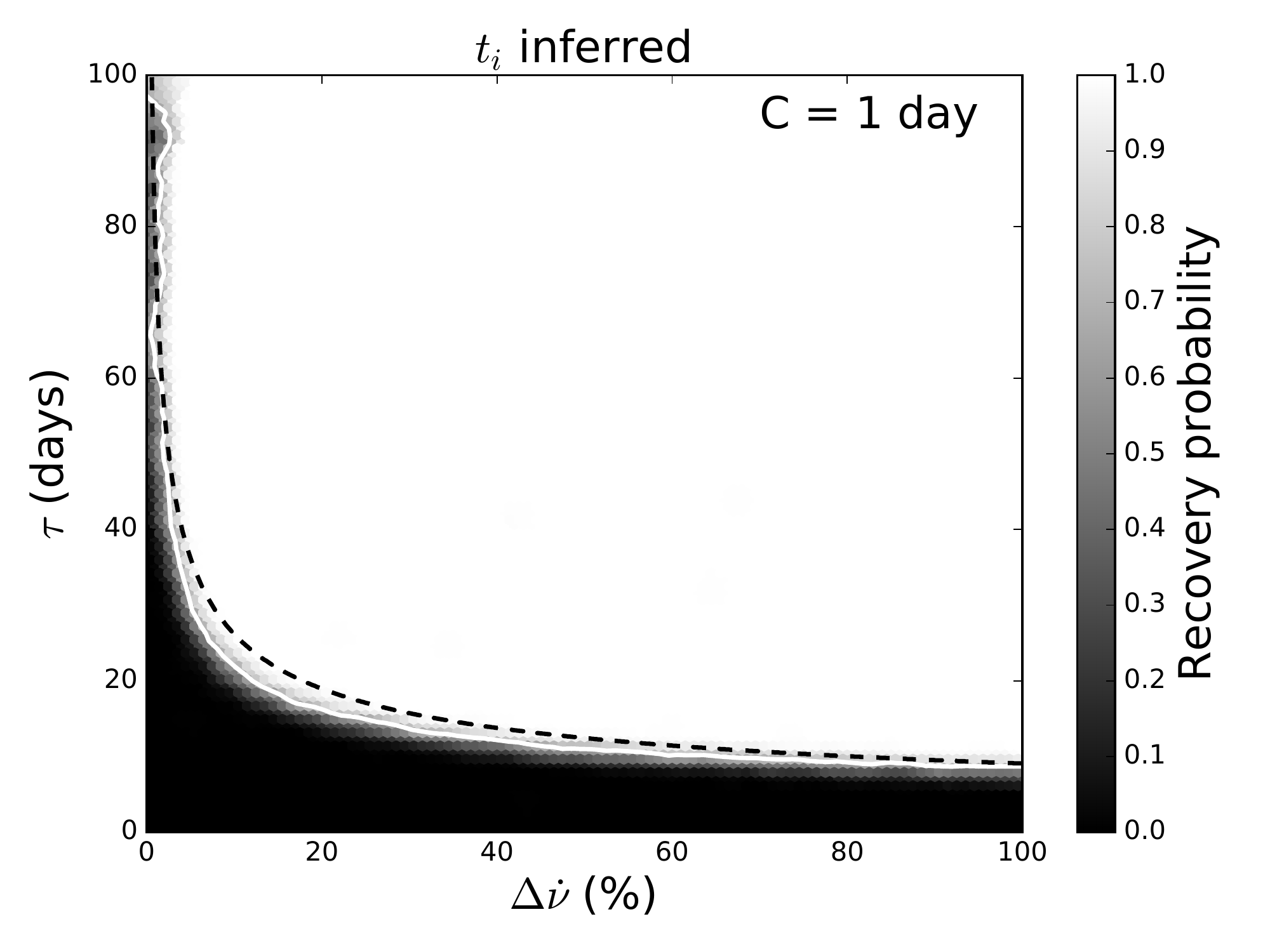}
    \includegraphics[width=\columnwidth]{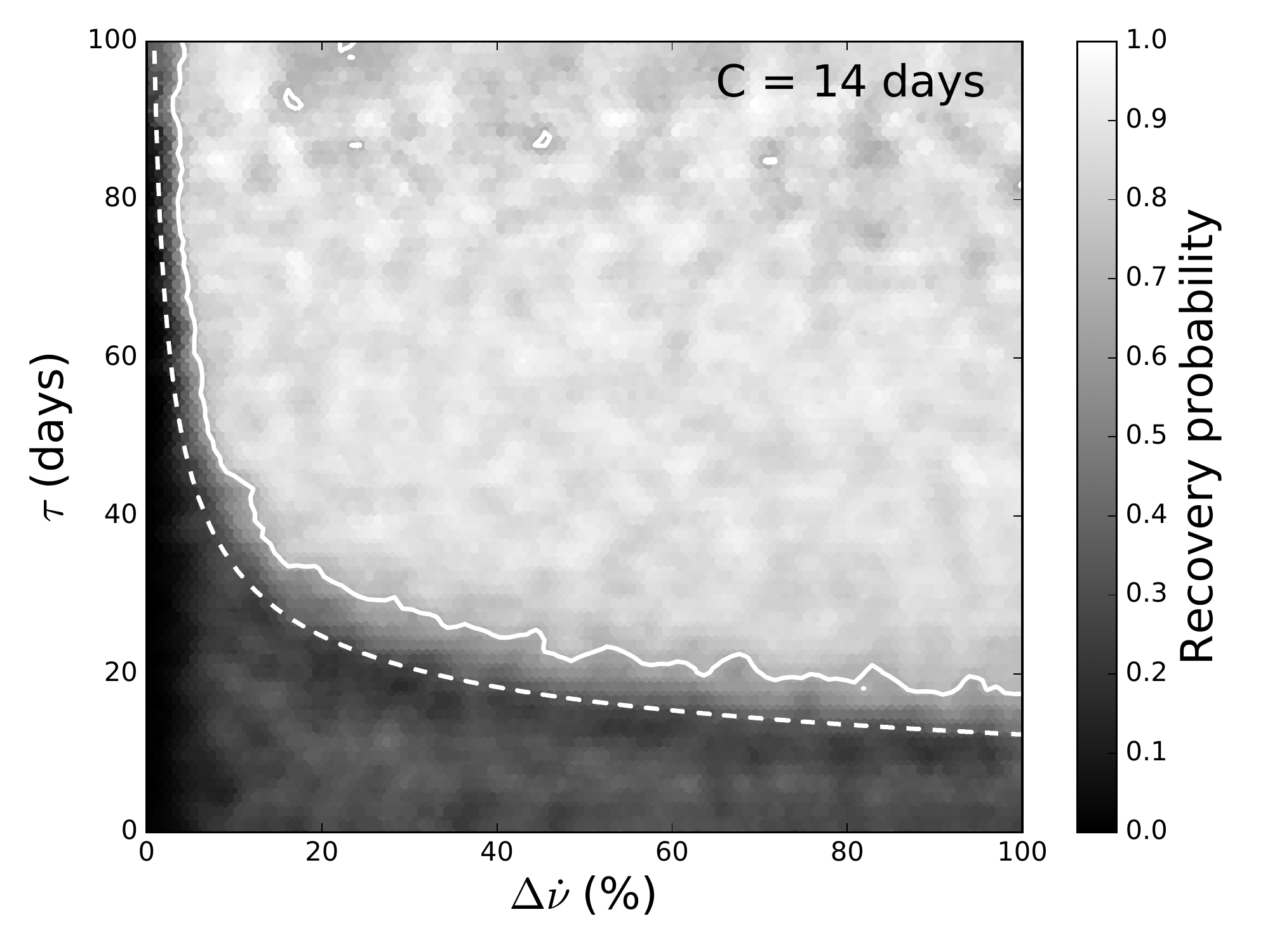}
    \includegraphics[width=\columnwidth]{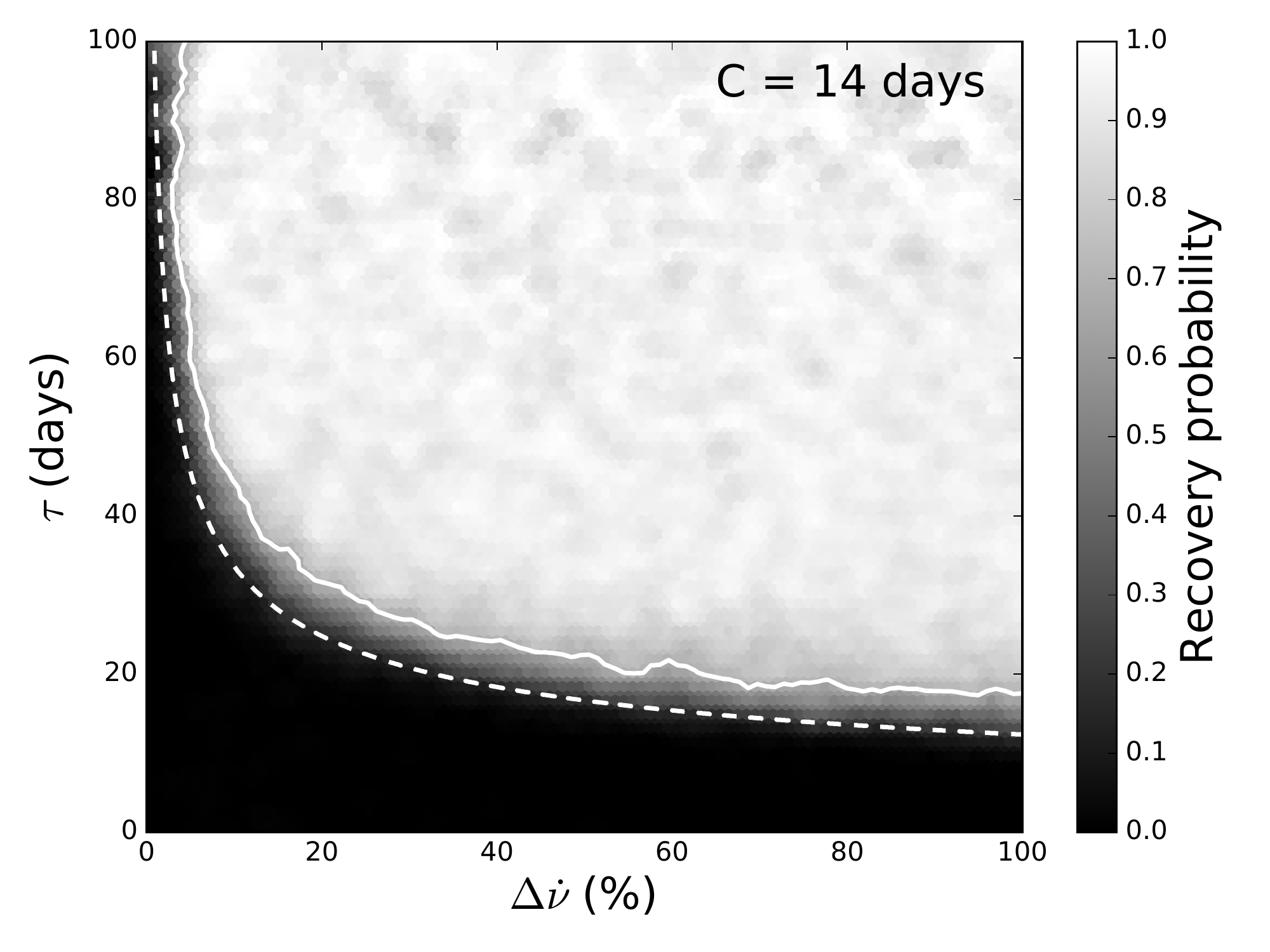}
    \caption{Heatmaps showing the fraction of simulations for which a recovery was made over the parameter space sampled.  Three transitions are simulated within each dataset. In this case we only show an analytical limit for the minimum detectable $\tau$, based on the first term of Equation \ref{chi2_3}. This is denoted by the dashed lines (black or white depending on the background colour). Note that in this case the vertical axis extends only to $\tau = 100$ days. The upper and lower plots represent $C = 1$ and 14 days respectively. As Figure \ref{rec_heatmaps_double_mode_no_prior} otherwise.}

    \label{rec_heatmaps_triple_mode_no_prior}
\end{figure*} 

Figure \ref{rec_heatmaps_triple_mode_no_prior} shows heatmaps representing the probabilities of detecting three $\dot{\nu}$ transitions as a function of their amplitudes and the interval between then. We only include maps for $C = 1$ and 14 days. These plots are constructed using the same procedures as Figures \ref{rec_heatmaps_double_mode_no_prior} and \ref{rec_heatmaps_double_mode_no_prior_2} and the average TOA error remains at $\bar{\sigma}_{\phi} = 1$ ms. Due to the difficulties described in Section \ref{three_trans_limit}, we do not include the lower limit described by Equation \ref{chi2_3}, though we do include the limit for which the first transition affects the residuals.  

There is much greater disparity in overall detection likelihood depending on whether or not emission-inferred transition epochs are used compared to the two-transitions case.  Without prior epoch constraints (Figure \ref{rec_heatmaps_triple_mode_no_prior}, left panels) there are no regions of the $\Delta \dot{\nu} - \tau$ parameter space in which there is a detection probability that is greater than 95 per cent  - even at the highest cadences.  At $C = 1$ day (Figure \ref{rec_heatmaps_triple_mode_no_prior}, top left panel) detection is less than 68 per cent likely for $\tau < 10$ days for the largest amplitude ($\Delta \dot{\nu}$ = 100 per cent) transitions.  This increases to $\tau = 20$ days for a more typical cadence of $C = 14$ days (Figure \ref{rec_heatmaps_triple_mode_no_prior}, lower left panel). At this cadence, transition amplitudes must exceed $\Delta \dot{\nu} \sim 3$ per cent for a greater than 68 per cent detection probability for the highest $\tau$. At monthly cadence (not shown), we found no regions in which detection probability is unambiguously greater than 68 per cent.   

Contrary to the two-transitions case, the use of emission-inferred transition epochs offers a significant increase in the probability of resolving transition parameters above the 68 per cent probability contour (Figure \ref{rec_heatmaps_triple_mode_no_prior}, right panels). Although the 68 per cent contour lies in approximately the same location regardless of whether they are used, the gradient in detection probability is significantly sharper when they are.  Simply blind-fitting for the $t_i$ increases the likelihood that the fitting routine finds a local minimum in the $\chi^2$ surface, resulting in a poorly optimised solution with a $\chi^2_\mathrm{r,trans} \gg 1$. Conversely, the ability to constrain $t_i$ allows one to refine a value for $\Delta \dot{\nu}$ after which we can further refine values for $t_i$ resulting in a good fit to the data.  Where $t_i$ is estimated in advance, for $C = 1,2$ days, detection probability exceeds 95 per cent when $\tau > \sim15$ days for $\Delta \dot{\nu} = 100$ per cent. This increases to $\sim$20 days for $C = 7$ days.  For $C = 14$ days, 95 per cent probability is not unambiguously achieved.  

Figure \ref{rec_heatmaps_triple_mode_no_prior} also shows that detectability appears more probable for low $\Delta \dot{\nu}$, low $\tau$ transition sets when emission inferred transition epochs are not used. When $\tau$ is long, all three transitions contribute to the overall distribution of the residuals in the dataset.  When $\tau$ is short, the distribution of residuals is dominated by the last transition only. In this case, as described in Section \ref{three_trans_limit}, it is possible that a single transition is able to model all three. The remaining two epochs are placed very close together elsewhere in the dataset such that their nett effect on the residuals is negligible. In such cases, transition sets are "detected" however the recovered values are highly discrepant from those simulated.  Therefore, where there are three transitions in a dataset, inferring transition epochs from profile variability, not only increases the probability of resolving discrete spin-down states but also rules out the detection of spurious transition parameters.  

\begin{figure*}
    \includegraphics[width=0.9\columnwidth]{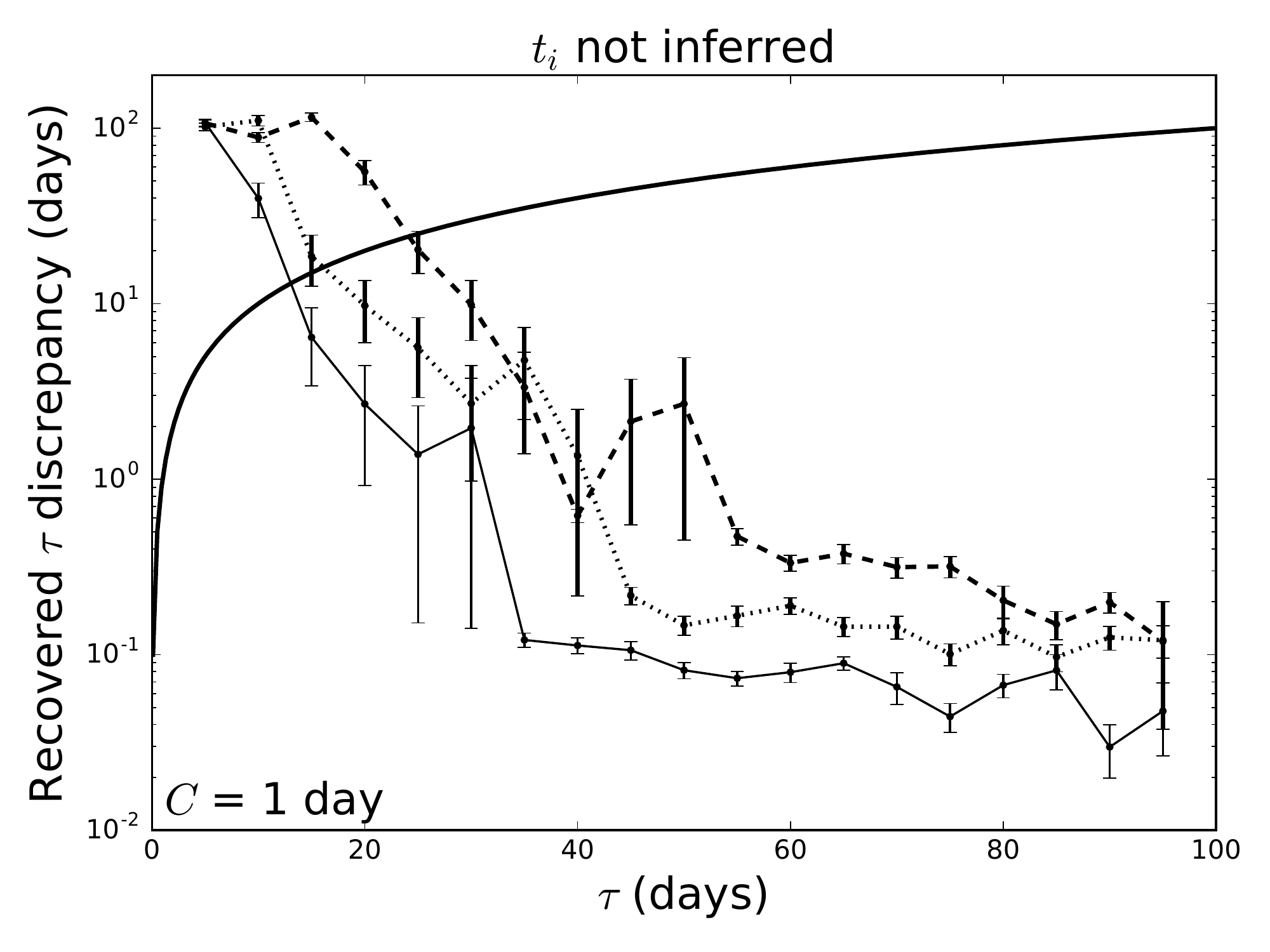}
    \includegraphics[width=0.9\columnwidth]{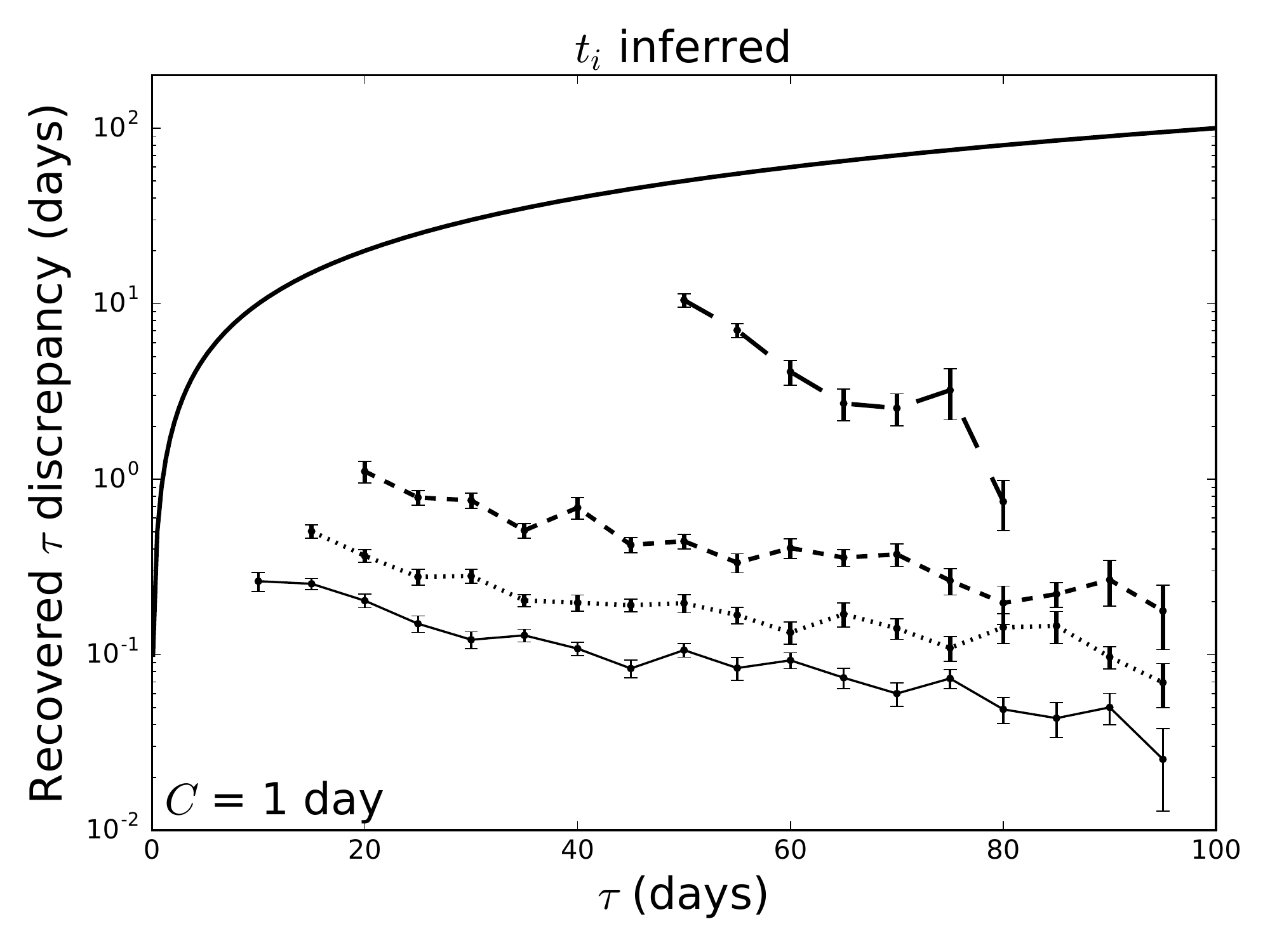} \\
    \includegraphics[width=0.9\columnwidth]{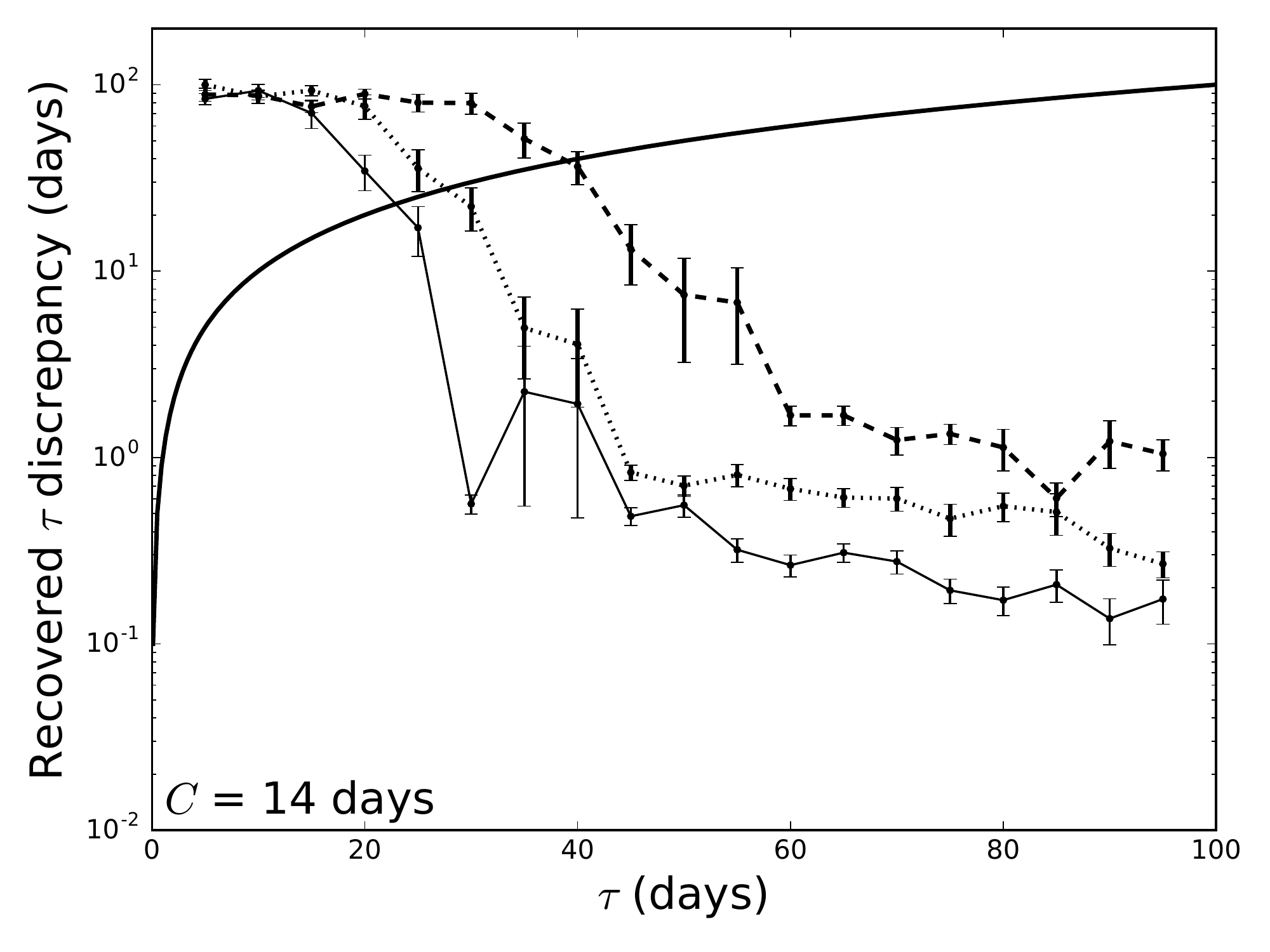}
    \includegraphics[width=0.9\columnwidth]{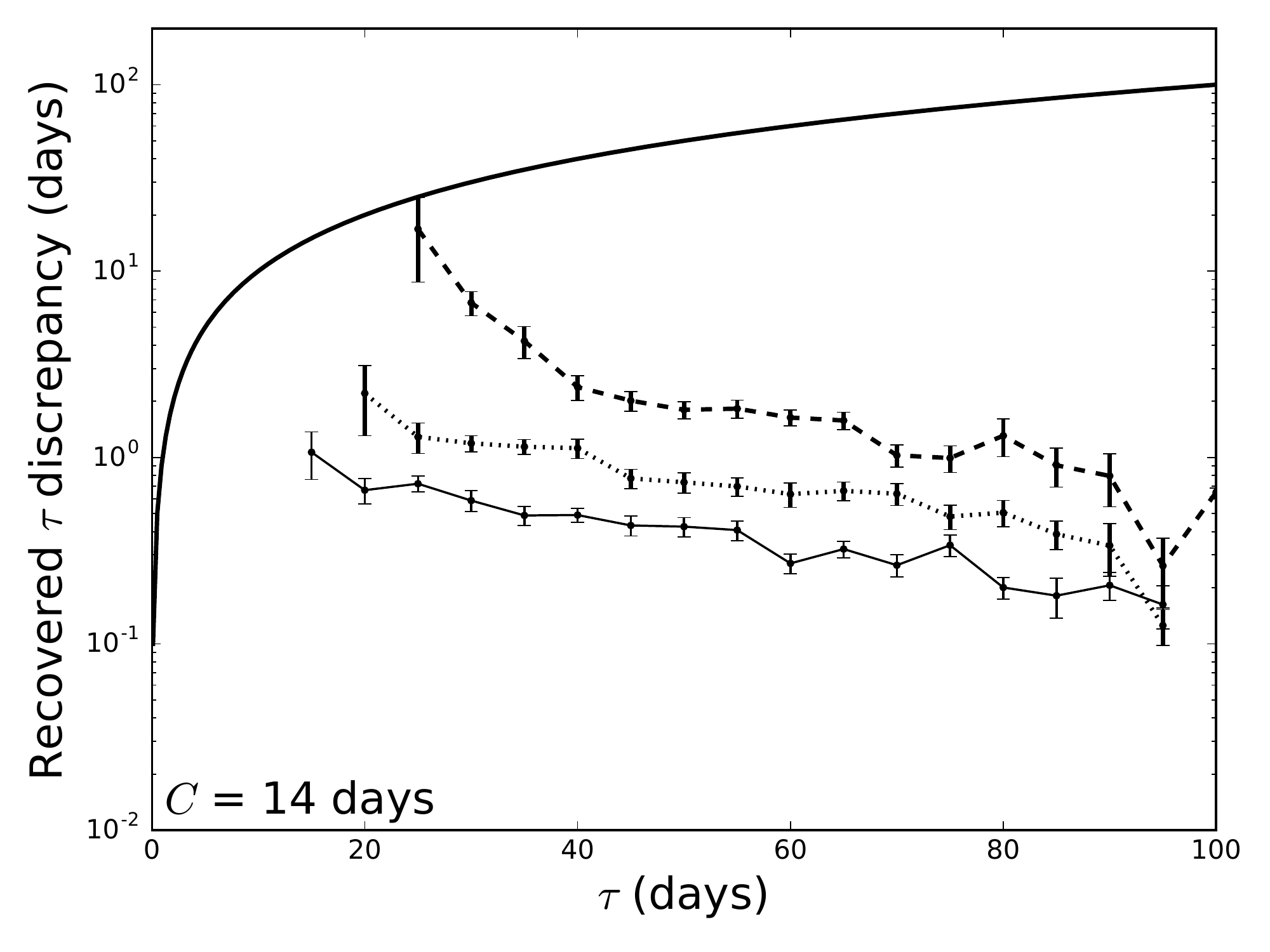} \\
    \caption{Average discrepancies in the recovered values of $\tau$ as a function of the simulated $\tau$ values where three transition exist in a dataset. As Figure \ref{ddiscs} otherwise.}
    \label{3ddiscs}
\end{figure*}

We show the differences between the simulated and recovered values of $\tau$ for three transitions in Figure \ref{3ddiscs} for the cadences of 1 and 14 days, both for the cases when emission-inferred transition epochs are used (right) and when they are not (left). For a given $\Delta \dot{\nu}$, the discrepancy in $\tau$ increases as cadence worsens, as expected. For example, when $\Delta \dot{\nu} = 50$ per cent, $\tau$ was recovered to within $\sim$0.1, 0.3, 0.6 and 1 days for $C = 1$, 7, 14 and 28 days respectively regardless of whether epochs were estimated a priori. Figure \ref{3ddiscs} clearly shows an excess of detections at low $\tau$ when $t_i$ are not constrained in advance (left panels).  In many cases $\Delta \tau$ is greater than the values in the simulation showing that the detections in this region do not represent the true transitioning behaviour of the pulsar. We also note that at low $\tau$, for a given cadence, the lines representing the $\Delta \tau$ for each value of $\Delta \dot{\nu}$ are less delineated from one another when emission-inferred transition epochs are not used. This arises for two reasons. Firstly we are more likely to make a discrepant detection at low $\tau$. Secondly, there are generally fewer detections overall at low $\tau$ than high $\tau$. The result is greater $\Delta \tau$ values in a smaller number of detections and therefore a greater level of noise in the distributions.

At the lowest cadence simulated ($C = 28$ days, not shown), apart from the tendency to spuriously detect low $\tau$ transitions without prior epoch estimates, the discrepancies become very similar.  At such a low cadence, not only are there fewer TOAs with which to fit a timing model (for the highest $\tau$, there are $\sim$3 TOAs per spin-down state) but estimates of $t_i$ become poorly constrained as there could be up to 14 days between a true transition epoch and the epoch estimated from profile changes.  Therefore as cadence worsens the practical benefit of estimating the transition epochs reduces as the estimates become little better than guesses. 

\section{Discussion}
\label{disc}

Timing noise is widespread across the pulsar population and has been shown, in some cases, to be at least partially attributable to a transitioning frequency derivative.  In a number of these cases, modulations of the frequency derivative correlate with the nature of the pulsed emission, either by complete cessation of detectable pulses \citep{klo+06} or by changes to the overall shape of the pulse profile \citep{lhk+10}.  Such pulsars potentially offer a method for correcting timing noise if changes to the integrated pulse profile can be used as a proxy for epochs at which $\dot{\nu}$ transitions occur. 

It is likely that the prevalence of timing noise is underestimated due to the current limitations of telescope sensitivity and constraints on attainable cadence.  Pulsars whose timing residuals are dominated by TOA uncertainties (white noise), may begin to exhibit detectable fluctuations of their rotational parameters as new, more sensitive facilities become available in the near future.  With this in mind, it is important to understand these current limitations and what the potential improvements are with the next generation of radio telescopes.

We have undertaken simulations of pulsar rotation where $\dot{\nu}$ transitions between two well-defined values a number of times within a dataset.  We have then attempted to detect these transitions using standard phase-coherent timing techniques.  From this we have been able to formulate a case study for a particular set of scenarios and have measured the probability of detection for a given set of transition parameters (switching timescale $\tau$, amplitude $\Delta \dot{\nu}$, and observing cadence $C$), as well as the average precision with which those parameters are recoverable.

Where two transitions of equal magnitude $|\Delta \dot{\nu}|$ and opposite direction exists in a dataset, we have defined the minimum detectable interval $\tau$ between them in terms of the expected $\chi^2$ of the residuals after the first transition (see Section \ref{two_trans_limit}). Our simulations of transition pairs have shown that a $1\sigma$ probability of resolving these transitions is consistent with these limits. They also show this consistency both when we assume that the nature of the pulsed emission yields estimates of transition epochs and when it does not. We have also demonstrated that the observing cadence, though having only a slight effect on the minimum detectable $\tau$, has a large impact on the probability of detection of a given set of transitions. For decreasing observing cadences, true detectability rapidly departs from this limit and only transitions which are much larger (or have longer transition timescales) are resolvable.  For studies of short timescale rotational variability, this means that cadence cannot be compromised when new more sensitive telescopes become available.  

We have also simulated our ability to resolve three transitions in a dataset, again demonstrating the strong impact of cadence on detection probability. We have also highlighted the difficulty in defining lower limits on detectable $\tau$ due to the degeneracy between timing models containing one and three transitions (see Section \ref{three_trans_limit}).    

\subsection{Do observed pulsed emission variations assist in resolving $\Delta \dot{\nu}$?}

Where two transitions are present in a dataset, we find no improvement in the detectability of spin-down changes for the majority of the parameter space simulated when we include prior estimates of the transition epochs in the trial timing solutions. For two transitions, blind fitting is sufficiently effective with the number of trial initial solutions used here. However, we have shown that there is an advantage when transitions are small and closely spaced in time.  Comparing Figures \ref{rec_heatmaps_double_mode_no_prior} and \ref{rec_heatmaps_double_mode_no_prior_2} shows that detectability extends lower in $\tau$ when transition epochs are not estimated in advance. This is especially notable at the lowest cadences simulated. These differences occur when the $\tau$ has a comparable timescale to the observing cadence. In this regime, there are too few data points in each spin-down state to resolve the individual transition epochs and this scenario extends to higher $\tau$ transitions with decreasing cadence.  When allowing a free choice of epochs at low $\tau$, there is a greater probability that the fit will find a solution that is not representative of the true transitioning behaviour.  In such cases, residuals may be well modelled locally but the transition parameters obtained are not representative of the true rotational behaviour of the pulsar. Good initial estimates of the transition epochs mitigates this as, where $\tau \approx C$, non-detections are more likely than "false" detections, so that misleading timing solutions are avoided.

PSR B1931+24 is a favourable case for detecting spin-down correlated emission modes as its periods of no emission are of long duration (spending $\sim$10 days in one spin-down state and then $\sim$40 days in another), $\Delta \dot{\nu} \approx 50$ per cent, and it was one of few pulsars to be monitored on an almost daily basis by the Lovell telescope. Using Equation \ref{chi2_2} to evaluate the expected $\chi^2$ for such transitions when the pulsar is observed daily we find $\langle \chi^2 \rangle \approx 6000$ for $T = 50$ days. For 50 post $t_1$ data points using a p-value of 0.003 $\chi^2_{\mathrm{crit}} \approx 80$, indicating that the timing residuals are significantly affected by this transitioning behaviour. We undertook individual simulations for the transitioning behaviour in PSR B1931+24 and found that, though the predicted timing noise was present in the residuals, the underlying transitions could not be found by fitting. To understand this we can scale PSR B1931+24's transition parameters to the values in our simulations as described at the end of Section \ref{model}. PSR B1931+24 spends 40 days in spin-down state A, undergoes a $\Delta \dot{\nu} = 6 \times 10^{-15}$ Hz s\textsuperscript{-1} (50 per cent) transition to state B where it remains for $\tau = 10$ days before transitioning back to state A for a further 40 days (totalling 90 days).  In this case $T = 40+10=50$ days and $C = 1$ day.  To make our simulations applicable to this scenario we can scale the dataset length of PSR B1931+24 by 5.6 times to 500 days. $T$ and $\tau$ are scaled likewise to 56 and 280 days respectively. We must then scale PSR B1931+24's $\Delta \dot{\nu}$ to $6 \times 10^{-15} / 5.6^2 = 2 \times 10^{-16}$ which corresponds to a 2 per cent change to our initial $\dot{\nu}$ (see Table \ref{properties}).  The cadence also decreases to once per 5.6 days.  Although this is not a cadence we simulated in Section \ref{two_trans}, our $C = 7$ days simulations are sufficiently close to approximate the detection probability for this scenario.  Inspection of Figure \ref{rec_heatmaps_double_mode_no_prior} (second panels) shows that transitions with $\Delta \dot{\nu} = 2$ per cent and $\tau = 56$ days have an approximately 25 per cent probability of detection whether or not $t_i$ is estimatable in advance. However, if $t_i$ are well constrained (i.e, $\tau \gg C$), it is possible to model the transitions without refinement of the initial epoch estimates. In other words we can fit for $\Delta \dot{\nu}$ only, leaving $t_i$ fixed at the estimates. In this sense, inferring $t_i$ in advance is beneficial for such borderline cases, provided the cadence is sufficiently high.  

Where three transitions are modelled, significant improvements are seen from the use of emission-inferred transition epochs at high cadence (see Figure \ref{rec_heatmaps_triple_mode_no_prior}).  As cadence worsens the two scenarios become less distinct due to increasingly poor constraints on $t_i$. At high values of $\Delta \dot{\nu}$ and $\tau$, there are much greater difficulty in resolving the transitions without prior estimates on $t_i$ as the fitting procedure commonly finds local minima that do not represent the true global minimum of the $\chi^2_{\mathrm{trans}}$ surface. As a result, the fit $\chi^2_{\mathrm{r,trans}}$ is not close enough to unity to constitute a detection however this could be addressed by using a larger number of trial solutions. At low $\tau$ a great number of detections occur in which the fit parameters are highly discrepant from those simulated. In these cases, we find that the effect of the transitions on the residuals is strongly dominated by the last transition only. This results in a detection for which two of the three transition epochs are located too close together in time to affect the residuals with all three transitions being represented by one only.  Though in these cases the fit $\chi^2_{\mathrm{r,trans}} \approx 1$, the recovered transition parameters are not correct.  The use of emission-inferred transition epochs mitigates both of these effects, resulting in high $\tau$ transitions being correctly resolved whilst inhibiting the "false" detection of low $\tau$ transition sets.    

We must address the point that in all of our simulations, the number of transitions in a dataset is known in advance and we fit only for that number.  In reality, when searching for transitions in a dataset one would fit for a smaller number of events and iteratively model any remaining structure.  If a pulsar exhibits emission-correlated spin-down modes, we are able to not only infer the epoch of transition but also infer how many of them there are - provided the pulsar is observed with sufficient cadence.  

\subsection{Pulsars with known variable emission}

\begin{center}
\begin{table*}

    \captionof{table}{The spin-down and emission characteristics of a number of pulsars which exhibit mode-switching, nulling or intermittency.  $\tau$ refers to the characteristic timescale of the emission changes and $T$ refers to the expected time between two identical transitions.  Except where underlined, the $\dot{\nu}$ variations occur on the same timescale as the emission variations.  Column 8 describes whether or not $\dot{\nu}$ variations are expected to be detectable at the current sensitivity of the receiver, according to Equation \ref{chi2_2} assuming spin-down is correlated with the emission mode.  The values of $\bar{\sigma}_{\phi}$ quoted are based on the order of the mean TOA uncertainty on the Lovell/Parkes data for that pulsar. PSRs B1828$-$11 and B1822$-$09 are examples of pulsars whose switching timescale is rapid but results in a longer term modulation.  Here the shorter modulation timescales are quoted for these pulsars (see text).}
    \begin{tabular}{ | l | l | l | l | l | l | l | l | l |}
    \hline
    Pulsar &      Emission variability &    $\tau$ & $T$ &  $\Delta \dot{\nu}$ (\%) & $\bar{\sigma}_{\phi}$ & $\bar{C}$ &   Detectable $\Delta \dot{\nu}$? & Reference  \\ \hline
    B0740$-$28 &    Mode-switching &          $\sim$50 d & 100 d &   $\sim 1$ &  20 $\mu$s & 8 d &                     Yes &  \cite{lhk+10} \\
    B0823+26 &    Nulling &                 $\sim$6 h & 1.25 d & $< 6$  &  100 $\mu$s & 3 d &      No  & \cite{ysw+12} \\
    B1540$-$06 &    Mode-switching &          $\sim$2 y &  $\sim$4 y &  2 &  100 $\mu$s & 19 d &                  Yes &          \cite{lhk+10} \\  
    J1634$-$5107 &  Intermittent &            $\sim 1$ d & $\sim$2 d & - &  1 ms & 10 d &                No &  \cite{ysw+15}\\
    J1717$-$4054 &  Nulling/Intermittent &    $\sim$10 m & $\sim 1$ h &  \underline{33} &   1 ms & 15 d &   No &\cite{ysw+15} \\
    B1822$-$09 &    Mode-switching &          $\sim$Minutes & $\sim$Minutes &  \underline{3} &  100 $\mu$s & 6 d &  No &   \cite{lhk+10} \\
    B1828$-$11 &    Mode-switching &          $\sim$Minutes & $\sim$Minutes  & \underline{1} &  50 $\mu$s  &   10 d       &   No  &          \cite{lhk+10} \\ 
    J1832+0029 &  Intermittent &            $\sim$600 d &$\sim$1400 d &  44 &  500 $\mu$s &  13 d  &   Yes &    \cite{llm12} \\   
    J1841$-$0500 &  Intermittent &            $\sim$600 d &$\sim$800 d  & 250 &  10 ms &  7 d&  Yes &   \cite{crc12} \\
    J1853+0505 &  Nulling/Mode-switching &  $\sim$Minutes & $< 1$ h & - & 10 ms & 50 d  & No &           \cite{ysw+15} \\
    B1859+07 &    Mode-switching &          $\sim$180 d & $\sim$350 d &  7  &  700 $\mu$s & 60 d  &  Yes &          \cite{psw+16}\\
    B1931+24 &    Intermittent &            $\sim$10 days &$\sim$50 d  &  50 &  1 ms & 1 d &      Yes &          \cite{klo+06}\\ 
    B2035+36 &    Mode-switching &          $> 5$ y & - &  13 &  200 $\mu$s & 13 d       &  Yes &          \cite{lhk+10} \\
    J2043+2740 &  Mode-switching &          $\sim$4 y & $>7$ y  &  6 &  30 $\mu$s  & 12 d                    & Yes &          \cite{lhk+10} \\
    \hline
    \end{tabular}
    \label{nullers}
\end{table*}
\end{center}

An accurate census on the number of pulsars that show variable emission is yet to be reached.  Many apparently stable pulsars may be transitioning between two or more similar integrated profile shapes that are indistinguishable from one another with current telescope sensitivity. Additionally, statistics on pulsar intermittency are ill-constrained due to observational biases against finding them.  Nevertheless, with the sample we have, integrated pulse intensity modulations are known to occur over a wide range of timescales, though in only a handful of cases is such behaviour known to correlate with $\dot{\nu}$.  Table \ref{nullers} shows the $\Delta \dot{\nu}$ (where known) and emission variability timescales for a selection of nulling/intermittent/mode-switching pulsars.  Where both $\Delta \dot{\nu}$ and the timescale over which the emission behaviour varies, $\tau$ are quoted, there is clear correlation between the value of the spin-down rate and the profile shape.  This is with the exception of those whose $\Delta \dot{\nu}$ values are underlined.      

\cite{lhk+10} showed that six mode-switching pulsars exhibit an emission-rotation correlation.  In all cases the $\dot{\nu}$ transitions are of low amplitude compared with those of PSR B1931+24 (by far the largest is in PSR B2035+36 with a $\Delta \dot{\nu}$ of 13 per cent).  All others are less than 6 per cent.  Three of these six undergo $\dot{\nu}$-variations on timescales of the order of 100s of days and the transitions are resolvable with relatively low cadence.  For example, PSR J2043+2740 was seen to undergo two large transitions $\sim1500$ days apart with amplitudes of $\Delta \dot{\nu} \sim \pm 6$ per cent respectively. Following the second transition back to the initial $\dot{\nu}$, the pulsar remained in this state for a further 1000 days until the end of the dataset. The Lovell telescope is currently able to time this source with an average TOA precision, $\bar{\sigma}_{\phi} \sim 30$ $\mu$s and it observes PSR J2043+2740 roughly once every 12 days.  Calculating the $\langle \chi^2 \rangle$ from Equation \ref{chi2_2} results in a value that is in excess of the critical value for $T/C$ data points. This is also the case when we ignore the second term in Equation \ref{chi2_2} and compare the result to $\chi^2_{\mathrm{crit}}$ for $\tau/C$ date points. This shows that these two $\dot{\nu}$ transitions are eminently resolvable.  

PSR B0740$-$28 is the most rapidly mode-switching source in the Lyne sample and exhibits the weakest correlation between its pulse shape and $\dot{\nu}$-state (of those that showed \emph{any} correlation). Its value of $\dot{\nu}$ oscillates by $\sim$0.7 per cent over a quasi-periodicity of $\sim$100 days.  Assuming the pulsar transitions back-and-forth between these $\dot{\nu}$ states every 50 days, current Lovell observing parameters for this source ($\bar{\sigma}_{\phi} \sim 20$ $\mu s$ and $\langle C \rangle = 8$ days) allow us to predict a $\langle \chi^2 \rangle$ from Equation \ref{chi2_2} of $\sim10^8$ - well in excess of $\chi^2_{\mathrm{crit}} = 31$ for $T/C = 13$ data points. In fact, evaluating the first term only results in $\langle \chi^2 \rangle = 10^6$ (whereas $\chi^2_{\mathrm{crit}} = 20$ for $\tau/C = 6$ data points).  

The mode-switching pulsars B1822$-$09 and B1828$-$11 were shown to exhibit $\dot{\nu}$-variations on timescales of years but single pulse analysis shows the emission behaviour changes on significantly shorter timescales (minutes/hours).  This however, does not rule out emission-rotation correlations.  As determination of the profile shape and spin-down rate at some reference epoch is achieved by averaging over all epochs within some characteristic windowing timescale, shorter timescale variations are smoothed out.  The time-averaged values of the pulse shape and $\dot{\nu}$ depend on the fraction of the time the pulsar spends in each state within the windowing period.  A slowly changing ratio of the two states produces the smooth, long-term variations in average $\dot{\nu}$ and profile shape which oscillates on timescales much longer than that of the true switching timescale. It is possible that $\dot{\nu}$ transitions, if occurring on the same timescales as the profile variations, may never be resolvable if they occur on timescales less than the time it takes to form an integrated pulse profile.

PSR J1717$-$4054 has a $\langle\dot{\nu}\rangle = -4.6737(7) \times 10^{-15}$ Hz s\textsuperscript{-1} and spends 75 per cent of the time in its \emph{off} mode. Although peak-to-peak $\dot{\nu}$ variations of up to 33 per cent have been measured \citep{ysw+15}, the observing cadence was insufficient to allow an accurate determination of any periodicity, compared to the switching timescale, in $\dot{\nu}$. If emission-correlated spin-down modes are occurring in PSR J1717$-$4054 then a peak-to-peak $\Delta \dot{\nu}$ variation of 33 per cent is clearly not sufficient to cause a detectable deviation of the TOAs from a timing solution containing a single frequency derivative at the quoted sensitivity.  In fact, were the pulsar observed with a TOA precision of 1 ns, 33 per cent variations in $\dot{\nu}$, occurring on a timescale of 2 hours would not be resolvable even when monitored almost continuously. 

No measurable $\Delta \dot{\nu}$ variations were detected in PSR J1853+0505 \citep{ysw+15} which undergoes emission changes on timescales of just minutes.  A TOA precision of 1 ns is required to resolve $\Delta \dot{\nu} = 1000$ per cent or larger occurring on the same timescale.  This very short timescale however would require that the pulsar be monitored continuously to enable measurement.

The radio emission properties of the nulling PSR B0823+26 were shown to be variable on both short (minutes) and long (hours) timescales \citep{ysw+12} with an average ON/OFF time of 1.4 days and 0.6 hours respectively. Though evidence of spin-down variations was not found, an upper limit of $\Delta \dot{\nu} < 6$ per cent has been established.  Currently the Lovell telescope can time B0823+26 to an average TOA precision $\bar{\sigma}_{\phi} \sim 100 \; \mathrm{\mu s}$.  According to Equation \ref{chi2_2} spin-down transitions of amplitude 6 per cent would not affect the residuals even if the pulsar were monitored continuously with this TOA precision.  Observing this source 8 times a day with 10 ns TOA precision would allow spin-down transitions with these parameters to be resolved according to Equation \ref{chi2_2}.

The detection of $\dot{\nu}$ variations in all pulsars listed in Table \ref{nullers} is consistent with the limits derived in Section \ref{limits}.  In some of the cases listed where spin-down variations are observed, there is a clear change in the emission behaviour of the pulsar - that being that the pulsar's detected emission ceases for a measurable period of time.  In many cases however, more subtle changes to the emission are observed, (i.e. mode-switching).  In such scenarios we are reliant on there being sufficient sensitivity to be able to resolve distinct profile shapes, such as those discussed in \cite{lhk+10}.  Often, timing noise is seen in pulsars with apparently stable pulse profiles indicating that either 1) the integrated profiles are intrinsically stable, 2) our ability to resolve profile variations is limited by the sensitivity of the receiver, 3) no profile variations occur at the particular observing frequency or 4) the line-of-sight does not cross an affected part of the emission region.  In these cases, assuming the timing noise can be at least partially attributed to spin-down variations, no emission-inferred transition epochs are available and a blind search for rotational irregularities, (such as the striding technique in \cite{lhk+10})  must be utilised.

\subsection{Sub-threshold $\dot{\nu}$-variations} 

Resolving short timescale $\dot{\nu}$ transitions requires that their amplitude is sufficient to cause a departure of the TOAs from the timing solution by the time the pulsar switches back to the modelled spin-down rate (or by the time the dataset ends).  A pulsar whose timing residuals are devoid of timing noise may be either 1) an intrinsically stable rotator with no variations in its timing behaviour or 2) transitioning between different spin-down rates but $\Delta \dot{\nu}$, $\tau$ or both, may not be large enough to cause the residuals to depart from the timing model on relatively short timescales.  If a pulsar is switching regularly between two different well-defined values of $\dot{\nu}$, and $\dot{\nu}$ and/or $\tau$ are small, one may measure a continuous $\dot{\nu}$ that is the average of the two values of $\dot{\nu}$.  The measured value of $\dot{\nu}$ would be biased towards the value of $\dot{\nu}$ in which the pulsar spends the majority of its time.  Our simulations of $\dot{\nu}$ transitioning pulsars clearly exhibit regions of parameter space for which the $\dot{\nu}$ variations are unmeasurable. In these cases we are able to model the TOAs using a single continuous value of $\dot{\nu}$ however we find the $\dot{\nu}$ value we measure can be discrepant from either of the individual true values by as much as 1 per cent.    

There is potential, with new generations of radio telescopes, that new examples of $\dot{\nu}$-variable pulsars will be revealed.  The Canadian Hydrogen Intensity Mapping Experiment (CHIME) is a transit telescope under construction and will offer daily monitoring of many pulsars, with all Northern sky pulsars observed once per $\sim$10 days. A pulsar will be seen by CHIME for up to 10 minutes per transit depending on the declination of the source \citep{n17}.  Similarly, the large collecting area and wide field-of-view of the Molonglo Observatory Synthesis Telescope (UTMOST) is set to increase the number of Southern hemisphere pulsars for which daily timing is possible - up to 500 at full sensitivity \citep{bjf+17}.  Such high cadence monitoring of a large number pulsars is unprecedented and depending on the telescopes' sensitivities, could reveal a much larger sample of sources whose frequency derivatives are not stable that undergo transitions on $\tau > 1$ day timescales.  For example, PSR J1634$-$5107 in which profile variations occur on a timescale of $\sim$10 days (see Table \ref{nullers}) may show coincident $\dot{\nu}$-variations when observed daily, depending on sensitivity.  


The Square Kilometre Array (SKA), when online, will deliver high sensitivity, hence is capable of detecting individual pulses for many pulsars, resulting in vast improvements on current precision timing efforts. However, our simulations show that a nominal observing cadence of once per 2 weeks is unlikely to offer any considerable improvement in the resolution of discrete pulsar spin-down states. Instead, pulsar observations dedicated to understanding the state-switching phenomena should be configured and scheduled based on known and predicted switching timescales (minutes to years).  The SKA is set to be highly configurable in this sense, offering the potential for high-cadence monitoring of multiple sources by sub-arraying (see \cite{sks+08} for a review). In particular, the wide field-of-view of SKA1-LOW will allow simultaneous monitoring of up to 16 pulsars at full sensitivity.  Cadence could be enhanced further by splitting the array into a number of sub-arrays each with a number of tied-array beams (TABS).  In this mode, a larger number of pulsars could be observed more frequently but the full sensitivity of SKA1-LOW is sacrificed in favour of cadence.  Conversely, using the full SKA1-LOW sensitivity to observe fewer pulsars necessarily means a compromise on cadence due to the large number of pulsars in routine timing programs and competition with other science goals. In such a mode, the transitions that are potentially detectable are shorter and smaller but cadence must be higher than the switching timescale for an individual transition to be resolvable.  The key point here is that there is no simple trade-off between cadence and sensitivity. The optimal scenario of frequent and highly sensitive monitoring could be achieved however, for a subset of pulsars as part of a dedicated program to resolve $\dot{\nu}$-variations.  

Frequently observing a large number of pulsar with a highly sensitive instrument will likely reveal new examples of sources which exhibit a variable $\dot{\nu}$. This would possibly include PSR B0823+26 (see Table \ref{nullers}) which may exhibit peak-to-peak $\dot{\nu}$-variations of up to 6 per cent if several TOAs per day can be obtained and sensitivity improves by a factor of 10.  When observed with the SKA, it is expected that many pulsars, for which the spread of timing residuals is currently dominated by TOA uncertainty, will begin to show low amplitude timing noise \citep{wex+15}, symptomatic of a varying frequency derivative.  This may include more examples similar to PSR B1828-11 (discussed above) whose rapid switching timescale shows longer term modulation of the time spent in any one state.  Such pulsars can then be added to a list of sources in a targeted $\Delta \dot{\nu}$ study and afforded the appropriate cadence in order to determine $\dot{\nu}$ irregularities.  In should be noted that high sensitivity does not necessarily mean that $\dot{\nu}$-variations of any small amplitude or timescale can be detected.  The formation of a TOA relies on there being a sufficient number of pulses from which to form a stable integrated pulse profile.  This typically takes several minutes and so this sets a fundamental limit on the achievable cadence for a given pulsar.  In such pulsars, we may not be sensitive to subtle pulse profile changes that occur on timescales less than the required integration time. Additionally, the measurement of $\dot{\nu}$, requires a number of TOAs to be present in a spin-down state and so the cadence needs to be several times greater than the switching timescale.  

In addition to the improved timing of known pulsars, the SKA is predicted to increase the number of known Galactic pulsars by an order of magnitude \citep{kbk+15}. This will reveal many new examples of nulling/intermittent and mode-switching pulsars, yielding improved statistics on $\dot{\nu}$-variations both with and without associated pulse profile changes.  Although small amplitude, short term variations may not be resolvable with the  improved sensitivity of the SKA unless the observing cadence is carefully configured, better limits on the radio emission in the off-states of known intermittent pulsars can be established, from which a better understanding of the magnetospheric conditions in each spin-down state may be possible.  

\section{Conclusions}

We have quantified our capability to detect $\Delta \dot{\nu}$ events in radio pulsars by simulating pulse times-of-arrival based on a simple two-state switching model. We have shown that detection capability is comparable with the analytically derived limit given by Equation \ref{chi2_2}.  Our simulations have also shown that current observing setups may be blind to a diversity of $\dot{\nu}$-switching phenomena.  It follows that where a telescope with superior sensitivity to current efforts is available (e.g. the SKA), cadence cannot be compromised in studies dedicated to resolving $\dot{\nu}$-transitions. Conversely, new facilities such as CHIME and UTMOST may reveal new examples of $\dot{\nu}$-variable pulsars that have switching timescales of the order of several days with their ability to track hundreds of sources on a daily basis.

We have also demonstrated that inference of the epochs at which $\dot{\nu}$-transitions occur from coincident changes to the pulse profile is advantageous only near the limits of detectability, where two transitions exist in a dataset.  Elsewhere in $\Delta \dot{\nu}-\tau$ parameter space, transitions can be detected blindly using a number of trial timing solutions whether or not pulse profile changes are detectable. Where three transitions exist, pulse profile changes allow us to avoid ambiguities in the number of transitions that have taken place. However we have noted that regardless of the transition amplitudes or timescales, mode-switching behaviour still reveals the number of transition that have occurred in a dataset, thereby reducing ambiguity when fitting for odd numbers of transitions.  

Future studies of state-switching pulsars with new generations of radio telescopes are promising, in that many existing pulsars will be revealed to show $\dot{\nu}$-variations on timescales as low as minutes and surveys will increase the number of pulsars known to exhibit these phenomena.  If we can measure the $\dot{\nu}$ value in each spin-down state of a pulsar there is the potential to reduce or perhaps eliminate timing noise, thereby making more pulsars available to high precision pulsar timing efforts, such as those being undertaken to detect gravitational waves.

\section*{acknowledgments}

The authors thank the anonymous reviewer for their helpful and constructive comments that improved the quality and clarity of this manuscript. Pulsar research at the Jodrell Bank Centre for Astrophysics is supported by a consolidated grant from the STFC in the UK.




\bibliographystyle{mnras}
\bibliography{journals,psrrefs,modrefs} 



\appendix

\section{Derivation of $\langle \chi^2 \rangle$ for one, two and three $\dot{\nu}$ transitions.}

\begin{figure}
    \includegraphics[width=\columnwidth]{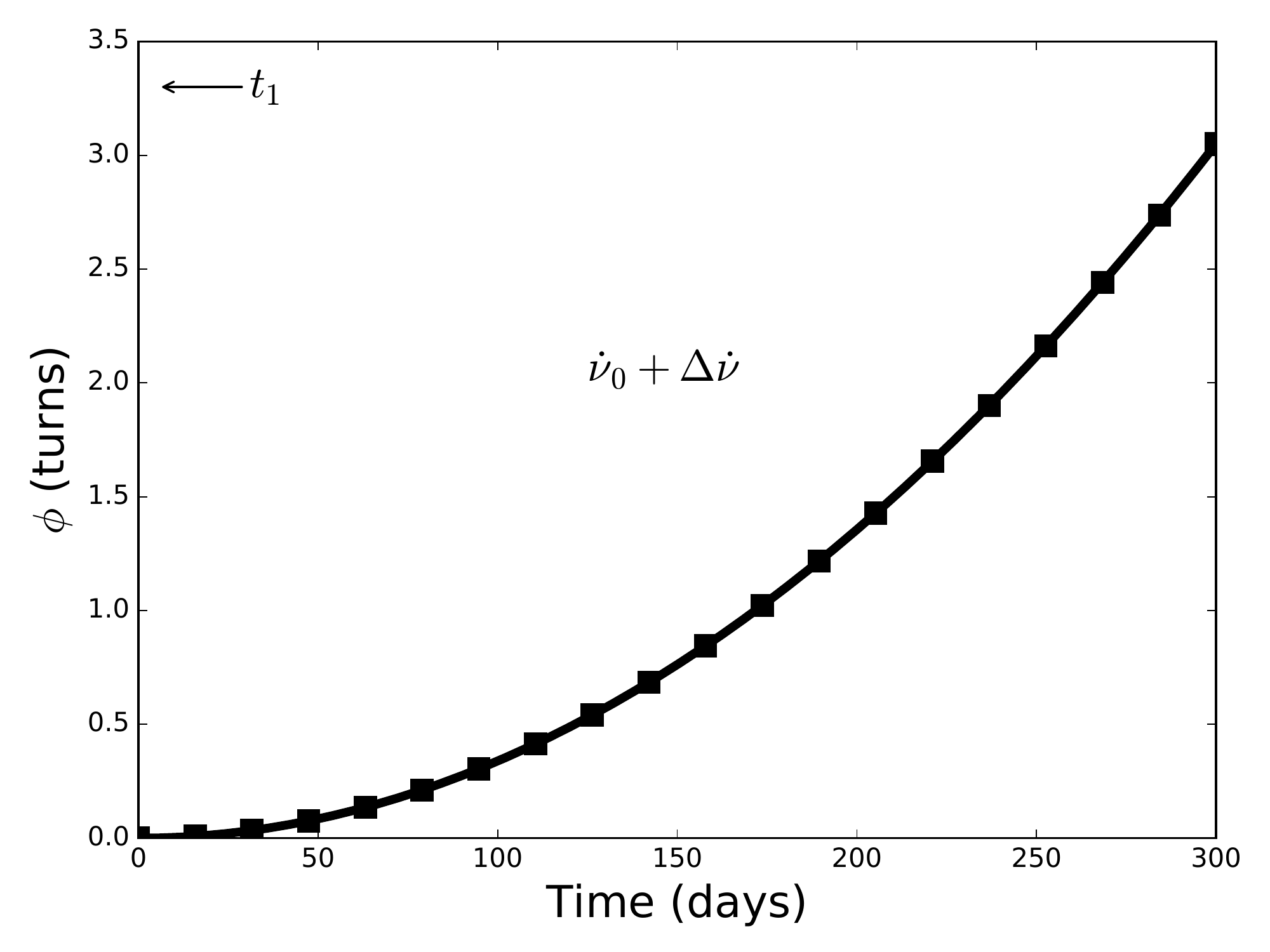} \\
    \includegraphics[width=\columnwidth]{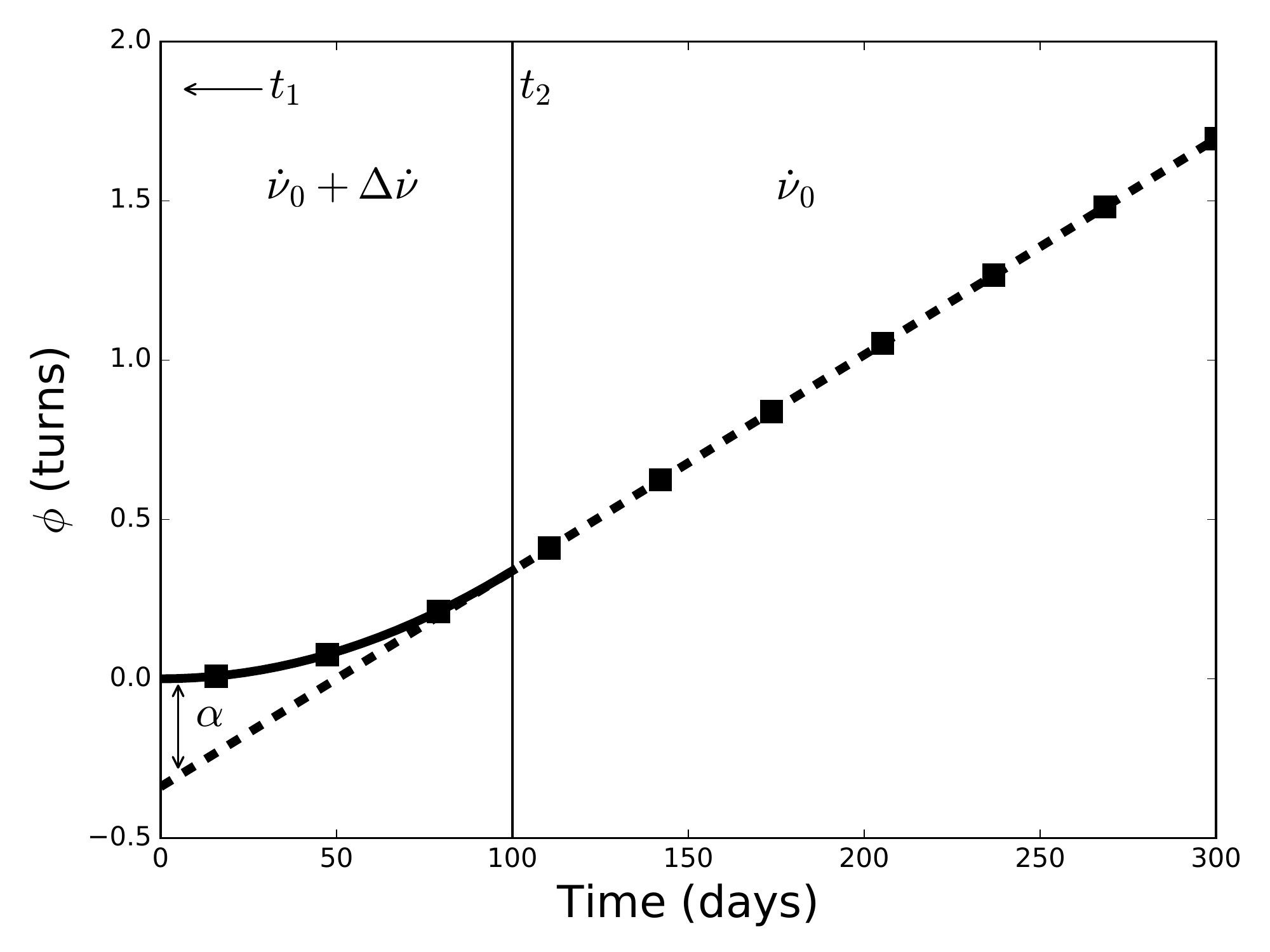} \\
    \includegraphics[width=\columnwidth]{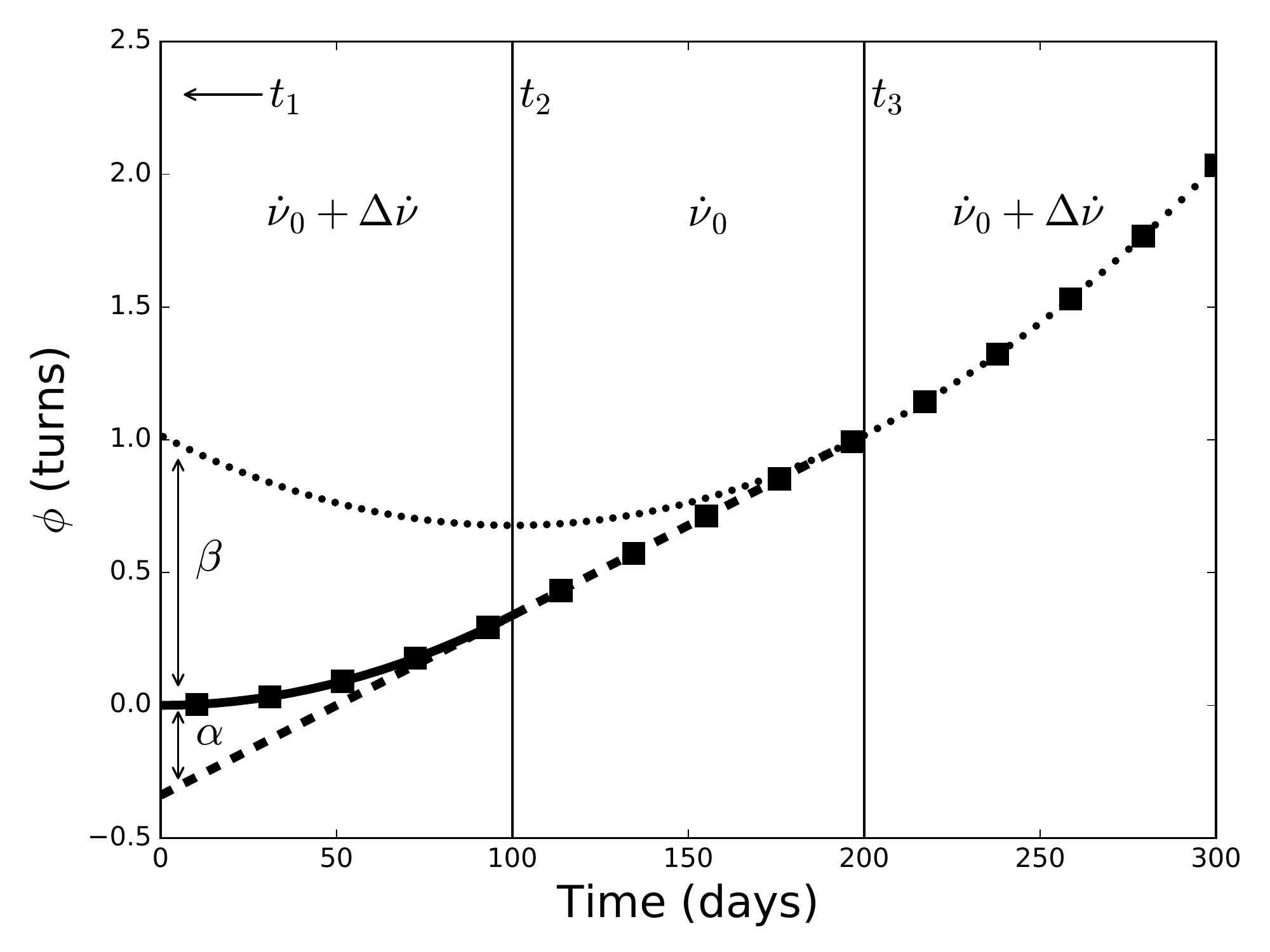}
    \label{functions}
    \caption{Plots showing the functions that underly the effects on the residuals of one (upper), two (centre) and three (lower) transitions. In all cases, the first transition occurs at $t_1$ ($t=0$). For one transition, the post $t_1$ residuals follow a parabolic function with no gradient term and a vertex at $t=0$. If the pulsar transitions back its initial $\dot{\nu}$ state at $t_2$, the residuals then follow a linear function (dashed line) until the dataset ends (centre) or a later transition occurs at $t_3$ (lower). In the latter case the post $t_3$ residuals follow a parabolic function again (dotted line).  The quantities $\alpha$ and $\beta$ are constants (see text).  The black squares are simulated timing residuals based on the relevant number of transitions for $\tau = 100$ days and $\Delta \dot{\nu} = 75$ per cent. the inital $\dot{\nu} = -1.21 \times 10^{-14}$ Hz s\textsuperscript{-1}}
\end{figure}

A pulsar's timing model predicts the arrival time of every one of its pulses. The expected value of the $\chi^2$ of the timing residuals is given by

\begin{equation}
    \label{wchi2} \langle \chi^2 \rangle = \sum_{\mathrm{i = 1}}^{N} \left( \frac{\phi_{\mathrm{i}} - \phi_{\mathrm{i,pred}}}{\bar{\sigma}_{\phi}}  \right)^2, 
\end{equation}

\noindent which, if the model accurately describes the pulsar's rotation, equates to zero. This is because each of the measured $N$ timing residuals $\phi_{\mathrm{i}}$ will be close to the value predicted by the timing model $\phi_{\mathrm{i,pred}}$ which itself, by definition, is always equal to zero.  If the pulsar exhibits some unmodelled rotational behaviour the magnitude of $\phi_{\mathrm{i}}$  will increase, causing the $\chi^2$ of the timing residuals to rise. As $\dot{\nu}$ transitions induce a quadratic increase in the magnitude of the residuals over time, we are able to predict the value of a residual at some time $t$ after a (set of) $\dot{\nu}$ transition(s). 

Where a single $\dot{\nu}$ transition occurs in a dataset (see Figure A1, upper panel), the value of a residual at time $t$ after the event is given by

\begin{equation}
    \phi = \frac{|\Delta \dot{\nu}| t^2}{2} \label{phase2}.
\end{equation}

\noindent Substituting Equation \ref{phase2} into Equation \ref{wchi2}, neglecting $\phi_{\mathrm{i,pred}}$ and integrating over all possible values of $t$, we obtain

\begin{equation}
    \label{wchi2sub} \langle \chi^2 \rangle = \int_{0}^T \left( \frac{|\Delta \dot{\nu}| t^2}{2 \bar{\sigma}_\phi} \right)^2 \frac{dt}{C} = \frac{|\Delta \dot{\nu}|^2 T^5}{20 \bar{\sigma}_\phi C}, \quad (T > 0),
\end{equation}

\noindent where $t=0$ refers to the epoch of the transition and $T$ is the total time covered since the transition. $C$ refers to the average interval between observation (the cadence) in seconds.

If after some time $\tau$ (at $t_2$, Figure A1, centre panel) the pulsar transitions back to the initial $\dot{\nu}$ (so that there are now two transitions in the dataset), the residuals will only follow the quadratic function in Equation \ref{phase} between $t_1 \leq t \leq t_2$, shown by the solid line in Figure A1.  Between $t_2$ and the end of the dataset the residuals follow a linear function (dashed line) that is given by $\Delta \nu t + \alpha$, where $\Delta \nu = |\Delta \dot{\nu}| \tau$ is the discrepancy in the pulsar's spin-frequency that occurs because the pulsar previously assumed a value of $\dot{\nu}$ that was not in the timing model.  The constant $\alpha$ is the value of the linear function at $t=0$ and is given by $-0.5 |\Delta \dot{\nu}| \tau^2$. Therefore the expected $\chi^2$ of the residuals for the case of two transitions is

\begin{equation}
     \begin{aligned}
     \label{wchi2sub2} \langle \chi^2 \rangle ={} & \frac{1}{\bar{\sigma}_\phi^2 C} \Bigg[\int_{0}^\tau \left(\frac{|\Delta \dot{\nu}| t^2}{2} \right)^2 dt + \int_{\tau}^T (\Delta \nu t + \alpha)^2 dt \Bigg] \\
                                   & = \frac{1}{\bar{\sigma}_\phi^2 C} \Bigg[ \frac{|\Delta \dot{\nu}|^2 \tau^5}{20} + \left( \frac{(\Delta \nu T + \alpha)^3}{3 \Delta \nu} - \frac{(\Delta \nu \tau + \alpha)^3}{3 \Delta \nu}\right) \Bigg], \\
                                   & (T > \tau).
     \end{aligned}                           
\end{equation}

\noindent Solving and substituting the expressions for $\alpha$ and $\Delta \nu$, Equation \ref{wchi2sub2} simplifies to 


\begin{equation}
    \label{wchi2sub2simp} \langle \chi^2 \rangle = \frac{|\Delta \dot{\nu}|^2 \tau^2}{60 \bar{\sigma}^2_{\phi} C} \left( -2\tau^3 + 15\tau^2 T - 30\tau T^2 + 20 T^3 \right), 
\end{equation}

\noindent Note that this reduces to Equation \ref{chi2_1} when $T=\tau$ (i.e., when no time has elapsed after $\tau$). To compute whether or not a pulsar's timing residuals have suffered a significant departure by the time $\tau$, without considering any of the $t > \tau$ data points, one can compute the contribution to $\langle \chi^2 \rangle$ only from the first integral of Equation \ref{wchi2sub2} and compare it to a $\chi^2_{\mathrm{crit}}$ for just $\tau/C$ data points.   

For three transitions we extend Equation \ref{wchi2sub2} to include a second period of time over which the residuals increase in amplitude quadratically. This is shown in Figure A1 (lower panel) as the region after the second vertical line ($t_3$) where a quadratic function (dotted line) meets the linear function at $t=2\tau$. In this case, the $\langle \chi^2 \rangle$ of the timing residuals is given by

\begin{equation}
    \begin{aligned}
    \label{wchi2sub3}  \langle \chi^2 \rangle ={} & \frac{1}{\bar{\sigma}_\phi^2 C} \Bigg[\int_{0}^\tau \left(\frac{|\Delta \dot{\nu}| t^2}{2} \right)^2 dt + \int_{\tau}^{2\tau} (\Delta \nu t + \alpha)^2 dt \\
                                  & + \int_{2\tau}^{\tau} \left( \frac{|\Delta \dot{\nu}|t^2}{2} - \Delta \nu t + \beta \right)^2 dt \Bigg], \\
                                  & (T > 2\tau).
    \end{aligned}
\end{equation}

\noindent Note that the upper limit on the second integral is now $2\tau$ as we impose in our simulations that equal time intervals elapse between each transition.  The quantity $\beta$ represents the point at which the function that describes the second parabola crosses $t=0$ (see Figure A1, lower panel) and is given by $(3/2)|\Delta \dot{\nu}| \tau^2$. Solving the integral and substituting in expression for $\alpha$, $\beta$ and $\Delta \nu$, the $\langle \chi^2 \rangle$ for the residuals due to three transitions is

\begin{equation}
    \label{wchi2sub3simp}  
         \begin{aligned}
                     \langle \chi^2 \rangle = & \frac{1}{\bar{\sigma}^2_{\phi} C} \Bigg( \frac{17}{15} |\Delta \dot{\nu}|^2 \tau^5 + \frac{9}{4}  |\Delta \dot{\nu}|^2 \tau^2 (T - 2\tau)  \\
                                            &  + \frac{5}{6} |\Delta \dot{\nu}|^2 \tau^2 (T^3 - 8\tau^3) - \frac{3}{2} |\Delta \dot{\nu}|^2 \tau^3 (T^2 - 4\tau^2)  \\
                                            & - |\Delta \dot{\nu}|^2 \tau (0.25T - 4\tau^4) + \frac{|\Delta \dot{\nu}|^2}{20} (T^5 - 32\tau^5) \Bigg). 
         \end{aligned}
\end{equation}

\noindent When $T=2\tau$, (i.e., no data is obtained after the final transition) the solution to Equation \ref{wchi2sub3simp} is the same as that obtained by setting $T=2\tau$ in Equation \ref{wchi2sub2simp} as the two scenarios are equivalent. Also note that when $\tau =0$, we recover Equation \ref{wchi2sub} as the residuals become affected only by the last transition (see Section \ref{three_trans_limit}). 

\bsp	
\label{lastpage}
\end{document}